\documentclass[11pt,a4paper]{article}

\pdfoutput=1
\usepackage{jheppub}
\usepackage[T1]{fontenc}
\usepackage[ansinew]{inputenc}
\usepackage[english]{babel}
\usepackage{color}
\usepackage{epsfig, palatino}
\usepackage{pstricks,pst-node,pst-tree}
\usepackage{epic}
\usepackage{mathrsfs}
\usepackage{ae} 
\usepackage{amsmath}
\usepackage{autobreak}
\usepackage{amssymb}
\usepackage{graphicx}
\usepackage{color}
\definecolor{darkblue}{cmyk}{0.9,0.9,0,0}
\usepackage{wasysym}
\usepackage{varioref}
\usepackage{makeidx}
\usepackage{simplewick}
\usepackage{array}
\usepackage{mathrsfs}
\usepackage{feynmf} 
\usepackage{tikz}
\usepackage{dsfont}
\usepackage{slashed}
\usepackage{float}
\usepackage{subcaption}
\usepackage{lscape}
\usepackage{afterpage}
\usepackage{soul}

\makeatletter
\DeclareRobustCommand*{\bfseries}{%
  \not@math@alphabet\bfseries\mathbf
  \fontseries\bfdefault\selectfont
  \boldmath
}

\makeatother
\usepackage[textsize=footnotesize]{todonotes}

\usetikzlibrary{decorations.pathmorphing}

\newcommand{\comment}[1]{}

\newcommand{\beq}{\begin{equation}}
\newcommand{\eeq}{\end{equation}}
\newcommand{\beqq}{\begin{equation*}}
\newcommand{\eeqq}{\end{equation*}}
\newcommand\beqa{\begin{eqnarray}}
\newcommand\eeqa{\end{eqnarray}}
\newcommand\beqaa{\begin{eqnarray*}}
\newcommand\eeqaa{\end{eqnarray*}}
\newcommand\bea{\begin{array}}
\newcommand\eea{\end{array}}

\newcommand{\neqa}{\nonumber\end{eqnarray}}

\renewcommand{\d}{\partial}

\newcommand{\<}{{\langle}}
\renewcommand{\>}{{\rangle}}

\newcommand{\cA}{{\cal A}}

\newcommand{\re}{\relax{\rm I\kern-.18em R}}

\renewcommand{\sp}{p\hspace{-.40em}/}

\def\XXint#1#2#3{{\setbox0=\hbox{$#1{#2#3}{\int}$}
\vcenter{\hbox{$#2#3$}}\kern-.5\wd0}}

\def\[{\left[}
\def\]{\right]}

\def\({\left(}
\def\){\right)}
\def\[{\left[}
\def\]{\right]}

\def\<{\langle}
\def\>{\rangle}

\def\i2{\frac{i}{2}}

\def\spi{\relax{\rm \pi\kern-0.5em /}}
\def\sA{\relax{\rm A\kern-0.5em /}}
\def\sp{\relax{\rm p\kern-0.5em /}}
\def\sd{\relax{\rm \d\kern-0.5em /}}
\def\sk{\relax{\rm k\kern-0.5em /}}
\def\sn{\relax{\rm n\kern-0.5em /}}
\def\sl{\relax{\rm l\kern-0.5em /}}
\def\sP{\relax{\rm P\kern-0.7em /}}
\def\sBethe{\relax{\rm \Bethe\kern-0.5em /}}

\def\2F1{\,_2{\rm F}_1}

\def \Ascr {\cA^{\,^{\kern-0.2em (\infty)}}}

\makeindex

\numberwithin{figure}{section}
\title{Wilson Loop Duality and OPE for Super Form Factors of Half-BPS Operators}

\author[a]{Benjamin Basso}
\author[a,b]{and Alexander G. Tumanov}

\affiliation[a]{Laboratoire de Physique de l'Ecole Normale Sup\'erieure, ENS, Universit\'e PSL, CNRS, Sorbonne Universit\'e, Universit\'e de Paris, 24 rue Lhomond, F-75005 Paris, France}
\affiliation[b]{Institut Philippe Meyer, Ecole Normale Sup\'erieure, 24 rue Lhomond, F-75005 Paris, France}

\abstract{We propose a dual Wilson loop description for the MHV super form factors of half-BPS operators in planar $\mathcal{N}=4$ super-Yang-Mills theory. In this description, the local operators are represented by on-shell states, made out of zero-momentum particles, that are absorbed by a null periodic super Wilson loop. We present evidence for this duality at weak coupling, by performing an explicit calculation of the Wilson loop matrix elements through one loop. At tree level, the interactions localize at the cusps of the loop, revealing a simple connection between the super form factors and the $m=2$ tree amplituhedron. At loop level, we show that the Wilson loop calculation reproduces the known results for the super form factors. Inspired by this duality, we extend the OPE program developed for the form factors of the Lagrangian to the super form factors of the higher-charge operators. We introduce non-perturbative axioms and conjectures for the main building blocks that govern the exchange of the lightest flux-tube excitations. These blocks appear as simple refinements of the form factor transitions introduced in earlier OPE studies. They are expressed at any value of the 't Hooft coupling in terms of the tilted Beisert-Eden-Staudacher kernel. We carry out checks of our conjectures up to two loops at weak coupling for three- and four-point form factors of half-BPS operators of various lengths, finding perfect agreement with perturbative data.}

\begin{document}
\maketitle

\newpage

\section{Introduction}

In the last decade, a lot of progress has been made in the study of scattering amplitudes in planar  $\mathcal{N}=4$ Super-Yang-Mills (SYM) theory. This progress was driven to a large extent by the discovery of the amplitude / Wilson loop duality, which relates, in the simplest case, the maximally-helicity violating (MHV) $n$-gluon amplitude and the vacuum expectation value of a null $n$-sided polygonal Wilson loop,
\begin{equation}
\mathcal{A}^{\textrm{MHV}}_{n}(p_{1}, \ldots , p_{n}) / \mathcal{A}_{n}^{\textrm{tree}}(p_{1}, \ldots , p_{n}) = \langle 0| W_{n}(x_{1}, \ldots , x_{n}) |0\rangle\, ,
\end{equation}
where $x_i = p_{i}-p_{i-1}$ are the dual coordinates of the vertices of the polygon and $p_i$ are the light-like gluon momenta. The duality was first observed at strong coupling, on the string theory side of the AdS/CFT correspondence, where it appeared as a manifestation of the T-duality symmetry of the string sigma model~\cite{Alday:2007hr}, mapping a color-ordered sequence of on-shell open string states at the horizon of the Anti-de-Sitter space to the contour of a null polygonal Wilson loop at the boundary. It was sharpened at weak coupling on the gauge theory side, through the study of the conformal properties of planar amplitudes in the dual (region momentum) coordinates $x_i$~\cite{Drummond:2007aua} and through various perturbative comparisons with the Wilson loop computations~\cite{Brandhuber:2007yx,Drummond:2007cf,Drummond:2007au,Bern:2008ap,Drummond:2008aq,Drummond:2008vq,Caron-Huot:2010ryg,Mason:2010yk}. Besides putting tight restrictions on the analytic properties of scattering amplitudes~\cite{Drummond:2007au,Drummond:2008vq,Caron-Huot:2011dec}, the duality enabled the development of new methods for their calculation at higher loops. Among them is the Operator Product Expansion (OPE) for null Wilson loops~\cite{Alday:2010ku,Basso:2013vsa}, which provides an integrable description of scattering amplitudes in terms of the excitations of the flux tube sourced by the Wilson loop.\par
A remarkable property of the T-duality transformation is that it is expected to be an exact symmetry of the full superstring sigma model in the planar limit~\cite{Berkovits:2008ic}. As a consequence, the duality is believed to hold true at any value of the 't Hooft coupling constant, $g^2 = g^{2}_{\textrm{{\tiny YM}}} N/(4\pi)^2$. It is also expected to provide a general state-operator dictionary relating on-shell matrix elements to null polygonal Wilson loops. In particular, the Wilson loop description should apply to the form factors of local operators $\mathcal{O}(x)$,
\begin{equation}
\mathcal{F}_{\mathcal{O}}(p_{1}, \ldots, p_{n}; q) = \int d^{4}x \, e^{-iq\cdot x} \langle p_{1}, \ldots , p_{n} |\mathcal{O}(x) |0\rangle\, ,
\end{equation}
interpolating between the closed and open string sectors of the theory. In contrast with amplitudes, for form factors the total momentum of the state is not conserved, but adds up to the total momentum of the operator, $\sum_{i=1}^{n}p_{i} = q$. It follows that the dual Wilson loop cannot be closed. Instead, it has been proposed~\cite{Alday:2007he,Maldacena:2010kp,Brandhuber:2010ad,Brandhuber:2011tv,Bork:2014eqa,Ben-Israel:2018ckc,Bianchi:2018rrj,Sever:2020jjx} that the T-duality transformation maps a form factor to an infinite periodic Wilson loop, whose period is given by the total momentum $q$.\par
The simplest and most studied example of this extended duality is given by the MHV form factors of the lowest protected operator, $\mathcal{T}_{2} \sim \textrm{Tr}\, \phi^2$, where $\phi$ is a complex scalar field, falling in the same half-BPS multiplet as the stress tensor and the on-shell chiral Lagrangian. These form factors were proposed to be dual to vacuum expectation values of periodic null Wilson loops. Evidence for this duality was gathered at strong coupling, using string theory~\cite{Alday:2007he,Maldacena:2010kp}, and at weak coupling, where a precise match was found through one loop~\cite{Brandhuber:2010ad}. More recently, this duality served as a basis for the development of the Form Factor OPE program~\cite{Sever:2020jjx,Sever:2021nsq,Sever:2021xga}, generalizing the Wilson loop OPE method used for amplitudes. This approach led to a wealth of new data for form factors, enabling their determination at higher loops using the powerful form factor function bootstrap~\cite{Dixon:2020bbt,Dixon:2021tdw,Dixon:2022xqh,Dixon:2022rse,Guo:2022qgv}.\par
Beyond the ${\rm Tr}\,\phi^2$ case, the dictionary between form factors and Wilson loops is still largely unknown. In this paper, we take the first step towards expanding this dictionary to more general operators by considering the next-to-simplest class of form factors, associated with the half-BPS operators, $\mathcal{T}_{k} \sim \textrm{Tr}\, \phi^{k}$, carrying R-charge $ k =2,3,\ldots\, $. Perturbative data for these form factors is available up to two loops in the MHV sector~\cite{Penante:2014sza,Brandhuber:2014ica}. In this paper, we will rewrite this data in dual coordinates, using the standard duality map, and show that the MHV form factors may be expressed in terms of suitable R-invariants, similar to those encountered in the study of Next-to-MHV (NMHV) amplitudes~\cite{Drummond:2008vq,Mason:2009qx}. In particular, at tree level, we will show that the MHV form factors of ref.~\cite{Penante:2014sza} provide a realization of the tree $m=2$ amplituhedron~\cite{Arkani-Hamed:2013jha,Arkani-Hamed:2017vfh}, with $k-2$ playing the role of the NMHV degree, in close analogy with recent findings for the correlation functions of half-BPS operators in twistor space~\cite{Caron-Huot:2023wdh}.\par
Based on these results, we will argue that form factors of half-BPS operators are dual to non-trivial matrix elements of periodic null super Wilson loops, in which the local operators are replaced by dual states built out of zero-momentum scalars. The supersymmetric extension of the Wilson loops~\cite{Caron-Huot:2010ryg,Mason:2010yk,Belitsky:2011zm} appears necessary for this construction, to match the R-charge of the zero-momentum scalars. We will see that this picture provides a natural explanation for the emergence of R-invariants. We will also perform a check of this duality at one loop.\par
Finally, building on this duality, we will extend the Form Factor OPE program to the family of half-BPS operators and present axioms and conjectures for the fundamental flux-tube transitions, describing the collinear limit of MHV form factors at any value of the coupling constant. We will verify these conjectures explicitly up to two loops and explain how to produce higher-loop predictions for the form-factor function bootstrap.\par
The paper is organized as follows. In section~\ref{sec:superFF}, we present the alternative descriptions for the MHV form factors of half BPS operators, both in momentum space and in dual coordinates. We carry out checks through one loop for the dual Wilson loop and detail the correspondence with the tree amplituhedron. Section~\ref{section:FFOPE_main} is dedicated to the flux-tube bootstrap of the form factor transitions, which are argued to follow from a mild modification of the earlier proposals for the stress-tensor multiplet. In section~\ref{section:matching_data}, we explain how to assemble all the pieces together and perform the comparison between OPE formulae and perturbative data. We conclude in section~\ref{section:conclusion} with comments and generalizations. A few appendices contain some details of the calculations.

\section{Super form factors and Wilson loops}\label{sec:superFF}

In this section, we present three different descriptions of form factors of BPS operators in planar $\mathcal{N}=4$ SYM. Firstly, we briefly recall their standard formulation in the spinor-helicity formalism. Secondly, we rewrite them in the dual coordinates, uncovering an underlying $m=2$ amplituhedron description. Lastly, we will reinterpret these results as matrix elements of periodic super Wilson loops with states composed of zero-momentum scalars.

\subsection{BPS operators and their form factors}\label{section:super_FF_def}

In $\mathcal{N}=4$ SYM, both the external particles and the local operators form non-trivial representations of the supersymmetry algebra. This situation naturally leads one to consider super form factors, generating helicity components of form factors of local operators and their super-descendants. In this paper, following refs.~\cite{Brandhuber:2011tv,Penante:2014sza}, we will consider such super form factors for the half-BPS operators.\par
The precise operator that we will study is the so-called chiral part of the half-BPS superfield $\mathcal{T}_{k}(x, \theta) = \mathcal{T}_{k}(x, \theta, \bar{\theta} = 0)$~\cite{Penante:2014sza,Eden:2011yp,Eden:2011ku}. It is defined by acting with the chiral supercharges $Q^{\alpha A}$ on the dimension-$k$ chiral primary operator,
\begin{equation}\label{eq:calT-k}
    \mathcal{T}_{k}(x, \theta) = e^{\theta_{\alpha A} Q^{\alpha A}}\cdot \textrm{Tr} \, \phi(x)^k = \textrm{Tr} \, \phi(x)^k + \theta_{\alpha A} \left[Q^{\alpha A}, \textrm{Tr} \, \phi(x)^k\right] + \ldots \, ,
\end{equation}
where $\theta_{\alpha A}$ are (anticommuting) Grassmann variables and where $\phi$ is a generic complex scalar field of the theory. Without loss of generality, we may choose $\phi = \phi_{34}$, where $\phi_{AB} = -\,\phi_{BA} = \frac{1}{2}\epsilon_{ABCD} \bar{\phi}^{CD}$ denote the 6 scalar fields of the theory, with $A, B = 1, \ldots , 4$. Here, $\epsilon_{ABCD}$ is the totally antisymmetric tensor, with $\epsilon_{1234} = 1$.\par
Since the chiral primary operator is annihilated by half of the supercharges, $\mathcal{T}_{k}(x, \theta)$ is a function of half of the Grassmann variables only. In particular, for our choice of $\phi$, $\mathcal{T}_{k}$ depends on $\theta_{\alpha a'}$, with $a' = 3,4$, but not on $\theta_{\alpha a}$, with $a=1,2$. Accordingly, the Taylor series in eq.~\eqref{eq:calT-k} terminates at order $\mathcal{O}(\theta^4)$, with the top component being a Lorentz invariant operator carrying dimension $k+2$ and R-charge $k-2$. When $k=2$, we are dealing with the chiral part of the stress-tensor multiplet and the top component is proportional to the on-shell chiral Lagrangian $\textrm{Tr}\, \mathcal{L}$~\cite{Eden:2011yp,Eden:2011ku}. At higher $k$, the top operator becomes significantly more complicated, but, based on its quantum numbers, we may schematically represent it as $\textrm{Tr} \, \mathcal{L}\, \phi^{k-2}$.\par
The super form factors are defined as the super Fourier transform of the on-shell matrix elements of $\mathcal{T}_{k}(x, \theta)$. They read~\cite{Brandhuber:2011tv,Penante:2014sza}
\begin{equation}\label{eq:Fourier-Fkn}
   \mathcal{F}_{k,n}\left(1,\ldots,n;q,\gamma\right) = \int d^{4}x d^{4}\theta \, e^{-iq_{\mu} x^{\mu} - \gamma^{\alpha a'}\theta_{\alpha a'}} \langle1,\ldots,n|\mathcal{T}_{k}(x, \theta)|0\rangle\ ,
\end{equation}
where $q$ and $\gamma$ are, respectively, the momentum and super-momentum of the operator. The external state is a generic product of (super) momentum eigenstates,
\begin{equation}
    \langle 1, \ldots , n | = \langle \Phi(p_{1}, \tilde{\eta}_{1}), \ldots , \Phi(p_{n}, \tilde{\eta}_{n})|\ ,
\end{equation}
carrying momentum $P = \sum_{i=1}^{n}p_{i}$ and super momentum $Q^{\alpha A} = \sum_{i=1}^{n} \lambda_{i}^{\alpha} \tilde{\eta}^{A}_{i}$. Here, $\lambda_i^{\alpha}$ and $\tilde{\lambda}_i^{\dot{\alpha}}$ are the spinor-helicity variables associated to the null momenta, $p_{i}^{\alpha\dot{\alpha}} = p_{i}^{\mu} \sigma_{\mu}^{\alpha\dot{\alpha}} = \lambda^{\alpha}_{i}\tilde{\lambda}_{i}^{\dot{\alpha}}$, with $\sigma^{\mu} = (I, \vec{\sigma})$ and where $\vec{\sigma}$ are the Pauli matrices. The various helicity components are generated through the super-wave function~\cite{Nair:1988bq}
\begin{equation}
\Phi(p,\tilde{\eta}) = g^+(p) + \tilde{\eta}^A\psi_A(p)+\frac{\tilde{\eta}^A\tilde{\eta}^B}{2!}\,\phi_{AB}(p)+\epsilon_{ABCD}\frac{\tilde{\eta}^A\tilde{\eta}^B\tilde{\eta}^C}{3!}\,\bar{\psi}^D(p)+\tilde{\eta}^1\tilde{\eta}^2\tilde{\eta}^3\tilde{\eta}^4\,g^-(p)\ ,
\end{equation} 
where $\tilde{\eta}^{A}$, with $A=1,2,3,4$, are Grassmann variables transforming in the fundamental representation of the R-symmetry group $SU(4)$. Here, $g^{+}$ denotes the positive helicity gluon, $\psi_A$ the gauginos, $\phi_{AB} = -\,\phi_{BA}$ the 6 scalars, etc.\par
The super form factors~\eqref{eq:Fourier-Fkn} share a lot of similarities with scattering amplitudes. In particular, like scattering amplitudes, they are subject to super Ward identities that require them to live on the support of fermionic delta functions~\cite{Brandhuber:2011tv}. The simplest example corresponds to the MHV form factor of the stress-tensor multiplet $\mathcal{T}_{2}$. At tree level, it is given by a Parke-Taylor-Nair-like term~\cite{Brandhuber:2010ad,Brandhuber:2011tv}
\begin{align}
\mathcal{F}^{\text{MHV tree}}_{2,n}\left(1,\ldots,n;q,\gamma^{+}\right) = \frac{\delta^{(4)}\left(q-\sum\limits_{i=1}^n\lambda_i\tilde{\lambda}_i\right)\delta^{(4)}\left(\gamma^+-\sum\limits_{i=1}^n\lambda_i\tilde{\eta}^+_{i}\right)\delta^{(4)}\left(\sum\limits_{i=1}^n\lambda_i\tilde{\eta}^-_{i}\right)}{\langle12\rangle\,\langle23\rangle\,\ldots\,\langle n1\rangle}\ ,
\end{align}
where $\langle i j\rangle = \epsilon_{\alpha\beta}\lambda_{i}^{\alpha}\lambda_{j}^{\beta}$ is the usual antisymmetric spinor product, with $\epsilon_{12} =- \epsilon_{21} = 1$. Here, we introduced the notations $\tilde{\eta}_{i}^{+} = \tilde{\eta}_{i}^{a'}$ and $\tilde{\eta}^{-}_{i} = \tilde{\eta}^{a}_{i}$, and similarly for the super momentum $\gamma$, with $\gamma^{-} \equiv 0$. Non-MHV form factors look similar, but they come dressed with polynomials of higher degrees in the Grassmann variables $\tilde{\eta}_{i}^{\pm}$~\cite{Brandhuber:2011tv,Penante:2014sza,Bork:2014eqa,Bianchi:2018rrj}. In general, to study form factors of the operators $\mathcal{T}_{k}$, it is convenient to strip off the above universal factor and consider the ratio
\begin{align}\label{FFnormalized}
W_{k,n}\left(1,\ldots,n;q,\gamma^+\right) \equiv \frac{\mathcal{F}_{k,n}\left(1,\ldots,n;q,\gamma^+\right)}{\mathcal{F}^\text{MHV tree}_{2,n}\left(1,\ldots,n;q,\gamma^+\right)}\ .
\end{align}
with $q \equiv \sum_{i=1}^{n}\lambda_{i}^{\alpha}\tilde{\lambda}_{i}^{\dot{\alpha}}$ and $\gamma^{+} \equiv \sum_{i=1}^{n} \lambda_i \tilde{\eta}_{i}^{+}$.\par
The MHV form factors stand for the contributions to~\eqref{FFnormalized} with the lowest possible degree in the $\tilde{\eta}$ variables. To determine their form, one may use that these form factors describe, in components, the decay of $\textrm{Tr}\, \phi^{k}$ into $k$ scalars and $n-k$ positive-helicity gluons, or any other supersymmetrically-equivalent process. Hence, they yield terms of degree $2\left(k-2\right)$ in the variables $\tilde{\eta}_{i}^{A}$, after stripping off the delta functions, with $A = 1,2$ by charge conservation. More precisely, $W^{\textrm{MHV}}_{k, n}$ is a homogeneous polynomial of degree $k-2$ in the $SU(2)$ invariant products $\tilde{\eta}^{-}_{i}\cdot \tilde{\eta}_{j}^{-} = \frac{1}{2}\epsilon_{ab} \tilde{\eta}^{a}_{i}\tilde{\eta}^{b}_{j}$, with $a, b = 1,2$.%
\footnote{In general, at N$^{k'}$MHV level, the form factor ratio $W_{k,n}$ has fermionic degree $2\left(k+2k'-2\right)$, of which $2\left(k+k'-2\right)$ are $\tilde{\eta}^{-}_i$, and $2k'$ are $\tilde{\eta}^{+}_i$.}
For illustration, in the $\mathcal{T}_{3}$ case, one has the simple tree-level formula~\cite{Penante:2014sza}%
\footnote{Note that our normalization differs by a factor of $(-1)^k$ as compared to the one in ref.~\cite{Penante:2014sza}. This sign ensures that the matrix elements $\langle \phi_{12}^k|\textrm{Tr}\, \phi_{34}(0)^k|0\rangle = 1$ at tree level, for any $k$.}
\begin{align}\label{F3n}
W_{3, n}^{\text{MHV tree}} &= -\sum\limits_{1\leqslant i<j \leqslant n}\langle ij\rangle\,\tilde\eta^-_{i}\cdot\tilde\eta^-_{j}\ .
\end{align}
The general formula for any $k$ may be found in ref.~\cite{Penante:2014sza}. We will provide an expression for it in the next section, after introducing dual variables to simplify it. In this paper, we will focus on the form factors in the MHV sector and, for the sake of brevity, will be omitting the MHV labels.

\subsection{Dualization and the $m=2$ amplituhedron}

To gain insight into the structure of the super form factors~\eqref{FFnormalized}, we may use the formalism developed for scattering amplitudes and introduce dual variables~\cite{Drummond:2008vq}%
\footnote{We use the same letters $(x, \theta)$ for the dual variables and for the arguments of the local operator $\mathcal{T}_{k}(x, \theta)$, keeping in mind that they parametrize two different objects in two different spaces.}
\begin{align}\label{dual_variables_transform}
\lambda^{\alpha}_{i} \tilde{\lambda}_{i}^{\dot{\alpha}} = x_i^{\alpha\dot{\alpha}} - x_{i-1}^{\alpha\dot{\alpha}}\ ,\qquad \lambda_i^\alpha\tilde{\eta}_i^{A} = \theta_i^{\alpha A} - \theta_{i-1}^{\alpha A}\ ,
\end{align}
alongside the momentum supertwistors~\cite{Hodges:2009hk,Mason:2009qx}
\begin{align}\label{eq:super-Z}
\mathcal{Z}_i = \left(\begin{array}{c}
    \lambda_i^\alpha\\
    \mu_{i}^{\dot{\alpha}}\\
    \eta_i^{A}
\end{array}\right) = \left(\begin{array}{c}
    \lambda_i^\alpha\\
    x_{i}^{\alpha\dot{\alpha}}\,\lambda_{i\alpha}\\
    \theta_i^{\alpha A}\,\lambda_{i\alpha}
\end{array}\right)\, ,
\end{align}
where the indices are lowered using the Levi-Civita tensor, $\lambda_{\alpha} = \epsilon_{\alpha\beta} \lambda^{\beta}$. In the amplitude case, the variables $(x_{i}, \theta_{i})$ play the role of the super-coordinates of the cusps of the dual polygonal Wilson loops, whereas the momentum twistors provide a convenient parametrization of their null edges. The same can be said for form factors~\cite{Brandhuber:2011tv,Bork:2014eqa}. However, one notes that in the latter case the dual variables are not invariant under $i\rightarrow i+n$. Instead, they undergo the transformations
\begin{equation}\label{eq:shifts}
    x_{i+n} = x_{i} + q\, , \qquad \theta_{i+n} \rightarrow \theta_{i} + \gamma\, , \qquad \mathcal{Z}_{i+n} = \mathcal{Z}_{i} + \left(\begin{array}{c} 0 \\ q^{\alpha\dot{\alpha}} \lambda_{i \alpha} \\ \gamma^{\alpha A} \lambda_{i \alpha}\end{array}\right)\, ,
\end{equation}
with the shifts arising from the (super) momentum of the operator, $q, \gamma \neq 0$. These shifts indicate that the Wilson loops dual to form factors are not closed but periodic, as discussed in the next section. However, for the time being, we will put aside the Wilson loop interpretation and blindly apply the dual map to the form factors, as a mere change of variables from $(\lambda, \tilde{\lambda}, \tilde{\eta})$ to $(\lambda, \mu, \eta)$. To perform this task, it will be more convenient to use the following non-local relations,
\begin{equation}\label{eq:duality-transformation}
\begin{aligned}
&\tilde{\lambda}_i^{\dot{\alpha}} = \frac{\langle i-1 i \rangle\,\mu_{i+1}^{\dot{\alpha}} + \langle i+1 i-1 \rangle\,\mu_{i}^{\dot{\alpha}} +\langle i i+1 \rangle\,\mu_{i-1}^{\dot{\alpha}}}{\langle i-1i\rangle\,\langle ii+1\rangle}\ ,\\
&\tilde{\eta}_i^A = \frac{\langle i-1 i \rangle\,\eta_{i+1}^A + \langle i+1 i-1 \rangle\, \eta^{A}_{i} +\langle i i+1 \rangle\,\eta_{i-1}^A}{\langle i-1i\rangle\,\langle ii+1\rangle}\ ,\\
&x_i^{\alpha\dot{\alpha}} = \frac{\lambda_{i}^\alpha\,\mu_{i+1}^{\dot{\alpha}}-\lambda_{i+1}^\alpha\,\mu_{i}^{\dot{\alpha}}}{\langle ii+1\rangle}\ ,\qquad \theta_i^{\alpha A} = \frac{\lambda_{i}^\alpha\,\eta_{i+1}^A-\lambda_{i+1}^\alpha\,\eta_{i}^A}{\langle ii+1\rangle}\ ,
\end{aligned}
\end{equation}
which are well known to be equivalent to eqs.~\eqref{dual_variables_transform} and~\eqref{eq:super-Z}.\par
Actually, only a small subset of these relations is needed here. Indeed, as one can see from eq.~\eqref{F3n}, the tree-level MHV form factors $W_{k, n}$ are holomorphic functions of the spinor helicity variables: they depend only on $\lambda$, not $\tilde{\lambda}$. Moreover, as said earlier, they depend only on $\tilde{\eta}^{-}$, not on $\tilde{\eta}^{+}$. Dualizing the tree-level MHV form factors is thus a straightforward exercise. With the $\lambda$ variables being common to both descriptions, we only need to translate between $\tilde{\eta}^{a}$ and $\eta^{a}$, with $a=1,2$, using the second equality in eq.~\eqref{eq:duality-transformation}.\par
Let us then perform this transformation for the Parke-Taylor-Nair stripped form factors of $\mathcal{T}_{3}$ in eq.~\eqref{F3n}. The simplest example is for $n=3$. As stressed in ref.~\cite{Penante:2014sza}, on the support of the fermionic delta functions, one may equivalently write
\begin{equation}
W_{3,3}^{\textrm{tree}} = -\sum_{1\leqslant i<j \leqslant 3} \langle ij\rangle\, \tilde{\eta}^{-}_{i}\cdot \tilde{\eta}^{-}_{j} \approx \frac{\langle 12\rangle \langle 23 \rangle}{\langle 13\rangle}\,\tilde{\eta}_{2}^{-} \cdot \tilde{\eta}_{2}^{-}\ .
\end{equation}
Upon dualization, it becomes
\begin{equation}\label{eq:baby-R}
W_{3,3}^{\textrm{tree}} = -\,(123)\ ,
\end{equation}
in terms of the (totally antisymmetric) 3-bracket
\begin{equation}\nonumber
(ijk) = \frac{\delta^{0|2}(\langle ij\rangle \eta^{-}_{k}+\langle jk \rangle \eta^{-}_{i} + \langle ki \rangle \eta^{-}_{j})}{\langle ij\rangle \langle jk \rangle \langle ki\rangle} = \frac{\prod_{a=1}^{2}(\langle ij\rangle \eta_{k}^a+\langle jk \rangle \eta_{i}^a + \langle ki \rangle \eta_{j}^a)}{\langle ij\rangle \langle jk \rangle \langle ki\rangle} \ .
\end{equation}
This object is the elementary building block for the MHV form factors of all half-BPS operators. Its form is reminiscent of the R-invariant entering the dual representation for NMHV tree amplitudes~\cite{Drummond:2008vq}. The sole difference is that here we are dealing with R-invariants for ``super-spinors"
\begin{equation}\label{eq:zi}
    z_{i} = \left(\begin{array}{c} \lambda_{i}^{\alpha} \\ \eta_{i}^{a}\end{array}\right)\, ,
\end{equation}
constructed from suitable ``chiral" components of the $4d$ supertwistors. They are annihilated by an $SL(2|2)$ subgroup of the full $SL(4|4)$ dual superconformal group, obtained by restricting the indices of the dual generators to $\alpha = 1,2$ and $a=1,2$.%
\footnote{To be precise, the subgroup is generated by the chiral dual supercharges $\mathcal{Q}^{\alpha}_{a} = \sum_{i=1}^{n} \lambda_{i}^{\alpha} \partial/\partial \eta^{a}_{i}$ and $\mathcal{S}_{\alpha}^{a} = \sum_{i=1}^{n} \eta_{i}^{a}\partial / \partial \lambda_{i}^{\alpha}$, with $a=1,2$, together with the bosonic generators appearing in their anti-commutators.}
The analogy with the scattering amplitudes is particularly clear, since the variables~\eqref{eq:zi} are strictly periodic, with $\lambda_{i+n} = \lambda_{i}$, by definition, and $\eta^{a}_{i+n} = \eta^{a}_{i}$, because $\gamma^{-} = 0$.\par
The R-invariant~\eqref{eq:baby-R} is, of course, not new. It appeared in the study of scattering amplitudes in $2d$ kinematics~\cite{Caron-Huot:2013vda}. It is also the basic object in the so-called $m=2$ amplituhedron~\cite{Arkani-Hamed:2013jha,Arkani-Hamed:2017vfh}, which is a toy model for the geometric description of higher-dimensional scattering amplitudes, see ref.~\cite{Elvang:2013cua} for a nice discussion of the relation between amplitudes and polytopes. More recently, the R-invariants~\eqref{eq:baby-R} made their appearance in the study of correlation functions of half-BPS operators in twistor space~\cite{Caron-Huot:2023wdh}. At tree level, these correlators bear a resemblance to the Wilson loop matrix elements that we will be considering, suggesting that their R-invariant structures may share a common origin.\par
\begin{figure}[t]
  \centering
  \hspace{20pt}
  \begin{minipage}[b]{0.6\textwidth}
    \includegraphics[width=\textwidth]{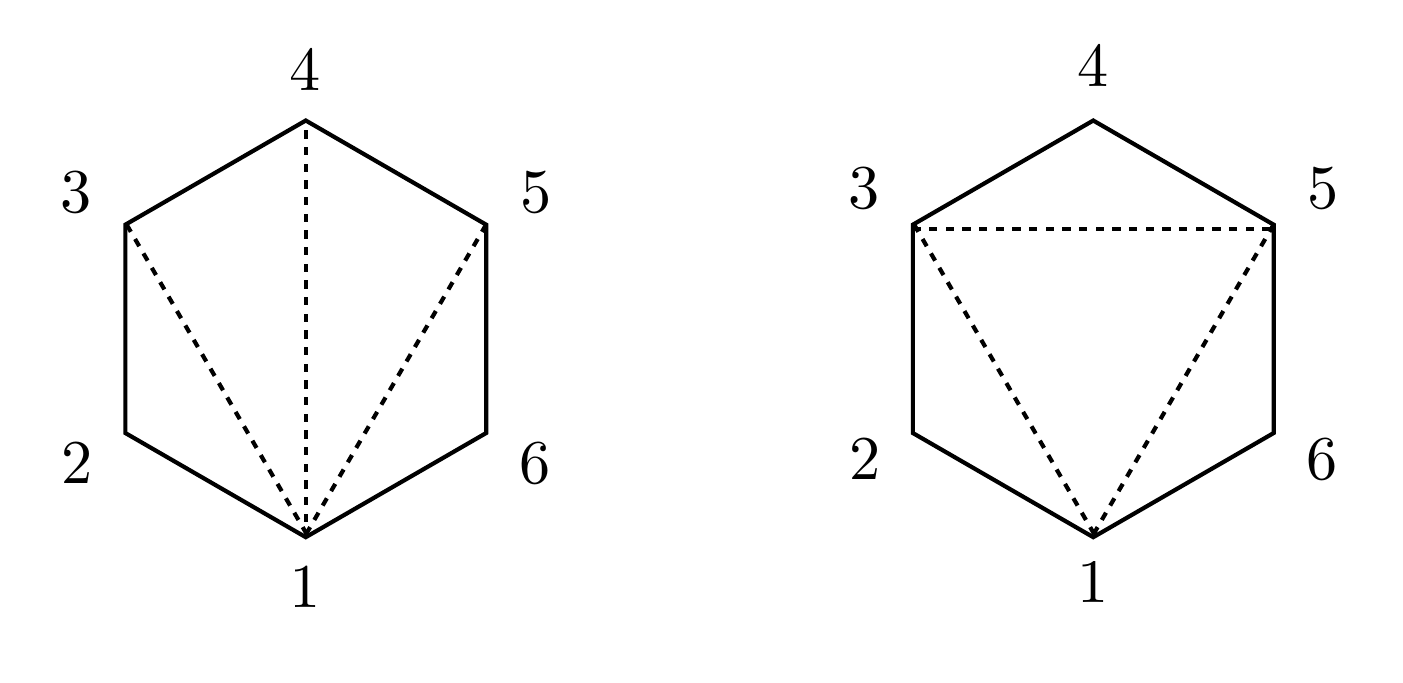}
  \end{minipage}
  \hspace{20pt}
  \caption{Triangulations of a 6-point form factor in spinor space corresponding to the identity $\sum_{i=2}^{5}(1ii+1) = (135)+\sum_{i=1,3,5}(ii+1i+2)$. This identity may be proven geometrically by flipping the triangulations for the square $(1345)$ using the 4-term identity, $(134)+(145) = (135)+(345)$.}
  \label{fig:triangulations}
\end{figure}

The connection to the $m=2$ amplitudes extends to the higher-point form factors of $\mathcal{T}_{3}$. The general expression for these form factors is the $m=2$ analog of the general formula for NMHV amplitudes~\cite{Drummond:2008vq,Mason:2009qx},
\begin{equation}\label{eq:W3n}
W^{\textrm{tree}}_{3, n} = -\sum_{i = 2}^{n-1} (1ii+1) = -\sum_{i=1}^{n} \left[\frac{\langle i+1 i-1 \rangle}{\langle i-1i \rangle\langle ii+1\rangle} \eta_{i}^{-}\cdot \eta_{i}^{-} + \frac{2}{\langle ii+1\rangle} \eta_{i}^{-}\cdot \eta_{i+1}^{-} \right] ,
\end{equation}
with $\eta_{i}^{-}\cdot \eta_{j}^{-} = \frac{1}{2}\epsilon_{ab}\eta_{i}^{a}\eta_{j}^{b}.$ As seen from the second equality, the result is invariant under cyclic permutations. It means that the choice of $1$ as a ``base point" in the intermediate sum is arbitrary. The equivalence between different representations can be established with the help of the $4$-term identity~\cite{Caron-Huot:2013vda,Caron-Huot:2023wdh}
\begin{equation}\label{eq:4-term-id}
(ijk) + (kli) = (lij) + (jkl)\ ,
\end{equation}
valid for any $4$ super-spinors, $z_i, z_j, z_k, z_l$. This identity may be interpreted geometrically as an equivalence relation between different triangulations of a plane convex polygon in spinor space~\cite{Hodges:2009hk,Elvang:2013cua}, see figure \ref{fig:triangulations} for an example of its use. The most general formula one may write along these lines is given for an arbitrary non-overlapping triangulation $\mathcal{T}_n$ of an $n$-sided polygon. It reads
\begin{equation}
 W^{\textrm{tree}}_{3, n} = -\sum_{T\in \mathcal{T}_n} (T_{1}T_{2}T_{3})\ ,
\end{equation}
where $(T_1, T_2, T_3)$ denote the spinors located at the vertices of the triangle $T \in \mathcal{T}_n$, with a common ordering for all triangles in the triangulation.

Another very democratic (and manifestly cyclic) representation may be found in the form
\begin{equation}
W^{\textrm{tree}}_{3, n} = -\sum_{i=1}^{n} (* ii+1)\ ,
\end{equation}
where $z_*$ is an arbitrary reference spinor. It reproduces eq.~\eqref{eq:W3n} when specializing to $z_* = z_1$ and using that $(iik) = 0$ for two identical spinors.\par
The higher-$k$ MHV form factors also take a simple form. To see that, recall that they were conjectured in ref.~\cite{Penante:2014sza} to obey a powerful recursion relation, expressing them in terms of form factors with less points or charges. In dual space, this relation reads concisely
\begin{equation}\label{eq:bcfw}
W^{\textrm{tree}}_{k, n}(1, \ldots , n) = W^{\textrm{tree}}_{k, n-1}(1, \ldots , n-1) - (1n-1n)\,W^{\textrm{tree}}_{k-1, n-1}(1, \ldots , n-1)\ .
\end{equation}
It is supplemented by the boundary condition
\begin{equation}\label{eq:BC}
W^{\textrm{tree}}_{2, n} = 1\ , 
\end{equation}
for the $n$-point form factor of the stress-tensor multiplet. Iterating eq.~\eqref{eq:bcfw} with~\eqref{eq:BC} uniquely determines the form factors for any $n$ and $k$. In particular, the recursion readily leads to the previous formula for $k=3$. A general solution for any $k$ was conjectured in ref.~\cite{Penante:2014sza} and checked to be cyclic symmetric for a wide range of $n$'s and $k$'s. Its form in the dual variables is easier to find by using eq.~\eqref{eq:bcfw} rather than by mapping the explicit formula in ref.~\cite{Penante:2014sza}. It yields the compact expression
\begin{equation}\label{eq:Wkn}
    W^{\textrm{tree}}_{k, n} = \frac{1}{(k-2)!} \left(W^{\textrm{tree}}_{3, n}\right)^{k-2}\ ,
\end{equation}
which is manifestly cyclic. To check that it fulfills eq.~\eqref{eq:bcfw}, one simply plugs eq.~\eqref{eq:bcfw} for $k=3$ in the right hand side of eq.~\eqref{eq:Wkn} and expands the product taking into account that the R-invariant is nilpotent, $(ijk)^2 = 0$. The latter condition also implies that the form factors vanish when $k>n$. This selection rule is fixed by R-charge conservation and thus must be observed to all loops. The ``extremal" cases are reached when $k=n$ and correspond to the so-called minimal (MHV) form factors~\cite{Brandhuber:2014ica}. They take the simple form
\begin{equation}
    W_{n,n}^{\textrm{tree}} = \frac{1}{(n-2)!}\,(W^{\textrm{tree}}_{3, n})^{n-2} = (-1)^{n}\prod_{i=2}^{n-1}(1ii+1)\ .
\end{equation}
Because this structure is unique, it determines the $\eta$-dependence of the minimal form factor to all loops~\cite{Brandhuber:2014ica}.\par
Formula~\eqref{eq:Wkn} is also known from the study of the $m=2$ tree amplituhedron, see eq.~(10.11) in ref.~\cite{Arkani-Hamed:2017vfh}, if we view $k-2$ as the helicity (NMHV) degree of the $m=2$ amplitude. One may conclude from it that the MHV form factors of all BPS operators exhaust the set of tree $m=2$ amplitudes.

\subsection{Periodic Wilson loops}\label{section:periodic_WL}

\begin{figure}[t]
  \centering
   \hspace{20pt}
  \begin{minipage}[b]{0.7\textwidth}
    \includegraphics[width=\textwidth]{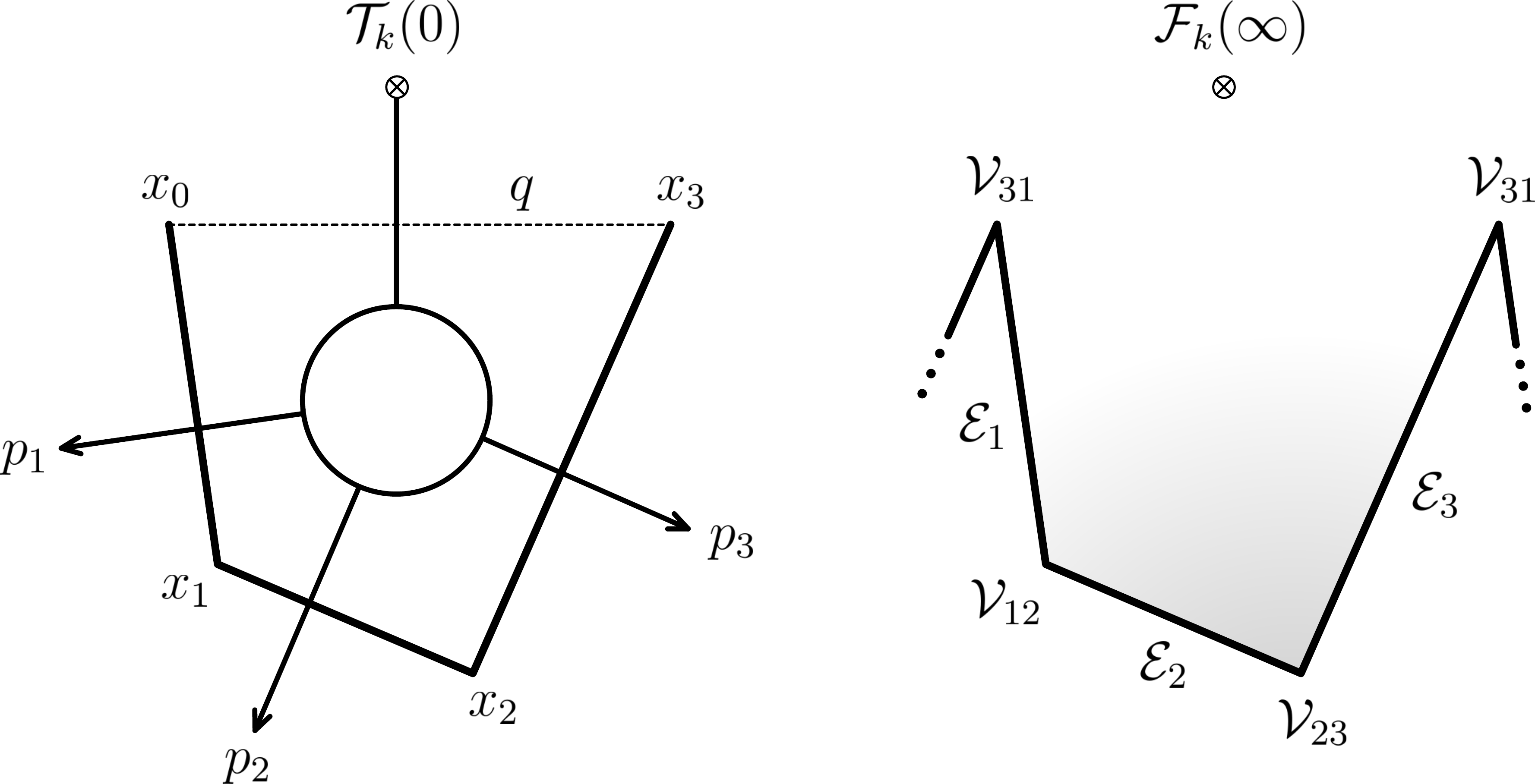}
  \end{minipage}
  \caption{Duality between form factors and periodic Wilson loops for $n=3$.}
  \label{fig:duality}
\end{figure}

The emergence of dual $SL(2|2)$ invariant structures in the form factors of $\mathcal{T}_{k}$ suggests seeking a dynamical interpretation in terms of a null polygonal Wilson loop $W_n$. It is achieved by following the same rules as for scattering amplitudes and substituting the non-local operator for the external super-particles,
\begin{equation}
    \Phi_{1}\ldots \Phi_{n} \qquad \Leftrightarrow \qquad W_{n}\ ,
\end{equation}
as sketched in the left panel of figure~\ref{fig:duality}. The resulting object is naturally defined in the T-dual space, parameterized by the coordinates $(x, \theta)$ or, alternatively, $(x, \eta)$. There are, however, a few important differences that one should address. The foremost one is that the total momentum of the external state is generically non-zero,
\begin{equation}
q = p_1+\ldots + p_n =  x_{n}-x_{0} \neq 0\ ,
\end{equation}
for form factors. As alluded to before, it implies that the ``loop'' in dual space does not close. Instead, to maintain cyclic symmetry, as well as gauge invariance, the loop is extended into an infinite periodic contour of period $q$, as shown in figure~\ref{fig:duality}. Evidence for this identification arose at strong coupling~\cite{Alday:2007he,Maldacena:2010kp}, where a prescription for calculating the form factors of light operators was given in terms of a periodic string hanging between the horizon and the boundary of the Anti-de-Sitter space. The prescription was made more precise at weak coupling in the case of the form factors of the stress-tensor multiplet~\cite{Brandhuber:2010ad,Brandhuber:2011tv,Bork:2014eqa}, which were verified to map to the vacuum expectation values of periodic Wilson loops through one loop.\par
These studies also provided a set of rules for handling the ``infinities'' coming from the infinite contour of the loop in the gauge theory. Namely, instead of compactifying the theory on a circle $S^{1}_{q}$, which would ensure the right periodicity properties but introduce an unrealistic tower of Kaluza-Klein modes, one should calculate the Wilson loop in the uncompactified theory using a restricted class of Feynman diagrams. The key selection criteria are 1) each diagram should fit within one period and 2) two diagrams are deemed equivalent if they are periodic images of one another.\par
This prescription was refined in refs.~\cite{Ben-Israel:2018ckc,Sever:2020jjx} by adopting a quantum worldsheet perspective, in which the periodic Wilson loop was mapped to a worldsheet theory subject to twisted boundary conditions, with the twist associated with a translation by $q$. Importantly, it follows from these considerations that, unlike in the planar scattering amplitudes case, the worldsheet topology is not a full disk, but a cylinder of radius $q$ ending on the contour of the loop on one end and stretching all the way to infinity on the other end.
The notion of planar diagrams gets twisted accordingly: a diagram is planar if it can be drawn on this cylinder, with no propagators crossing over each other.

The second key difference between amplitudes and form factors is that in the latter case one should also accommodate the local operators in the dual picture. Since the T-duality transformation interchanges states and operators by sending the former to the boundary and the latter to the horizon, the natural choice is to introduce non-trivial states in the dual picture to represent the operators. Specifically, we expect the duality to identify the $n$-point form factors of $\mathcal{T}_k$ in momentum space as normalized in eq.~\eqref{FFnormalized} with the matrix elements of a $n$-sided null periodic Wilson loop $W_n$ in dual space,
\begin{equation}\label{FF-WL_duality}
    W_{k,n} =  \langle \mathcal{F}_{k}| W_{n} |0\rangle\ ,
\end{equation}
for some non-trivial state $\langle \mathcal{F}_{k} |$. As alluded to before, in general, it is not clear how to assign a dual state to an operator. For the stress-tensor multiplet, the dual state is expected to be the vacuum, $\langle \mathcal{F}_{2} | = \langle 0|$, giving rise to a duality with the vacuum expectation value of a null periodic Wilson loop. For higher $k$, the state should carry R-charge and thus cannot be that trivial. In the following, we will explicitly determine the states that correspond to half-BPS operators for any $k$.

\subsection{States and operators}

Prior to discussing the dual state, let us review the definition of the Wilson loop operator that we will be using.\par
Since super form factors exhibit non-trivial dependence on the $\eta$ variables, already in the MHV sector, the right operator to consider is not the bosonic Wilson loop itself, but its supersymmetric extension~\cite{Caron-Huot:2010ryg,Mason:2010yk,Belitsky:2011zm} tailored for the $\eta$ components of the super-amplitudes. It reads~\cite{Caron-Huot:2010ryg}
\begin{equation}\label{eq:SWL-operator}
    W_{n} \equiv \frac{1}{N}\,\textrm{Tr}\left[P\ldots\mathcal{V}_{01}\,e^{i\int_{0}^{1}dt \mathcal{E}_{1}} \,\mathcal{V}_{12}\,e^{i\int_{0}^{1}dt \mathcal{E}_{2}}\,\mathcal{V}_{23}\,\ldots\, e^{i\int_{0}^{1}dt \mathcal{E}_{n}}\,\mathcal{V}_{nn+1}\,\ldots\right]\ ,
\end{equation}
with $P$ denoting the path ordering and $N$ the number of colors.%
\footnote{We follow the ordering introduced in ref.~\cite{Caron-Huot:2010ryg}, which is opposite to the standard textbook path ordering convention for Wilson loops.}
The infinite nature of the loop should be understood in accordance with the periodization procedure outlined earlier. This super Wilson loop is built out of the super connection
\begin{equation}\label{eq:calE-i}
\mathcal{E}_{i}(t) = \frac{1}{2}\,\tilde{\lambda}_{i}^{\dot{\alpha}}\lambda_{i}^{\alpha} A_{\dot{\alpha}\alpha} + \frac{i}{2}\, \tilde{\lambda}_{i}^{\dot{\alpha}}\bar{\psi}_{\dot{\alpha} a}\,\eta_{i}^{a} + \frac{i}{4 \langle i-1 i\rangle}\,\tilde{\lambda}_{i}^{\dot{\alpha}} \lambda_{i-1}^{\alpha} D_{\dot{\alpha}\alpha}\bar{\phi}_{ab}\,\eta_{i}^{a} \eta_{i}^{b}\ ,
\end{equation}
defined locally on each edge, $x(t) = x_{i-1}+t\,p_{i} \in (x_{i-1}, x_{i})$, using the kinematic variables. It is expressed in terms of the fundamental fields of the theory, in the adjoint representation of the gauge group $U(N)$. Following the common rule for the dual theory~\cite{Caron-Huot:2010ryg,Drummond:2010km}, the R-charges of the gaugino and scalar fields have been conjugated, using
\beq
\psi_{\alpha a} \rightarrow \bar{\psi}_{\dot{\alpha}a}\ , \qquad \phi_{ab} \rightarrow \bar{\phi}_{ab}\ , \qquad \bar{\psi}_{\dot{\alpha}}^{a} \rightarrow \psi_{\alpha}^{a}\ ,
\eeq
as compared to the notations used in section~\ref{section:super_FF_def}. The vector field $A$ is self-conjugate. It is defined as $A_{\dot{\alpha}\alpha} = A_{\mu} \bar{\sigma}^{\mu}_{\dot{\alpha}\alpha}$, and similarly for the covariant derivative, $D_{\mu} = \partial_\mu +i [A_{\mu}, .]$, following the same convention as for momenta, $p_{\dot{\alpha}\alpha} = p_{\mu}\bar{\sigma}^{\mu}_{\dot{\alpha}\alpha}$, where $\bar{\sigma}^{\mu} = (I, -\vec{\sigma})$.\par
Generically, the super connection $\mathcal{E}_{i}$ contains terms of degree up to four in $\eta_i^{A}$. For our considerations, however, we may restrict ourselves to the $SU(2)$ sector spanned by $\eta_{i}^a$ with $a=1,2$, leading to the truncated expression~\eqref{eq:calE-i}. We proceed similarly for the vertex $\mathcal{V}_{ii+1} = \mathcal{V}_{ii+1}(x_{i})$ located at the intersection of edge $i$ and $i+1$. In the $SU(2)$ sector, it reads
\begin{equation}\label{scalar_vertex}
\mathcal{V}_{ii+1} = 1 + \mathcal{V}^{(0)}_{ii+1} + \tfrac{1}{2} [\mathcal{V}^{(0)}_{ii+1}]^2\ ,
\end{equation}
where
\begin{equation}\label{eq:corner-vertex}
\mathcal{V}^{(0)}_{ii+1} = -\,\bar{\phi}_{ab} \left[\frac{\eta^{a}_{i}\eta^{b}_{i+1}}{\langle ii+1\rangle} + \frac{\langle i+1i-1\rangle \eta^a_{i}\eta^b_{i}}{\langle i-1i\rangle\langle ii+1\rangle} \right]\ .
\end{equation}
We stress that this expression is complete in the subspace of interest. The general formula for the corner insertion may be found in ref.~\cite{Caron-Huot:2010ryg} through the first few orders in the $\eta$ expansion and its complete expression was derived in refs.~\cite{Groeger:2012xz,Groeger:2012hqk}. As observed in ref.~\cite{Caron-Huot:2010ryg}, the vertex insertions are not independent of the super connections and may be redistributed along the edges of the loop, see also refs.~\cite{Mason:2010yk,Belitsky:2011zm} for alternative descriptions. However, for our study, it will be convenient to keep them manifest, since, as we will see, the interactions localize on them at weak coupling.\par
Let us proceed with the construction of the state dual to the operator $\mathcal{T}_{k}$. As mentioned earlier, our proposal is to choose a state made out of zero-momentum scalars. The key reason for this is that it ensures that the matrix element of the Wilson loop is invariant under all (super) translations,
\begin{equation}
x_i^{\alpha\dot{\alpha}} \rightarrow x_i^{\alpha\dot{\alpha}} + c^{\alpha\dot{\alpha}}\ , \qquad \theta_i^{\alpha A} \rightarrow \theta_i^{\alpha A} + \xi^{\alpha A}\ .
\end{equation}
Indeed, recall that these shifts are not actual symmetries of the amplitudes or form factors. They do not act on them \textit{at all}. Instead, they are a redundancy of the dual description, which must drop out from any ``physical" quantity. The only way to achieve this is by choosing a state that is annihilated by the corresponding dual generators,
\begin{equation}\label{eq:PQ}
\langle \mathcal{F}_{k}|\mathcal{P}^{\alpha\dot{\alpha}} = 0\ , \qquad \langle \mathcal{F}_{k}|\mathcal{Q}_{A}^{\alpha} = 0\ .
\end{equation}
In contrast, the R-symmetry generators act canonically on both sides of the duality, up to the conjugation mentioned earlier. Hence, the quantum numbers of the state should match the ones carried by the form factor. For $k=2$, after stripping off the fermionic delta functions, we are left with an R-charge singlet, so the most natural candidate is the vacuum state. For $k=3$, the R-charge is non-zero and the most natural choice is a zero-momentum scalar state.\par
To be precise, for $k=3$, we choose a single-trace state $\langle \mathcal{F}_{3}| = \langle \, \textrm{Tr}\, \phi^{12}|$, which we normalize such that
\begin{equation}
\langle\, \textrm{Tr}\, \phi^{12} | \bar{\phi}^{a}_{AB} (x)|0\rangle = (\delta_{A}^{1}\delta_{B}^{2}-\delta_{A}^{2}\delta_{B}^{1}) \times \textrm{Tr}\, T^{a}\ ,
\end{equation}
for $A, B = 1,2,3,4$. Here, $T^{a}$ ($a = 1, \ldots , N^2$) are the generators of the $u(N)$ algebra. It is important to work with the $u(N)$ algebra, as opposed to $su(N)$, to define the single-scalar state. In the $su(N)$ gauge theory, the trace would be empty. This distinction may sound dubious at first sight, given that $u(1)$ states are usually not particularly interesting, due to them being non-interacting. However, as alluded to before, the rules for computing periodic Wilson loops are not quite the usual ones and, as we will see later on, they prevent the decoupling of the $u(1)$ sector. The ``singlet" scalar appears as non-trivial in this context as the higher-particle states; see also ref.~\cite{Cavaglia:2020hdb} for similar twisted consideration. That being said, we will suppress the color indices in the following, bearing in mind that at large $N$ we are only interested in the single-trace term. When focusing on the $SU(2)$ sector of interest, we will simply write
\begin{equation}
\langle \phi^{12} | \bar{\phi}_{ab} (x)|0\rangle = \epsilon_{ab}\ ,
\end{equation}
for the fundamental matrix element.\par
Similarly, at higher $k$, the minimal solution is to pick a state containing $k-2$ zero-momentum scalars, $\langle \mathcal{F}_{k}| = \langle \, \textrm{Tr}\, \phi^{12} \ldots \phi^{12} |$. It is manifestly consistent with translation invariance. It also ensures that supersymmetry is preserved. The point is that the $\mathcal{Q}$'s carry dimension, whereas zero-momentum particles have no scale. As such, they cannot undergo supersymmetry transformations.\par
The generic $k$ state is defined such that
\begin{equation}
\langle\, \textrm{Tr}\, (\phi^{12})^{k-2} | \bar{\phi}^{a_{1}}_{A_{1}B_{1}} (x_{1})\,\ldots\,  \bar{\phi}^{a_{\ell}}_{A_{\ell}B_{\ell}}(x_{\ell})|0\rangle = \delta_{k-2, \ell} \prod_{i=1}^{\ell} (\delta_{A_{i}}^{1}\delta_{B_{i}}^{2} -\delta_{A_{i}}^{2}\delta_{B_{i}}^{1}) \times \textrm{Tr}\, T^{a_{1}}\ldots T^{a_{\ell}}\ ,
\end{equation}
or simply
\begin{equation}
    \langle (\phi^{12})^{k-2} | \bar{\phi}_{a_{1} b_{1}} (x_{1})\,\ldots\,\bar{\phi}_{a_{\ell}b_{\ell}}(x_{\ell})|0\rangle = \delta_{k-2, \ell}\,\epsilon_{a_{1}b_{1}} \ldots \epsilon_{a_{\ell}b_{\ell}}\ ,
\end{equation}
in the $SU(2)$ sector.

\subsection{Tree-level checks}

\begin{figure}[t]
  \centering
  \hspace{20pt}
  \begin{minipage}[b]{0.4\textwidth}
    \includegraphics[width=\textwidth]{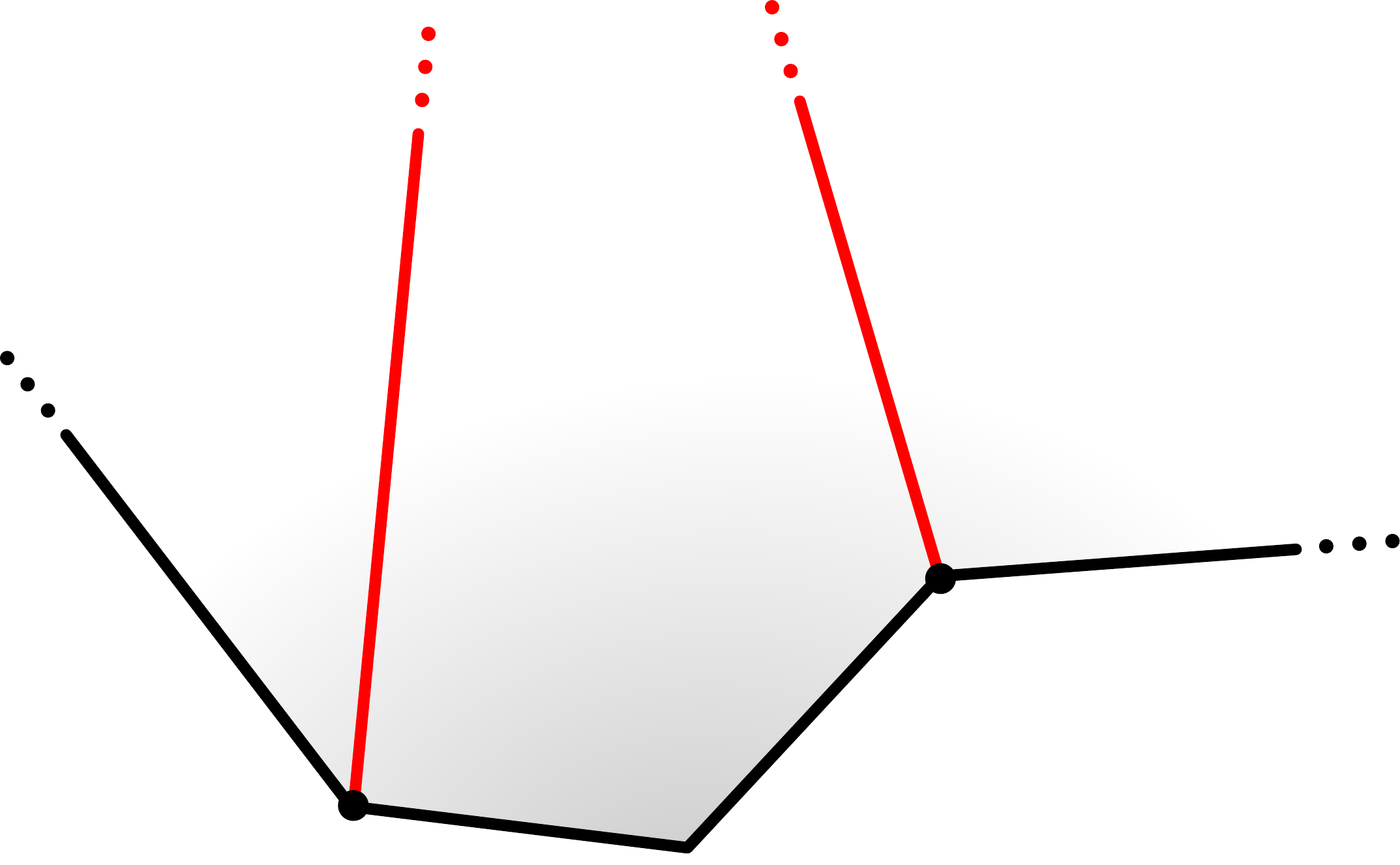}
  \end{minipage}
  \hfill
  \begin{minipage}[b]{0.4\textwidth}
    \includegraphics[width=\textwidth]{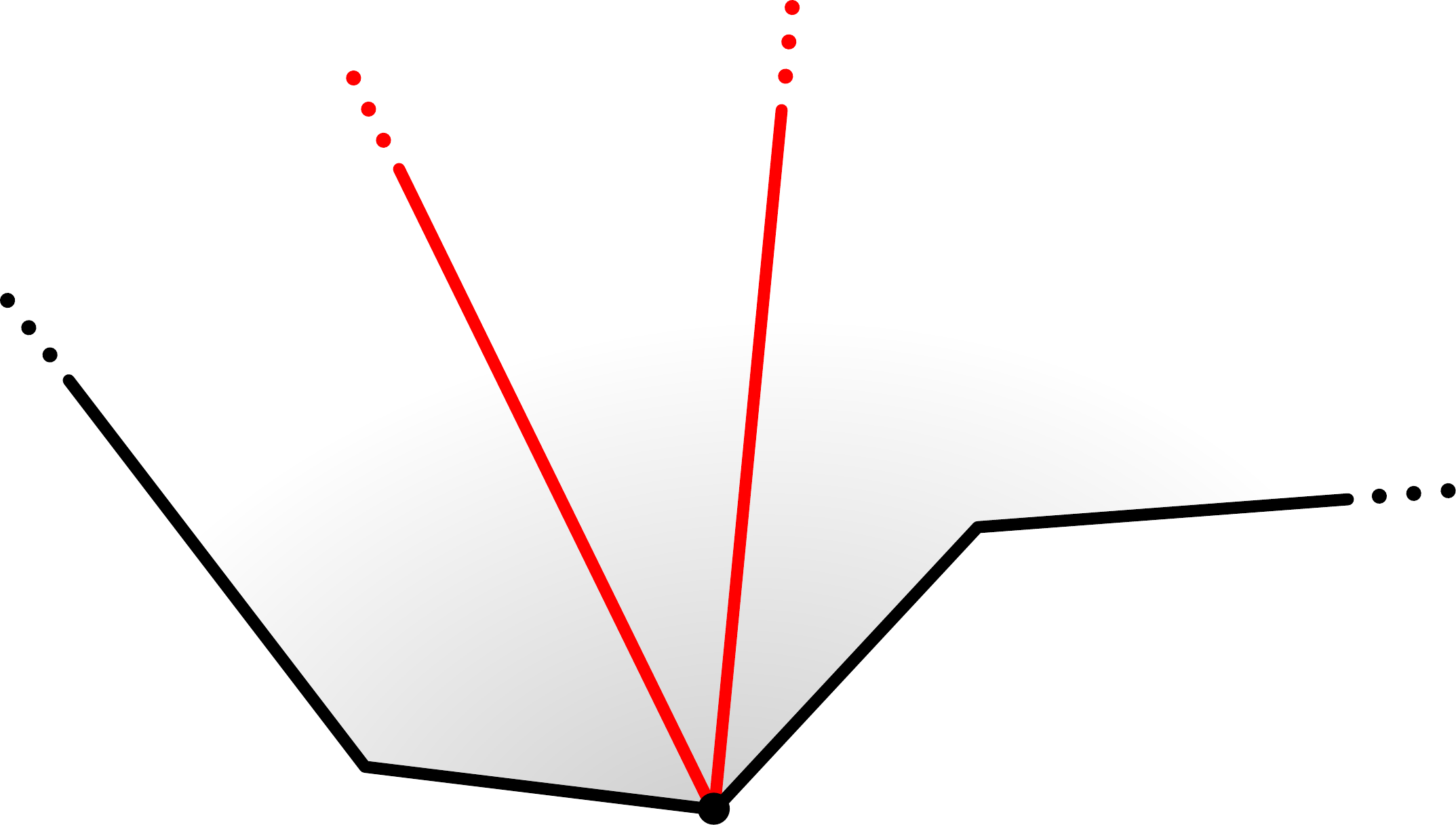}
  \end{minipage}
  \hspace{20pt}
  \caption{Two ways how scalars can connect to the vertices of the super Wilson loop. Each vertex $\mathcal{V}_{ii+1}$ can emit either $1$ or $2$ scalars.}
  \label{fig:born_level}
\end{figure}

Given the states, one may calculate the matrix elements of the super Wilson loop. Since the external particles all have zero momentum, they cannot be produced along the edges of the loop at tree level; the only edge term with the right R-charge in eq.~\eqref{eq:calE-i} is the covariant derivative of the scalar field, which kills the state at weak coupling,
\begin{equation}
\langle \phi^{12} | D_{\dot{\alpha}\alpha}\bar{\phi}_{ab}(x) |0 \rangle_{\textrm{tree}}  = 0\ .
\end{equation}
Therefore, the scalars can only be produced at the corners of the loop, using the vertex~\eqref{scalar_vertex}. Restricting the super Wilson loop in eq.~\eqref{eq:SWL-operator} to a single period, we get
\begin{equation}
    W_{k, n}^{\textrm{tree}} = \langle \phi^{12} \ldots \phi^{12} | \prod_{i=1}^{n} \mathcal{V}_{ii+1}|0\rangle \ ,
\end{equation}
with periodic boundary conditions, $\mathcal{V}_{nn+1} = \mathcal{V}_{n1}$. The calculation of the matrix element is straightforward. In particular, for $k=3$, one immediately gets
\begin{equation}\label{W3ntree}
    W_{3, n}^{\textrm{tree}} = \sum_{i=1}^{n} \langle \phi^{12}|\mathcal{V}^{(0)}_{ii+1}|0\rangle = -\,\sum_{i=1}^{n} \left[\frac{2\eta_{i}^{-}\cdot \eta^{-}_{i+1}}{\langle ii+1\rangle} +\frac{\langle i+1i-1\rangle \eta^{-}_{i}\cdot \eta^{-}_{i}}{\langle i-1i\rangle \langle ii+1\rangle}\right] .
\end{equation}
It agrees perfectly with the form-factor data for $\mathcal{T}_{3}$, see eq.~\eqref{eq:W3n}.\par
For higher $k$ we should distribute $k-2$ scalars on the $n$ corners of the loop. Given the structure of the vertex insertion~\eqref{eq:corner-vertex}, no more than two scalars can be emitted from a given corner, as shown in figure \ref{fig:born_level}. For instance, for $k=4$, we find
\begin{equation}
\begin{aligned}
    W_{4,n}^{\textrm{tree}} &= \sum_{1\leqslant i<j \leqslant n} \langle \phi^{12}\phi^{12} |\mathcal{V}^{(0)}_{ii+1}\mathcal{V}^{(0)}_{jj+1} |0\rangle + \frac{1}{2}\sum_{i=1}^{n} \langle \phi^{12}\phi^{12} | [\mathcal{V}^{(0)}_{ii+1}]^2 |0\rangle \\
    &= \frac{1}{2}\left[W_{3, n}^{\textrm{tree}}\right]^2\ .
\end{aligned}
\end{equation}
The general formula for any $k$ is easily inferred from these particular examples and verified to match with the form-factor prediction~\eqref{eq:Wkn} for higher-charge BPS operators.\par
It is worth pointing out that the periodic nature of the Wilson loop is irrelevant here. The tree algebra is blind to it and would apply just as well to the calculation of the matrix element of a \textit{closed} Wilson loop. The reason for this is that we are focusing on the MHV sector, which is only sensitive to the variables $z_{i} = z_{i+n}$ at tree level, see eq.~\eqref{eq:zi}.

\subsection{One-loop analysis}

To probe the periodic nature of the Wilson loop and the rules for handling it at the quantum level~\cite{Brandhuber:2010ad,Bork:2014eqa,Ben-Israel:2018ckc}, we will now consider our matrix elements at one loop. There are two classes of diagrams to consider. The first one comprises familiar interactions between edges of the (super) Wilson loop. The second class is less standard, as it corresponds to diagrams ``renormalizing" the outgoing state. The latter diagrams are superficially IR divergent, but their contributions cancel out in the end, in line with the non-renormalization property of the half-BPS operators. We defer their study to appendix \ref{app:IR-div} and focus here on the first-class diagrams.

\begin{figure}[t]
  \centering
  \begin{minipage}[b]{0.80\textwidth}
    \includegraphics[width=\textwidth]{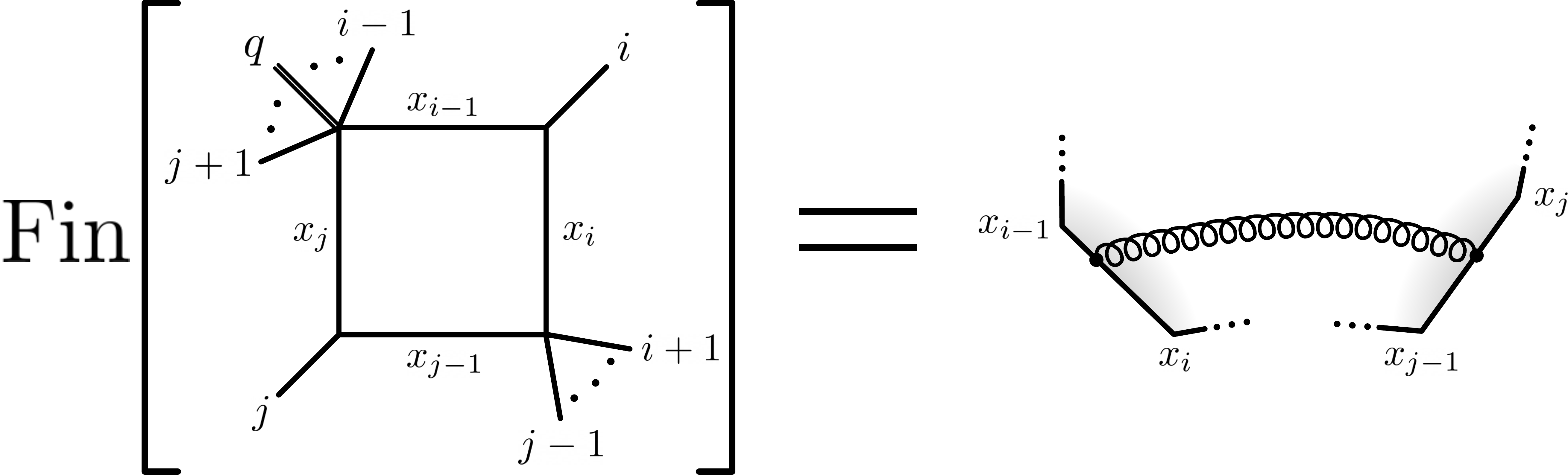}
  \end{minipage}
  \caption{The gluon exchange between non-adjacent edges is given by the finite part of the two-mass easy box, given explicitly by (\ref{easy_box_expr}).}
  \label{fig:easy_box}
\end{figure}

The most basic diagram from the first class is the gluon exchange between two edges $i$ and $j$. It is common to all null polygonal Wilson loops and its analysis was discussed at length in refs.~\cite{Drummond:2007aua,Brandhuber:2007yx,Drummond:2007cf} in the context of the duality with scattering amplitudes. Its calculation is most easily done in Feynman gauge, in which gluon and scalar propagators are proportional to each other, up to the Lorentz and R-symmetry structures. After summing over the colors, one gets
\begin{equation}
\langle A_{\mu}(x)A_{\nu}(y) \rangle = -\,2\pi^2 g^{2} \eta_{\mu \nu}\,\Delta(x-y)\ ,
\end{equation}
with $g^{2} = \lambda/(4\pi)^2$ the 't Hooft coupling, $\eta_{\mu\nu}$ the Minkowski metric, in the mostly-minus signature, and with
\beq
\Delta(x) = \int \frac{d^Dk}{(2\pi)^D} \frac{i}{k^2+i0}\,e^{-ikx} = \frac{\Gamma(\tfrac{D}{2}-1)}{4 \pi^{\frac{D}{2}}[-x^2]^{\frac{D-2}{2}}}\ ,
\eeq
being the free massless propagator in dimension $D$. The continuation to dimension $D$ is introduced here to deal with the UV divergences of the Wilson loop. In Feynman gauge, the divergences only show up for adjacent edges. For non-neighbouring edges, the contour integral is finite and may be done directly in $D=4$. It yields~\cite{Brandhuber:2007yx}
\begin{equation}\label{eq:int-ij}
    i^2\int_{i} \int_{j}\,  p_{i}^{\mu}p_{j}^{\nu}\langle A_{\mu} A_{\nu} \rangle = g^2 F_{ij}\ ,
\end{equation}
where $F_{ij}$ coincides with the finite part of the two-mass easy box diagram, as shown in figure \ref{fig:easy_box},
\begin{align}\label{easy_box_expr}
    F_{ij} &= \textrm{Li}_{2}\left[1-\frac{x^2_{i-1, j}}{x^2_{i-1,j-1}}\right] + \textrm{Li}_{2}\left[1-\frac{x^2_{i-1, j}}{x^2_{i,j}}\right] + \textrm{Li}_{2}\left[1-\frac{x^2_{i, j-1}}{x^2_{i-1,j-1}}\right] + \textrm{Li}_{2}\left[1-\frac{x^2_{i, j-1}}{x^2_{i,j}}\right] \nonumber\\
    & - \textrm{Li}_{2}\left[1-\frac{x^2_{i-1, j}x^2_{i, j-1}}{x^2_{i-1,j-1}x^2_{i,j}}\right] + \frac{1}{2}\log^{2}{\left[\frac{x^2_{i-1,j-1}}{x^2_{i,j}}\right]}\ ,
\end{align}
with $x^2_{ab} = (x_a-x_b)^2 = (p_{a+1}+p_{a+2}+ \ldots + p_{b})^2$ and $\textrm{Li}_{2}(x) = \sum_{k=1}^{\infty} x^k/k^2$ the dilogarithm. For adjacent edges, the contour integral~\eqref{eq:int-ij} has UV divergences controlled by the cusp anomalous dimension~\cite{Korchemsky:1987wg}%
\footnote{Up to an overall constant $c(\epsilon) = \pi^{\epsilon}\Gamma(1-\epsilon) = 1 +\mathcal{O}(\epsilon)$ absorbed here in the definition of the coupling.}
\begin{equation}
    F_{ii+1} = -\,g^2 \frac{(-s_{ii+1})^{\epsilon}}{\epsilon^2}\ ,
\end{equation}
with $s_{ii+1} = (p_i+p_{i+1})^2$ and $D=4-2\epsilon$. Lastly, recall that the integral vanishes, $F_{ii} \propto p_i^2 = 0$, for identical edges, $j=i$.

\begin{figure}[t]
  \centering
  \begin{minipage}[b]{0.3\textwidth}
    \includegraphics[width=\textwidth]{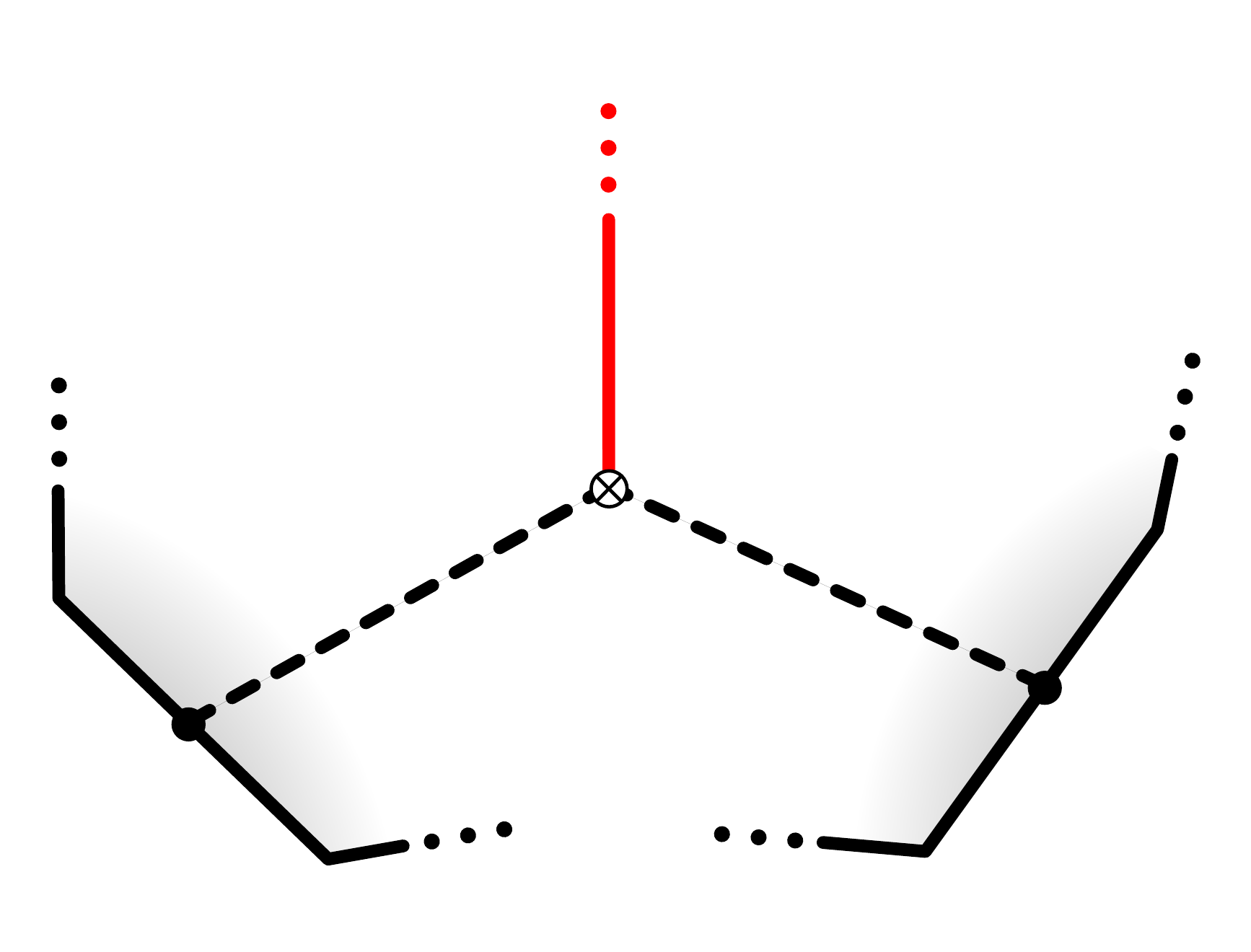}
  \end{minipage}
  \hfill
  \begin{minipage}[b]{0.3\textwidth}
    \includegraphics[width=\textwidth]{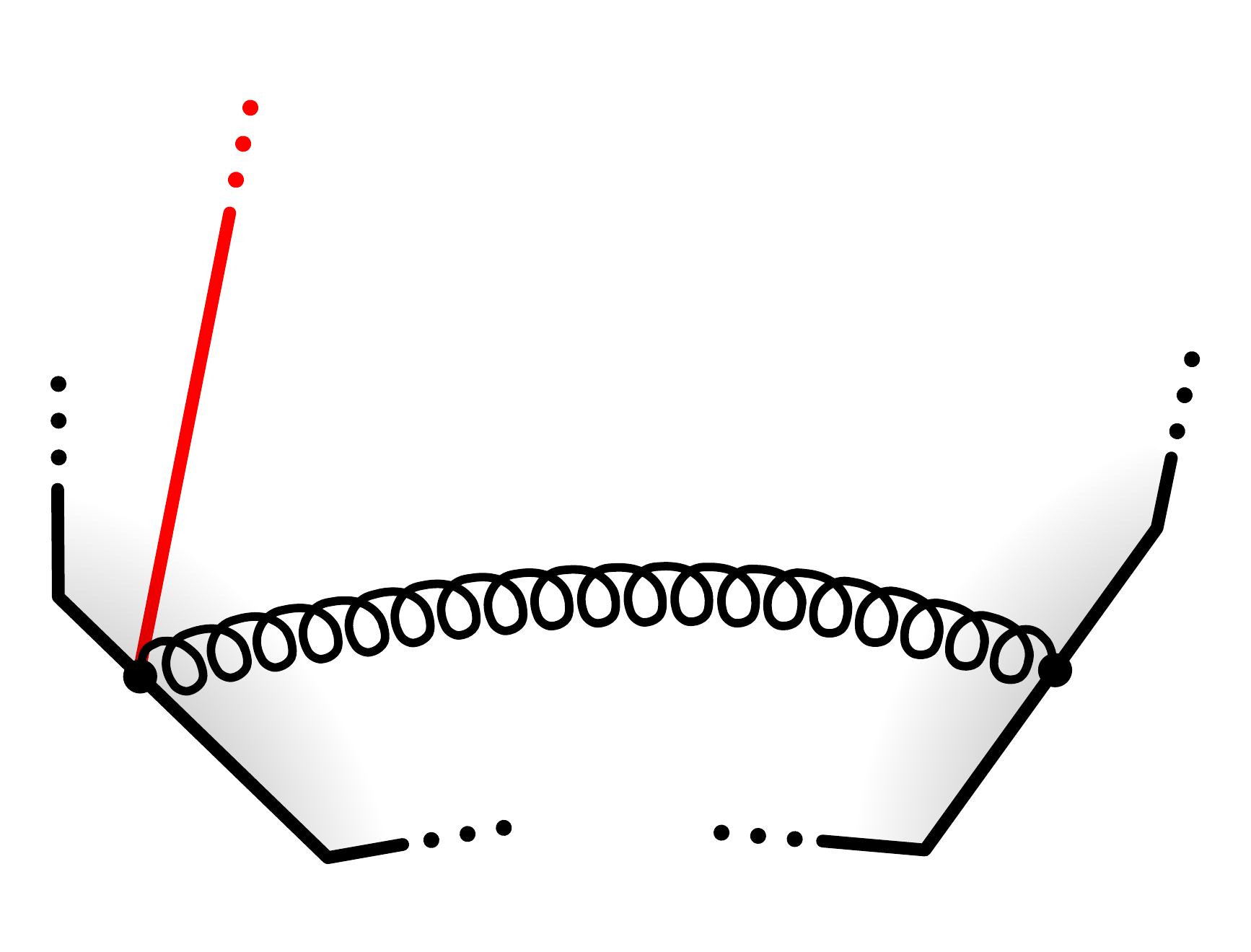}
  \end{minipage}
  \hfill
  \begin{minipage}[b]{0.3\textwidth}
    \includegraphics[width=\textwidth]{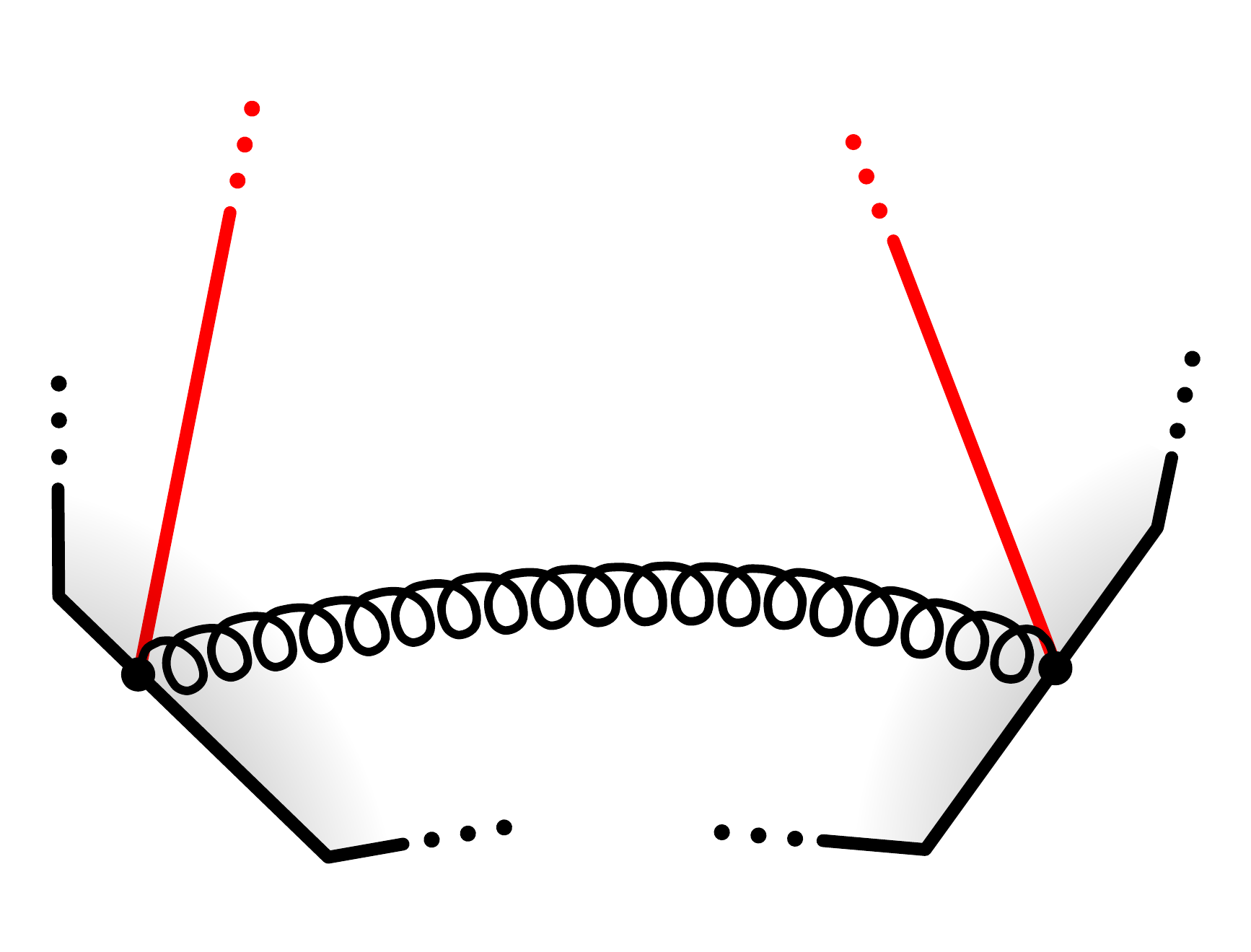}
  \end{minipage}
  \caption{Diagrams producing long-range interactions between $\eta$ variables, with dashed/wavy lines corresponding to fermions/gluons.}
  \label{fig:one_loop}
\end{figure}

For the super Wilson loop, we should add diagrams generated by the superpartners of the gauge connection in eq.~\eqref{eq:calE-i}. In the presence of scalars in the outgoing state, the latter generate contributions proportional to powers of $\eta_{i}^{-}$ and $\eta_j^{-}$, as shown in figure~\ref{fig:one_loop}. Remarkably, all these diagrams are proportional to the basic gluon integral~\eqref{eq:int-ij}.\par
Take, for instance, the correlator between fermions, which produces a term $\propto \eta_i^{a}\eta_j^{b}$. Since the fermions have same chirality, their two-point function must vanish. In the vacuum, this is true to all-loop order. However, in a background of zero-momentum scalars, the two-point function picks up a one-loop contribution through the Yukawa interactions, as shown in figure~\ref{fig:one_loop}. It reads
\begin{equation}
    \langle \bar{\psi}_{\dot{\alpha} a}(x)\,\bar{\psi}_{\dot{\beta} b}(y) \rangle = 2\pi^2 g^2 \epsilon_{\dot{\alpha}\dot{\beta}}\,\Delta(x-y)\,\bar{\phi}_{ab}\ ,
\end{equation}
where the field $\bar{\phi}_{ab}$ is meant to be contracted with a zero-momentum scalar. After contracting with spinor variables and performing the contour integral, we arrive at
\begin{equation}
    -\,\frac{\bar{\phi}_{ab}}{\langle ij \rangle}\,\eta_{i}^{a}\eta_{j}^{b} \times g^2 F_{ij}\ .
\end{equation}
A similar analysis applies to the terms $\propto (\eta_i)^2, (\eta_j)^2$ or $(\eta_i)^2(\eta_j)^2$ coming from the covariant derivative, $D_{\dot{\alpha}\alpha}\bar{\phi}_{ab} \rightarrow i[A_{\dot{\alpha}\alpha},\bar{\phi}_{ab}]$, which may be produced on edge $i$ and $j$ separately or on both edges simultaneously, as shown in figure~\ref{fig:one_loop}.\par
Putting all the terms together, we find that the super-correlator can be cast into the form
\begin{equation}
i^2 \int_{i}\int_{j} \langle \mathcal{E}_i \mathcal{E}_j\rangle  = g^2 F_{ij} \times \tilde{\mathcal{V}}_{ij}\ ,
\end{equation}
where $\tilde{\mathcal{V}}_{ij}$ stands for the induced vertex
\begin{equation}\label{eq:delta-V}
\tilde{\mathcal{V}}_{ij} = 1 + \mathcal{V}^{(1)}_{ij} + \tfrac{1}{2} [\mathcal{V}^{(1)}_{ij}]^2\ ,
\end{equation}
with
\begin{equation}
  \mathcal{V}^{(1)}_{ij} = -\,\frac{\bar{\phi}_{ab}}{\langle ij \rangle} \bigg[ \eta_{i}^{a}\eta_{j}^{b} + \frac{1}{2} \frac{\langle ji-1 \rangle}{\langle i-1 i \rangle} \eta_{i}^{a}\eta_{i}^{b}+\frac{1}{2}\frac{\langle ij-1\rangle}{\langle j-1j \rangle} \eta_{j}^{a}\eta_{j}^{b}\bigg]\ .
\end{equation}
It is worth noting that this expression reduces to the tree vertex~\eqref{eq:corner-vertex} when the edges are adjacent,
\begin{equation}
    \tilde{\mathcal{V}}_{ii+1} = \mathcal{V}_{ii+1}\ .
\end{equation}
This property ensures the universality of the cusp divergences, which are required to take the same form as for a (closed) bosonic Wilson loop. At one loop, they must appear in front of the tree-level result,
\begin{equation}
W^{\textrm{1-loop}}_{k, n} = g^2 \sum_{i=1}^{n} F_{ii+1} \times W^{\textrm{tree}}_{k, n} + \textrm{Finite}\ ,
\end{equation}
for any $k$.\par
\begin{figure}[t]
  \centering
  \hspace{20pt}
  \begin{minipage}[b]{0.40\textwidth}
    \includegraphics[width=\textwidth]{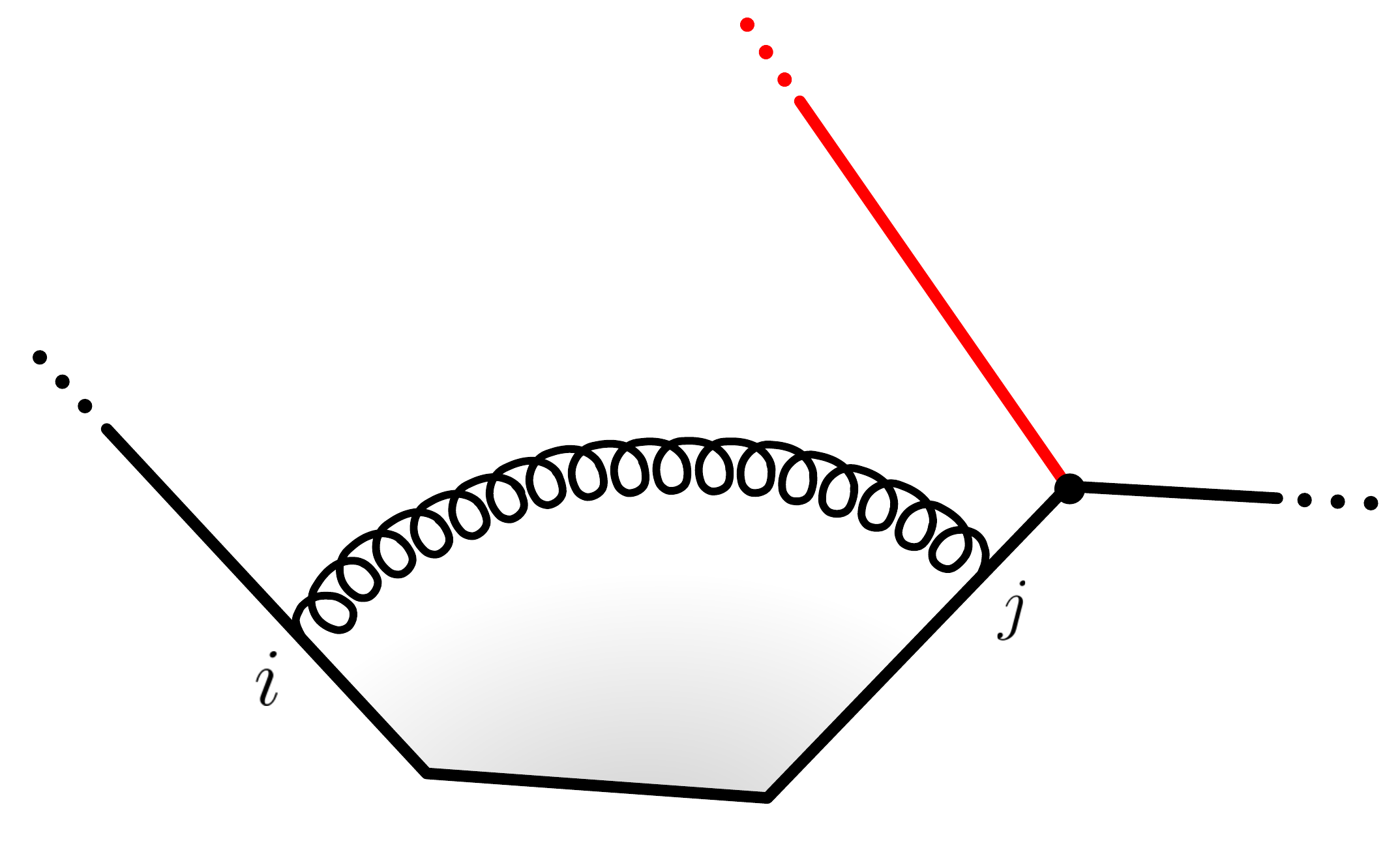}
  \end{minipage}
  \hfill
  \begin{minipage}[b]{0.40\textwidth}
    \includegraphics[width=\textwidth]{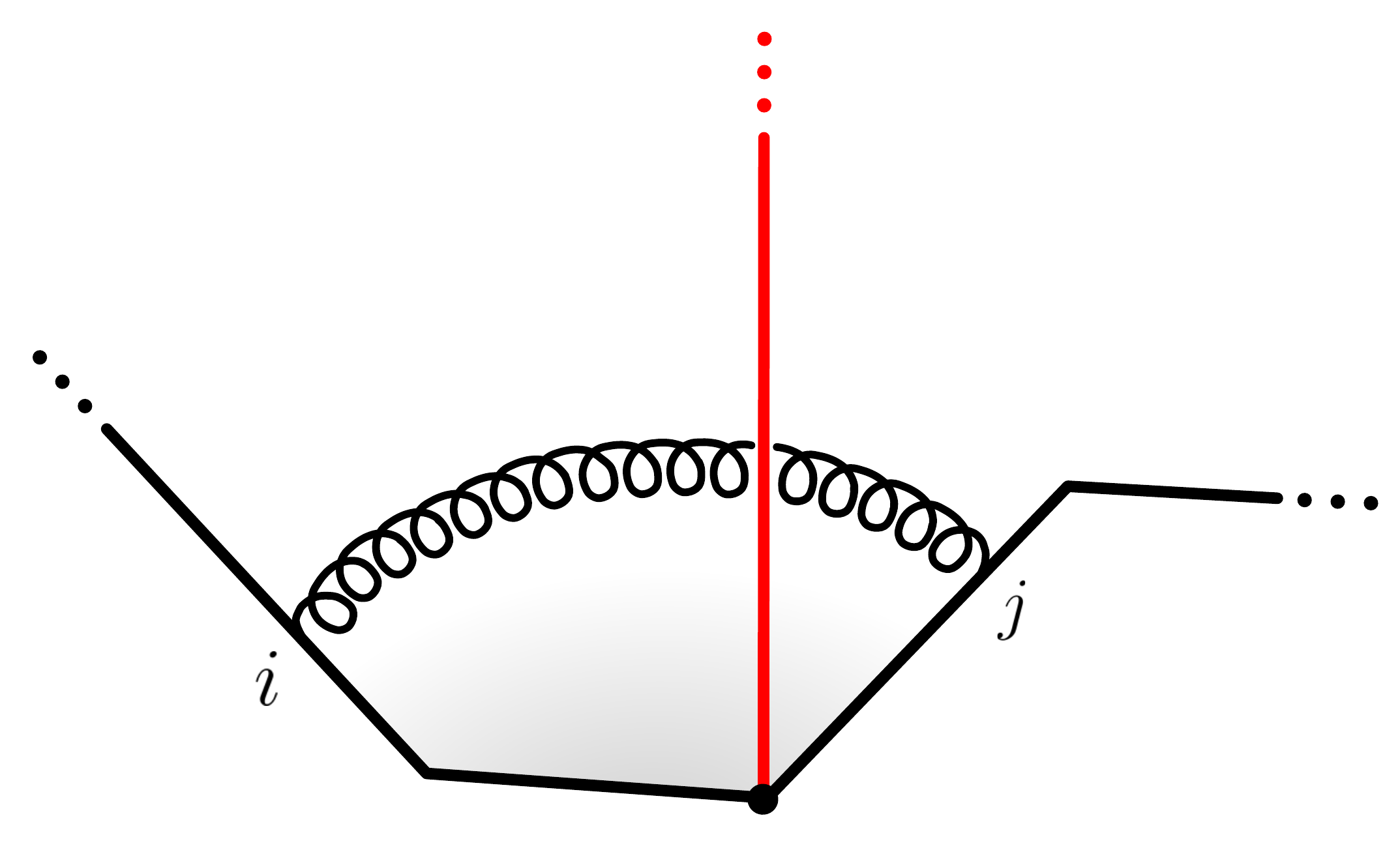}
  \end{minipage}
  \hspace{20pt}
  \caption{One-loop diagrams with a scalar emitted at a cusp and a gluon exchange between edges $i$ and $j$. Despite both diagrams being of the same order at large $N$, only the diagram in the left panel is allowed for a periodic Wilson loop and contributes to $W_{3,n}$. Taking into account diagrams with scalar and gluon propagators crossing over each other, like in the right panel, would lead to the decoupling of the singlet state, $\textrm{Tr}\, \phi^{12}$, and to the factorization of the super Wilson loop.}
  \label{fig:allowed_forbidden}
\end{figure}

Equipped with the vertex~\eqref{eq:delta-V}, we may proceed with the calculation of the matrix elements with the zero-momentum scalars. Let us illustrate the calculation for $k=3$. There are two types of contributions to this process at one loop. Firstly, we have the interactions between edges coming from the induced vertex (\ref{eq:delta-V}). Secondly, we have the tree-level contribution~(\ref{W3ntree}) dressed by gluon exchanges between edges. Collecting all the terms proportional to the same $F_{ij}$ we find
\begin{equation}
\langle \phi^{12}|\mathcal{V}^{(1)}_{ij}|0\rangle + \sum\limits_{l=j}^{i+n-1}\langle \phi^{12}|\mathcal{V}^{(0)}_{ll+1}|0\rangle = W^{\textrm{tree}}_{3, i+n-j+1} (j, j+1, \ldots , i+n)\ .
\end{equation}
Note that the sum over the position of the corner vertex $\mathcal{V}_{ll+1}$ has to be restricted from $l=j$ to $l=i+n-1$. This is to avoid the intersections of the propagators, as explained in figure \ref{fig:allowed_forbidden}. Finally, we should sum over inequivalent diagrams $F_{ij}$. For a closed Wilson loop, it entails summing over all pairs of edges, that is, $1\leqslant i<j \leqslant n$. For a periodic loop, each pair gives rise to two diagrams, for the two ways of measuring the distance between edges. In other words, the range of the sum is doubled, with one set of terms corresponding to $F_{ij}$ with $i<j$ and the other to $F_{ji+n}$. Taking it into account, we arrive at
\begin{equation}
W^{\textrm{1-loop}}_{3, n} = g^2 \sum_{i=1}^{n}\sum_{j=i+1}^{i+n-1} F_{ij} \times W^{\textrm{tree}}_{3, i+n-j+1} (j, j+1, \ldots , i+n)\ .
\end{equation}

The general formula for any $k$ may be derived by going along the same lines. It reads as the one above, but with $3\rightarrow k$. The expression may be further simplified by restricting the sum to $j < i+n-k+2$, using the fact that $W_{k, n'} = 0$ for $k>n'$. In particular, for $k = n$, that is, for the minimal form factors, only the divergent terms remain,
\begin{equation}
W^{\textrm{1-loop}}_{n, n}/ W^{\textrm{tree}}_{n,n} = g^2 \sum_{i=1}^n F_{ii+1} = -\,g^2 \sum_{i=1}^n \frac{(-s_{ii+1})^{\epsilon}}{\epsilon^2}\ .
\end{equation}
This result, as well as the general one for $k < n$, can be compared with the form-factor expressions derived in ref.~\cite{Penante:2014sza} using the generalized unitarity method~\cite{Bern:1994zx,Bern:2011qt}. They are in agreement with each other, up to the divergent parts. The latter only match after flipping the sign of the dimensional regulator, $\epsilon \rightarrow -\epsilon$.%
\footnote{One must also renormalize the coupling constants differently with $\mathcal{O}(\epsilon)$ terms.} This mild difference is not a novelty of the form-factor analysis and was previously observed for amplitudes~\cite{Drummond:2007aua}. It relates to the scale inversion of the T-duality transformation~\cite{Alday:2007hr} or, equivalently, to the fact that the divergences have different nature on each side of the duality.

\section{Form factor OPE}\label{section:FFOPE_main}

The previous analysis lends support to the idea that super form factors of half-BPS operators admit a dual description in terms of matrix elements of super Wilson loops. The super Wilson loops have the exact same structure as the ones used for the calculation of super amplitudes. The key difference lies in the presence of a state at infinity, which in our case carries a non-trivial R-charge.\par
It follows from this identification that the super form factors of half-BPS operators should admit an exact description in the collinear limit in terms of the OPE for null (super) Wilson loops~\cite{Alday:2010ku}. A great advantage of this approach is that it enables one to leverage the integrability method for calculating observables at finite coupling, as demonstrated for the scattering amplitudes in refs.~\cite{Alday:2010ku,Basso:2013aha}.\par
The essential building blocks for carrying out the OPE of form factors were unveiled recently in refs.~\cite{Sever:2020jjx,Sever:2021nsq,Sever:2021xga} for the stress-tensor multiplet, corresponding to $k=2$ in our notations. In this section, we explain how to extend these analyses to the form factors of any half-BPS operator.

\subsection{OPE description}

\begin{figure}[t]
\centering
\includegraphics[width=.5\textwidth]{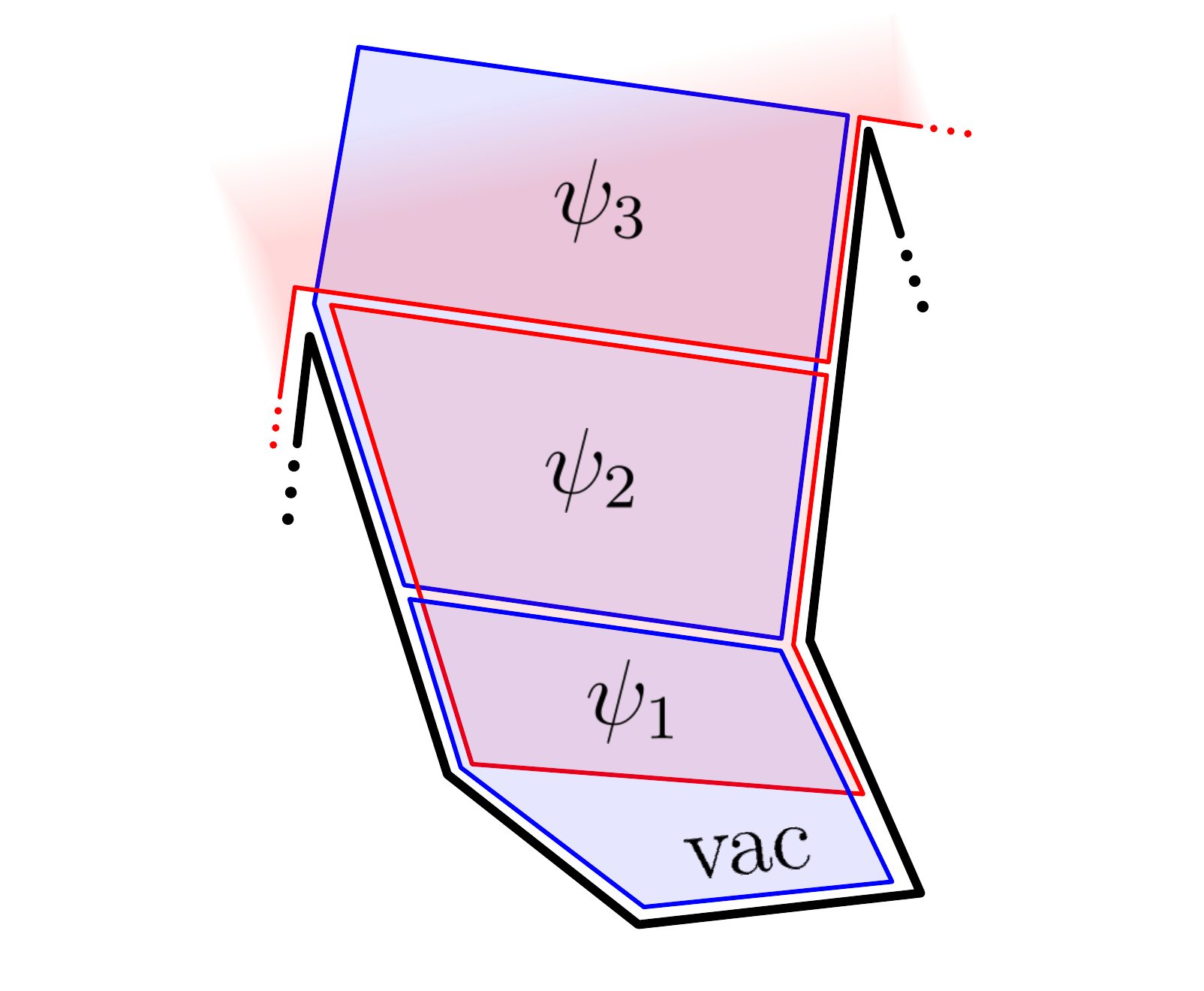}
\caption{Decomposition of a null periodic Wilson loop into a sequence of pentagons and a two-sided periodic Wilson loop, here for $n=5$. Every two consecutive pentagons overlap on a null square, which supports a flux-tube state. The state begins as the vacuum at the bottom and ends as a non-trivial state at the top, where it gets absorbed by the local operator.}
\label{fig:FFOPE}
\end{figure}

The core idea of the OPE approach is to decompose null polygonal Wilson loops into sequences of elementary building blocks, called pentagon transitions. All pentagon transitions are associated with null pentagon Wilson loops, which collectively cover the polygon. When considering a periodic Wilson line, one focuses on a given period and introduces a ladder of pentagons to cover it all the way up. This ladder ends with a two-sided periodic Wilson loop that overlaps with the final pentagon. This geometric decomposition is associated with a form-factor expansion in which flux-tube excitations are produced at the bottom of the picture and evolved successively across all intermediate OPE channels (null squares). The procedure is illustrated in figure \ref{fig:FFOPE}.\par 
The exact object that is subject to the OPE decomposition is a finite ratio of polygon Wilson loops, denoted $\mathcal{W}_{k,n}$, which we explicitly define in the following section. It is parameterised by $3n-7$ variables determined by the geometry of the Wilson loop, as described in refs.~\cite{Basso:2013vsa,Sever:2020jjx}. They are interpreted as OPE time ($\tau$), space ($\sigma$) and angle ($\phi$) variables. To the pentagon decomposition introduced above we associate a sum over the flux-tube eigenstates propagating on top of the Wilson loop geometry,
\begin{equation}\label{FFOPE_def}
\mathcal{W}_{k,n}=\sum\limits_{\psi}\left(\prod\limits_{j=1}^{n-2}e^{-E_j\tau_j+ip_j\sigma_j+im_j\phi_j}\right)P(0|\psi_1)\,\ldots\,P(\psi_{n-3}|\psi_{n-2})\,F_{\mathcal{T}_k}(\psi_{n-2})\ ,
\end{equation}
where in the final channel, the angular variable $\phi_{n-2}$ is equal to zero. Here, $E_{j}, p_{j}, m_{j}$ are the flux-tube energy, momentum and angular momentum of the state $\psi_{j}$, respectively. Each pentagon in the decomposition is associated with a pentagon transition $P(\psi_i|\psi_j)$, which, along with the corresponding integration measures $\mu(\psi_i)$, have been extensively studied and bootstrapped at finite coupling \cite{Basso:2013vsa,Basso:2013aha,Basso:2014koa,Basso:2014hfa,Basso:2015rta,Basso:2015uxa,Belitsky:2014sla,Belitsky:2014lta,Belitsky:2016vyq}.\par
The two-sided periodic Wilson loop is associated with the form factor transition $F_{\mathcal{T}_k}(\psi_i)$, which is the only part in eq.~(\ref{FFOPE_def}) that carries the dependence on the local operator. These objects have been introduced in ref.~\cite{Sever:2020jjx} and bootstrapped in the case of the stress-tensor multiplet $\mathcal{T}_2$ in ref.~\cite{Sever:2021xga}. The only non-zero contributions to the form factor transitions in this case come from states that are neutral under the $U(1)_\phi\times SU(4)_R$ symmetries of the null square, where the $U(1)_\phi$ factor encodes rotations in the plane transverse to the square and $SU(4)_R$ is the R-symmetry group.
This happens because under the duality~\eqref{FF-WL_duality} the operator $\mathcal{T}_2$ maps to the vacuum, $\langle\mathcal{F}_2| = \langle 0|$, which is invariant under the symmetries of the null square.\par
More generally, in the case of the operator $\mathcal{T}_k$, the states $\psi_{n-2}$ in the last OPE channel should have the same $U(1)_\phi\times SU(4)_R$ charges as the dual asymptotic state $\langle\mathcal{F}_k|$. As we know from section \ref{section:periodic_WL}, this state consists of $k-2$ identical scalars. As a result, $\psi_{n-2}$ should have no angular momentum, $m_{n-2} = 0$, and should transform in the representation with Dynkin labels $[0,k-2,0]$ of $SU(4)_{R}$. The lowest-energy states fulfilling these selection rules are easy to characterize: They are made out of $\ell = k-2$ identical flux-scalar scalars $\phi$. The associated form factor transition is a function of the rapidities $u_{1}, \ldots , u_{\ell}$, parametrizing the energy and momentum, $E = \sum_{i=1}^{\ell}E_{\phi}(u)$ and $p = \sum_{i=1}^{\ell}p_{\phi}(u_{i})$, of the scalars. We denote it as
\begin{equation}
F_{\phi^{\ell}}(u_{1}, \ldots , u_{\ell}) = \langle \mathcal{F}_{k} | \phi(u_{1})\, \ldots , \phi(u_{\ell})\rangle\, .
\end{equation}
Its knowledge is enough to determine the leading OPE contribution to the Wilson loop $\mathcal{W}_{k, n}$ at any value of the coupling, for any $n$ and $k$.

\subsection{Form factor transition axioms}\label{section:axioms}

\begin{figure}[t]
\centering
\includegraphics[width=.6\textwidth]{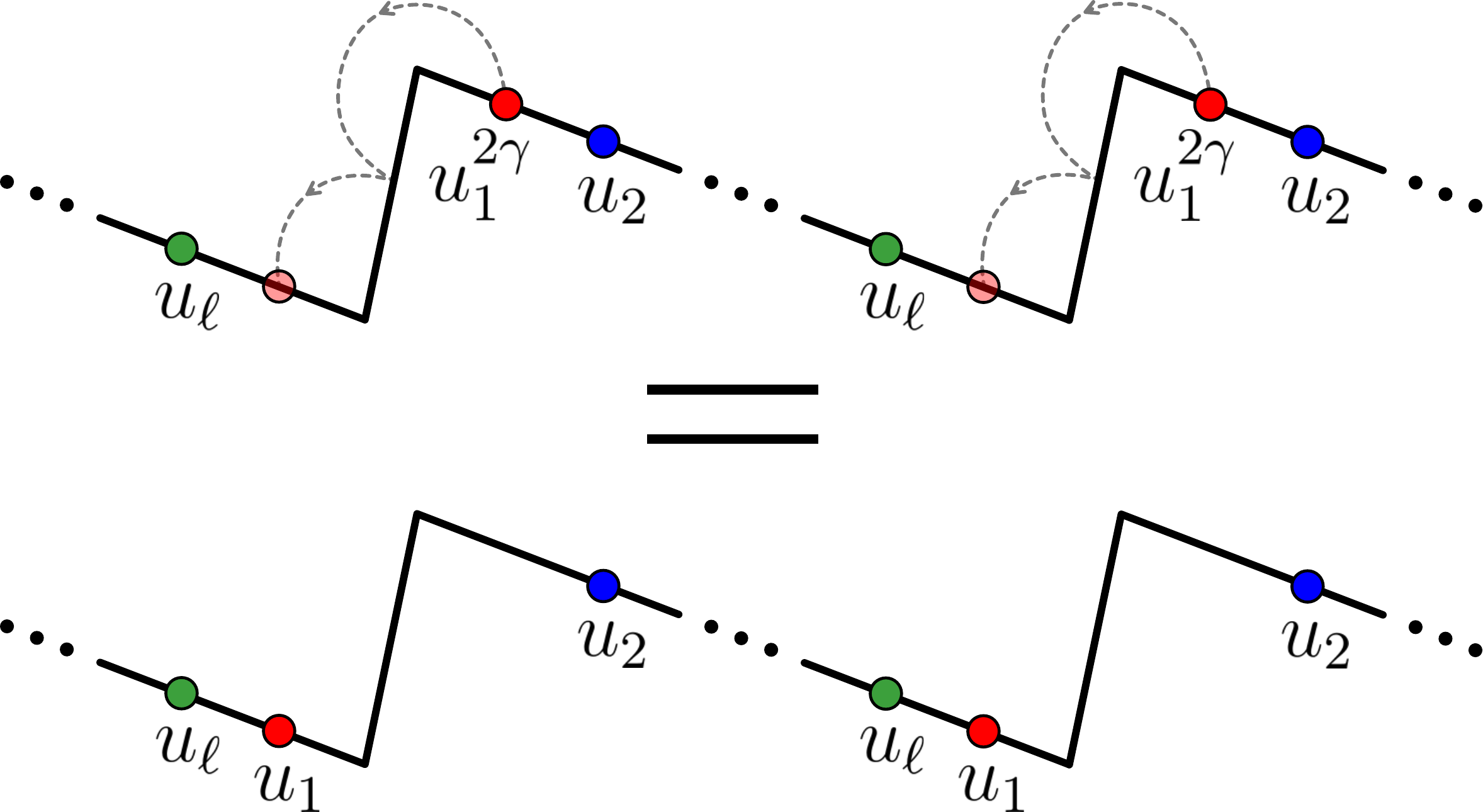}
\caption{Crossing symmetry of the form factor transition. Two consecutive mirror transformations applied to the first excitation shifts it to the last position.}
\label{fig:crossing}
\end{figure}

The form factor transition $F_{\phi^\ell}(u_1,\ldots,u_\ell)$ is subject to a set of non-perturbative axioms, which mimic those proposed for the stress-tensor multiplet in refs.~\cite{Sever:2020jjx,Sever:2021xga}. They may be used to bootstrap this function at finite coupling for any choice of the rapidities. Note that since all $\ell$ scalars in this transition are identical, the R-symmetry matrix part is trivial. It implies that we can suppress all R-indices and treat $F_{\phi^{\ell}}$ as a scalar quantity. It leads us to the following set of axioms:

\begin{description}
\item[\textbf{Watson}] Adjacent excitations may be permuted using the flux-tube S-matrix,
\begin{equation}\label{axiom:Watson}
F_{\phi^\ell}\left(\ldots,u_i,u_{i+1},\ldots\right) = S_{\phi\phi}(u_i,u_{i+1})\,F_{\phi^\ell}\left(\ldots,u_{i+1},u_i,\ldots\right)\ ,
\end{equation}
where $S_{\phi\phi}(u, v)$ is the scattering phase between scalars in the symmetric channel.
\item[\textbf{Reflection}] Periodic Wilson loops are invariant under spacetime reflections. It implies that form-factor transitions stay invariant under a flip of the order of the excitations, accompanied by a sign flip of their rapidites,
\begin{equation}
F_{\phi^\ell}\left(u_1,\ldots,u_\ell\right) = F_{\phi^\ell}\left(-u_\ell,\ldots,-u_1\right).
\end{equation}
\item[\textbf{Mirror}] The flux-tube theory is invariant under the mirror transformation $u\rightarrow u^{\gamma}$ exchanging energy and momentum~\cite{Alday:2007mf,Basso:2011rc}, $E_{\phi}(u^{\gamma}) = ip_{\phi}(u)$ and $p_{\phi}(u^{\gamma}) = iE_{\phi}(u)$. This transformation transports the excitations to the next edge of the Wilson loop,
\begin{equation}
F_{\phi^\ell}\left(u_1^\gamma,\ldots,u_\ell^\gamma\right) = F_{\phi^\ell}\left(u_1,\ldots,u_\ell\right)\ ,
\end{equation}
leaving the transition invariant.
\item[\textbf{Crossing}]
Applying the mirror transformation twice to the first excitation translates it to the periodic image of the same edge,
\begin{equation}\label{axiom:crossing}
F_{\phi^\ell}\,(u_1^{2\gamma},\ldots,u_\ell) = F_{\phi^\ell}\left(u_2,\ldots,u_\ell,u_1\right)\ ,
\end{equation}
as illustrated in figure \ref{fig:crossing}.
\item[\textbf{Square limit}]
Unlike the R-singlet transitions studied in ref.~\cite{Sever:2021xga}, the symmetric transition $F_{\phi^\ell}$ must vanish in the kinematic limit where two rapidities coincide, $u_i \to u_j$,
\begin{equation}\label{axiom:square-limit}
\lim\limits_{u_i\to u_j}F_{\phi^\ell}\,(u_1,\ldots,u_\ell) = 0\ .
\end{equation}
This condition is due to identical scalars obeying Fermi-like statistics, $S_{\phi\phi}(u, u) = -1$.
\end{description}

\subsection{Tilted transitions}\label{sec:tilted-Pmu}

Up to minor differences, the axioms obeyed by the symmetric transitions are identical to the ones that govern the singlet transitions analyzed in ref.~\cite{Sever:2021xga}. In particular, the crossing axiom~(\ref{axiom:crossing}) is common to both singlet and non-singlet transitions. The solution to this equation that has been found in ref.~\cite{Sever:2021xga} uses the so-called octagon kernel~\cite{Belitsky:2019fan,Basso:2020xts}, which also plays a role in various other contexts, see e.g.~refs.~\cite{Coronado:2018cxj,Kostov:2019stn,Belitsky:2019fan,Bargheer:2019exp,Basso:2020xts,Belitsky:2020qrm,Basso:2022ruw}. Here, we will repurpose this solution to also describe the non-singlet states, while providing a natural generalization of the building blocks that enter its construction.\par
The key object that controls the coupling dependence of all the Wilson loop OPE constituents is the Beisert-Eden-Staudacher (BES) kernel~\cite{Beisert:2006ez}. It may be cast into the form of a semi-infinite matrix, with elements
\begin{align}\label{eq: BES kernel}
\mathbb{K}_{ij}=2j(-1)^{ij+j}\int\limits_0^\infty\frac{dt}{t}\frac{J_i(2gt)J_j(2gt)}{e^t-1}\ ,
\end{align}
where $i, j = 1, 2, \ldots$ and $J_{i}$ is the $i$-th Bessel function of the first kind. In ref.~\cite{Basso:2020xts}, a tilted version of the kernel (\ref{eq: BES kernel}) has been introduced, with an angle-like parameter $\alpha$ making the kernel act differently on the spaces of even and odd Bessel functions,
\begin{align}\label{eq: tilted BES kernel}
\mathbb{K}(\alpha)=2\cos\alpha\left[\begin{array}{cc}
             \cos\alpha\,\mathbb{K}_{\circ\circ} &\sin\alpha\,\mathbb{K}_{\circ\bullet}\\
            \sin\alpha\,\mathbb{K}_{\bullet\circ} &\cos\alpha\,\mathbb{K}_{\bullet\bullet}
            \end{array}\right]\ .
\end{align}
Here, $\mathbb{K}_{\circ\circ}$ is the overlap of odd Bessel functions with odd Bessel functions, $\mathbb{K}_{\circ\bullet}$ is the overlap of odd Bessel functions with even Bessel functions, etc. Setting $\alpha=\pi/4$ returns the original BES kernel. Setting $\alpha=0$ gives the octagon kernel, for which there is no mixing between even and odd components. Hence, the auxiliary parameter $\alpha$ allows us to interpolate between the two kernels entering the construction of the form factor transitions.\par
The other important ingredients are the infinite vectors $(\kappa^{u})_{j}, (\tilde{\kappa}^{u})_{j}$, with $j = 1,2, \ldots$, carrying the information about the rapidity $u$ of the scalar excitation. These objects too may be equipped with the parameter $\alpha$ such as to cover all cases at once. The natural deformation that fits this purpose is
\begin{align}\label{eq: source terms}
\kappa^u_\alpha = 2\cos{\alpha}\left[\begin{array}{cc}
            \sin{\alpha}\,\kappa^u_{\phi,\circ}\\
            \cos{\alpha}\,\kappa^u_{\phi,\bullet}
            \end{array}\right] ,\qquad \tilde{\kappa}^v_\alpha = 2\cos{\alpha}\left[\begin{array}{cc}
            \cos{\alpha}\,\tilde{\kappa}^u_{\phi,\circ}\\
            \sin{\alpha}\,\tilde{\kappa}^u_{\phi,\bullet}
            \end{array}\right] ,
\end{align}
where
\begin{equation}
\begin{aligned}
\kappa_{\phi,j}^u=-\int\limits_0^\infty\frac{dt}{t}\,J_j(2gt)\,\frac{\cos(ut)\,e^{t/2}-J_0(2gt)}{e^t-1}\ ,\,\,\, \tilde{\kappa}_{\phi,j}^u = (-1)^{j+1}\int\limits_0^\infty\frac{dt}{t}\,J_j(2gt)\,\frac{\sin(ut)\,e^{t/2}}{e^t-1}\ .
\end{aligned}
\end{equation}
By contracting kernel and vectors together, we define the two so-called $f$-functions used to construct pentagon and form factor transitions~\cite{Basso:2013pxa,Basso:2013aha,Sever:2021xga}. They read
\begin{equation}\label{tiltedf}
\begin{aligned}
&f_a^{[\alpha]}(u,v)=\frac{1}{\cos^2{\alpha}}\left[\kappa^u_\alpha\,\mathbb{Q}\,\frac{1}{1+\mathbb{K}(\alpha)}\,\tilde{\kappa}^v_\alpha-\tilde{\kappa}^u_\alpha\,\mathbb{Q}\,\frac{1}{1+\mathbb{K}(\alpha)}\,\kappa^v_\alpha \right] , \\
&f_s^{[\alpha]}(u,v)=\frac{1}{\cos^2{\alpha}}\left[\kappa^u_\alpha\,\mathbb{Q}\,\frac{1}{1+\mathbb{K}(\alpha)}\,\kappa^v_\alpha-\tilde{\kappa}^u_\alpha\,\mathbb{Q}\,\frac{1}{1+\mathbb{K}(\alpha)}\,\tilde{\kappa}^v_\alpha\right] ,
\end{aligned}
\end{equation}
where
\begin{equation}
  \frac{1}{1+\mathbb{K}(\alpha)} = 1- \mathbb{K}(\alpha) + \mathbb{K}(\alpha)^2 -\ldots \ .
\end{equation}
Here, $\mathbb{Q}$ is a diagonal matrix with elements $\mathbb{Q}_{ij} = i \left(-1\right)^{i+1}\delta_{ij}$, defined such as to make the product $\mathbb{Q}\, [1+\mathbb{K}(\alpha)]^{-1}$ symmetric. It immediately follows that the functions $f_{s}$ and $f_{a}$ are, respectively, symmetric and antisymmetric under the exchange of rapidities, $f_s^{[\alpha]}(v,u) = f_s^{[\alpha]}(u,v)$ and $f_a^{[\alpha]}(v,u) = -\,f_a^{[\alpha]}(u,v)$.\par
With these functions at hand, we can now define a tilted version of the pentagon transition,
\begin{equation}\label{tiltedP}
\begin{aligned}
&P_{\phi\phi}^{[\alpha]}(u|v) = \frac{\Gamma\left(iu-iv\right)}{g^2\,\Gamma\left(\frac{1}{2}+iu\right)\Gamma\left(\frac{1}{2}-iv\right)}\,\exp\left[J_\phi(u) + J_\phi(-v) +if^{[\pi/4]}_a(u,v) + f^{[\alpha]}_s(u,v)\right],
\end{aligned}
\end{equation}
and an associated tilted measure,
\begin{equation}\label{tiltedMu}
\mu_\phi^{[\alpha]}(u) = \frac{\pi g^2}{\cosh{\pi u}}\,\exp\left[-J_\phi(u)-J_\phi(-u) - f^{[\alpha]}_s(u,u)\right],
\end{equation}
defined canonically through the residue of~\eqref{tiltedP} at $v=u$,
\begin{equation}
    \underset{v=u}{\rm res}\, P^{[\alpha]}_{\phi\phi}(u|v) = \frac{i}{\mu^{[\alpha]}_{\phi}(u)}\ .
\end{equation}
Here, $J_\phi(u)$ is a fixed integral given by
\begin{align}\label{junk}
J_\phi(u) = \frac{1}{2}\int\limits_0^\infty \frac{dt}{t}\left(J_0(2gt)-1\right)\frac{J_0(2gt)+1-2\,e^{t/2-iut}}{e^t-1}\ .
\end{align}
Notice that in the definition~\eqref{tiltedP} only the symmetric $f$-function is tilted, whilst the antisymmetric one stays at the same angle $\alpha=\frac{\pi}{4}$. This choice ensures that the transition (\ref{tiltedP}) satisfies the same Watson axiom for all values of the parameter $\alpha$,
\begin{align}\label{eq:PalphaS}
P_{\phi\phi}^{[\alpha]}(u|v) = S_{\phi\phi}(u,v)\,P_{\phi\phi}^{[\alpha]}(v|u)\ .
\end{align}
The same is true for the mirror axiom, $P^{[\alpha]} (u^{\gamma}| v^{\gamma}) = P^{[\alpha]}(u|v)$ for any $\alpha$, as shown in appendix~\ref{appendix: crossing}.
It implies, in particular, that the measure~\eqref{tiltedMu} is mirror symmetric, $\mu^{[\alpha]}(u^{\gamma}) = \mu^{[\alpha]}(u)$. However, the transformation properties of the transition under mirror moves of a single rapidity are heavily dependent on the value of $\alpha$. For the two specific values of $\alpha$ that are relevant for the form factor construction, we have
\begin{align}\label{eq:crossing-alpha}
P_{\phi\phi}^{[\pi/4]}(u^{-\gamma}|v) = P_{\phi\phi}^{[\pi/4]}(v|u)\ ,\qquad P_{\phi\phi}^{[0]}(u^{-2\gamma}|v) = P_{\phi\phi}^{[0]}(v|u)\ ,
\end{align}
where $u^{-\gamma}$ stands for the inverse mirror rotation.\par
The transition (\ref{tiltedP}) naturally interpolates between the two cases of interest: $\alpha=\frac{\pi}{4}$ and $\alpha=0$. When $\alpha$ is set to its untilted value of $\frac{\pi}{4}$, all the objects introduced above get reduced to their original values, used to construct pentagon transitions and measures,
\begin{align}
P_{\phi\phi}(u|v) = P_{\phi\phi}^{[\pi/4]}(u|v)\ ,\qquad \mu_{\phi}(u) = \mu_{\phi}^{[\pi/4]}(u)\ ,
\end{align}
For $\alpha=0$, we denote
\begin{align}\label{QandNu}
Q_{\phi\phi}(u|v) = P_{\phi\phi}^{[0]}(u|v)\ ,\qquad \nu_{\phi}(u) = \mu_{\phi}^{[0]}(u)\ .
\end{align}
Below, we use these building blocks to construct non-perturbative expressions for scalar form factor transitions that satisfy the axioms of section \ref{section:axioms}.
\subsection{Form factor transitions and measures}
One can see that the transition $Q_{\phi\phi}(u|v)$ is proportional to the singlet form-factor transition $F_{\phi\bar{\phi}}(u,v)$ defined in ref.~\cite{Sever:2020jjx}.
This is because the function $f_s^{[\alpha]}$ is tailored to match the difference of the form-factor building blocks, $f_{5}$ and $f_6$, of ref.~\cite{Sever:2020jjx}, when $\alpha=0$,
\begin{align}
f_s^{[0]}(u,v) = 2f_5(u,v) - 2f_6(u,v)\ .
\end{align}
To be precise, $F_{\phi\bar{\phi}}$ is identical to $Q_{\phi\phi}$, up to a simple overall factor,
\begin{align}\label{eq:singlet-F-to-Q}
F_{\phi\bar{\phi}}(u,v) = -\,\frac{4}{\left(u-v-2i\right)\left(u-v-i\right)}\,\sqrt{\frac{\nu_\phi(u)\,\nu_\phi(v)}{\mu_\phi(u)\,\mu_\phi(v)}}\,Q_{\phi\phi}(u|v)\ .
\end{align}
The rational function of $u-v$ takes into account that $F_{\phi\bar{\phi}}$ satisfies the singlet Watson relation, rather than the symmetric one,
\begin{align}
F_{\phi\bar{\phi}}(u,v) = S_{\phi\bar{\phi}}(u,v)\,F_{\phi\bar{\phi}}(v,u)\ ,
\end{align}
with the flux-tube scattering phases, in the singlet ($\phi\bar{\phi}$) and symmetric ($\phi\phi$) channels, obeying
\begin{equation}
    S_{\phi\bar{\phi}}(u,v) = \frac{\left(u-v+2i\right)}{\left(u-v-2i\right)}\frac{\left(u-v+i\right)}{\left(u-v-i\right)}\,S_{\phi\phi}(u,v)\ .
\end{equation}
The remaining factors are required to match the normalization used in ref.~\cite{Sever:2020jjx}.\par
The normalization of the transition is directly tied to that of the measure. Namely, we have a freedom of simultaneously rescaling
\begin{align}\label{eq:scaling-sym}
F_{\phi\bar{\phi}}(u,v) \to f(u)\,f(v)\,F_{\phi\bar{\phi}}(u,v)\quad\text{and}\quad \mu_\phi(u)\to f^{-1}(u)\,\mu_\phi(u)\ ,
\end{align}
as one can see by looking at how form factor transitions and measures are assembled into the OPE integrals in refs.~\cite{Sever:2020jjx,Sever:2021xga}. The choice made in ref.~\cite{Sever:2020jjx} is to use the same scalar measure $\mu_{\phi}$ in \textit{all} OPE channels, including the final one. Our analysis suggests that a nicer normalization for the form factor transitions is found after performing a rescaling by
\begin{equation}
    f_{\textrm{nice}}(u) = \sqrt{\frac{\mu_\phi(u)}{\nu_\phi(u)}}\ ,
\end{equation}
which maps eq.~\eqref{eq:singlet-F-to-Q} to the simpler relation
\begin{align}
F_{\phi\bar{\phi}}(u,v)_{\textrm{nice}} = -\,\frac{4\,Q_{\phi\phi}(u|v)}{\left(u-v-2i\right)\left(u-v-i\right)}\ .
\end{align}
Let us stress that this rescaling affects the definition of the integration measure in the final OPE channel. According to eq.~\eqref{eq:scaling-sym} the latter has to be replaced by
\begin{align}\label{measuregeneral}
\mu_{\phi}(u) \longrightarrow \sqrt{\mu_\phi(u)\,\nu_\phi(u)}\ ,
\end{align}
while measures in all other channels remain untouched.\par
Having set our normalization scheme, we can now present a symmetric scalar form factor transition that is consistent with all the axioms of section \ref{section:axioms}. Following the parallelism with the pentagon transitions, we pick a minimal solution that is totally factorized in terms of the elementary ($\alpha=0$) transition $Q_{\phi\phi}$. Namely, our main conjecture is that
\begin{align}\label{FFtransgeneral}
F_{\phi^\ell}(u_1,\ldots,u_\ell) = \prod\limits_{i<j}^{\ell} \frac{1}{Q_{\phi\phi}(u_j|u_i)}\ .
\end{align}
It is readily seen to meet all the requirements given in section~\ref{section:axioms}. In particular, the Watson and crossing axioms, eqs.~\eqref{axiom:Watson} and~\eqref{axiom:crossing}, follow from eqs.~\eqref{eq:PalphaS} and~\eqref{eq:crossing-alpha}, respectively, using the unitarity of the S-matrix, $S_{\phi\phi}(u, v)\,S_{\phi\phi}(v,u)=1$, and the mirror symmetry, $Q_{\phi\phi}(u|v^{2\gamma}) = Q_{\phi\phi}(u^{-2\gamma}|v) = Q_{\phi\phi}(v|u)$. One can also verify that the transition vanishes whenever two rapidities coincide, $u_{i} \rightarrow u_{j}$, in agreement with eq.~\eqref{axiom:square-limit}.\par
Mind that the solution~\eqref{FFtransgeneral} implies that the one-point function is equal to $1$,
\begin{equation}
F_{\phi}(u)=1\ .
\end{equation}
This equation is similar to the condition used in ref.~\cite{Basso:2013aha} to normalize charged pentagon transitions. It may be viewed as an alternative way of defining our normalization scheme.%
\footnote{In comparison, the one-point function would read $F_{\phi}(u) = 1/f_{\textrm{nice}}(u)$ in the normalization used in ref.~\cite{Sever:2021xga}.}
Nonetheless, it does not imply that the absorption of a single scalar by the BPS operator is trivial, just that the dynamical information is contained in the tilting of the measure of integration~\eqref{measuregeneral}, controlled by $\mu_{\phi}$ and $\nu_{\phi}$.\par
The transition (\ref{FFtransgeneral}), along with the measure (\ref{measuregeneral}), are sufficient to construct the leading OPE contribution to MHV form factors of any BPS operator $\mathcal{T}_k$, at any value of the coupling constant $g^2$.\par
Subleading contributions are more involved. They comprise more flux-tube particles that may form states with the right quantum numbers in multiple ways. Nonetheless, we expect them to enjoy the same factorization properties as the singlet multi-particle transitions studied in ref.~\cite{Sever:2021xga}. Namely, general transitions should be obtained by dressing the basic ones~(\ref{FFtransgeneral}) with any number of two-particle singlet transitions, using proper matrix parts to account for the R-symmetry structures. We leave their detailed study to a future investigation.

\section{Matching data}\label{section:matching_data}

In this section, we put our conjectures to the test at weak coupling, through a comparison with direct diagrammatic results. At first, we will outline a few necessary steps needed to make this comparison possible, by recalling how to deal with the UV divergences of the super Wilson loop and how to address its helicity components in the OPE formalism.

\subsection{Finite ratios and dual conformal symmetry}

\begin{figure}[t]
\centering
\includegraphics[width=.75\textwidth]{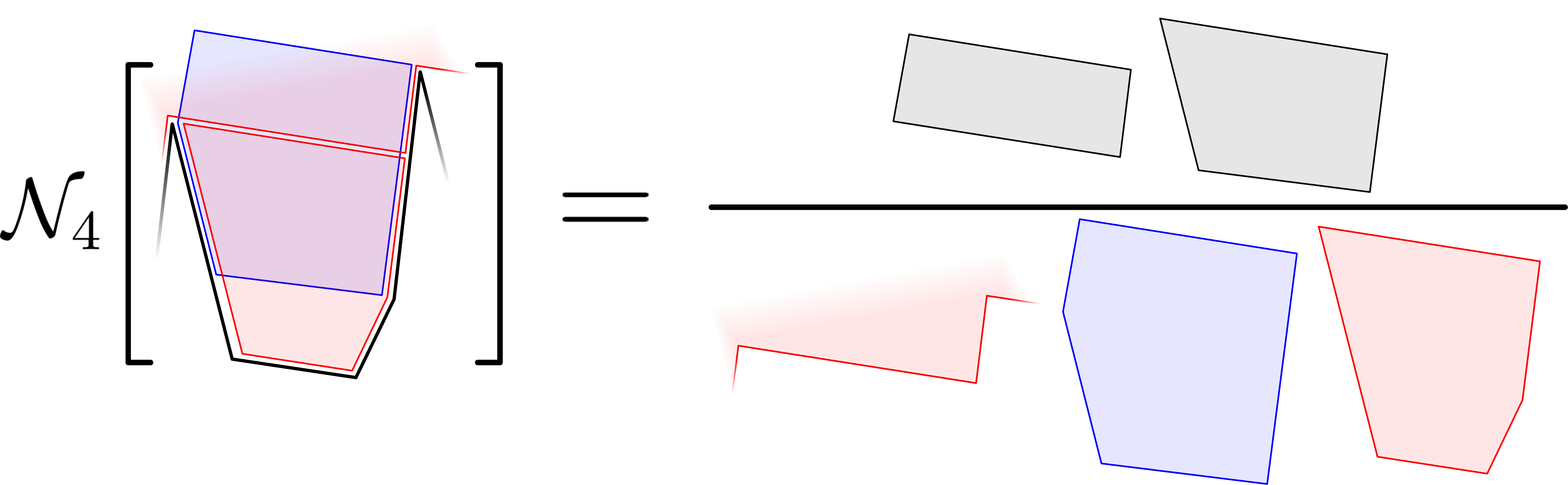}
\caption{Operator-independent universal framing factor $\mathcal{N}_n$ for the $n=4$ case. It is given by the product of all the internal squares of the polygon, divided by the product of the internal pentagons and the two-sided periodic polygon.}
\label{fig:ratio_def}
\end{figure}

The OPE approach requires us to regularize the UV divergences of the super Wilson loop. Thankfully, form factors of half-BPS operators with same external states and, by extension, their super Wilson loop duals, all have same divergent behavior. This feature allows us to define a universal OPE regularization factor $\mathcal{N}_n$, which is independent of the operator and identical to the one used in ref.~\cite{Sever:2020jjx} for the stress-tensor multiplet $(k=2)$. The construction of this factor is illustrated in figure \ref{fig:ratio_def}. It leads us to the following definition of the finite OPE ratio, that we will be using to match perturbative results,
\begin{align}\label{OPE_ratio}
\mathcal{W}_{k,n}\left(1,\ldots,n;q\right) = |q|^{2-k} \,\mathcal{N}_n(1,\ldots,n)\,W_{k,n}\left(1,\ldots,n;q\right)\ ,
\end{align}
with the Parke-Taylor-Nair stripped form factor defined in eq.~(\ref{FFnormalized}) and $|q| =\sqrt{-q^2}$. Another convenient way to remove UV divergences is by defining a ratio function, similar to the one introduced for non-MHV amplitudes \cite{Drummond:2008vq},
\begin{align}\label{eq:ratio-function}
\mathcal{R}_{k,n}\left(1,\ldots,n;q\right) =  |q|^{2-k}\frac{W_{k,n}\left(1,\ldots,n;q\right)}{W_{2,n}\left(1,\ldots,n;q\right)}\ .
\end{align}
Besides being finite, this ratio has the nice property of being invariant under cyclic permutations and reflections of the kinematic data, like the original bare form factor. This is not the case for the OPE ratio~\eqref{OPE_ratio} designed to study form factors around a particular multi-collinear limit. The drawback is that the ratio function~\eqref{eq:ratio-function}, unlike the OPE ratio~\eqref{OPE_ratio}, does not have a clear flux-tube interpretation beyond the leading order in the collinear expansion. Nonetheless, one may always go from one to the other using
\begin{equation}
\mathcal{R}_{k,n} = \mathcal{W}_{k, n} / \mathcal{W}_{2, n}\ .
\end{equation}
Unlike the Wilson loops dual to amplitudes, those dual to form factors are not scale invariant. In the original units, $W_{k, n}$ has mass dimension $k-2$, with $k$ coming from the dimension of the half-BPS operator and the offset from the various factors that we stripped off. Its dimension is opposite in the T-dual units.\footnote{Recall that in the T-dual picture the kinematic variables may be assigned the mass dimensions $[\lambda] = 1/2, \, [\tilde{\lambda}] = -3/2,\, [\eta] = 0,$ from where it follows that the R-invariant $(ijk)$ has the dimension of a length. Being an homogeneous polynomial of degree $k-2$ in the $(ijk)$'s, $W_{k, n}$ has dimension $2-k$.} The factors $|q|^{2-k}$ in eqs.~\eqref{OPE_ratio} and~\eqref{eq:ratio-function} have been introduced to absorb this dimension and make the ratios~\eqref{OPE_ratio} and~\eqref{eq:ratio-function} scale invariant.

It is worth mentioning that one can think of these ratios as being fully dual conformal invariant in the sense of refs.~\cite{Ben-Israel:2018ckc,Bianchi:2018rrj,Bork:2014eqa}. To see that, recall first that the geometry of a closed super Wilson loop may be specified by a set of $n$ momentum supertwistors $\mathcal{Z}_i = (Z_{i} | \eta_{i})$, defined as in eq.~\eqref{eq:super-Z}, with dual conformal transformations acting linearly on the (bosonic components) $Z_i$'s as $SL(4)$ transformations. Dual conformal symmetry implies that the vacuum expectation value of the super Wilson loop, or, more precisely, its finite part, is invariant under global $SL(4)$ transformations of all the twistors,
\begin{equation}
    Z_{i} \rightarrow \mathcal{K}\cdot Z_{i}\ ,
\end{equation}
for any $\mathcal{K} \in SL(4)$. As a result, it may be expressed in terms of $SL(4)$-invariants 4-brackets
\begin{equation}
    \langle i j k l \rangle = \textrm{Det}\, [Z_{i}, Z_{j}, Z_{k}, Z_{l}]\ ,
\end{equation}
built from the $n$ twistors. In the periodic Wilson loop case, apart from the original set of $n$ twistors describing the loop in a given period, we are also required to consider their images in other periods. For this purpose, one introduces a shift operator $P \in SL(4)$ that implements the translation by $q$ in momentum twistor space, see eq.~\eqref{eq:shifts}. Periodic images are defined by acting with $P$, or its inverse, on the twistors, with
\begin{equation}
    Z^{\pm m}_i = P^{\pm m}\cdot Z_i
\end{equation}
denoting the $i$-th twistor in the $\pm m$-th period. The vev of the periodic Wilson loop may then be viewed as a dual conformal invariant object if, apart from transforming the original set of twistors $Z_i$, $i=1\ldots n$, one also acts on the periodicity operator $P$. Namely,
\begin{align}\label{eq:dual-conf-sym}
\mathcal{W}(\mathcal{K}\cdot Z_1,\ldots,\mathcal{K}\cdot Z_n;\mathcal{K}\cdot P\cdot \mathcal{K}^{-1}) = \mathcal{W}(Z_1,\ldots,Z_n;P)\ ,
\end{align}
for any $\mathcal{K} \in SL(4)$.\par
This relation extends to the non-trivial matrix elements dual to the form factors of $\mathcal{T}_k$, after observing that the 2-brackets entering these quantities can be reexpressed in terms of 4-brackets of twistors and shifted images. Precisely, for any $i, j$, one has
\begin{equation}\label{eq:2-to-4}
\langle ij\rangle^2 = \frac{\langle i\, j\, i^+\, j^+\rangle }{q^2}\ ,    
\end{equation}
with $\langle i\, j\, i^+\, j^+\rangle  = \textrm{Det}\, [Z_{i}, Z_{j}, P\cdot Z_{i}, P\cdot Z_{j}]$.
The overall dependence on $q^2$ drops out in our ratio~\eqref{OPE_ratio}, leaving an expression in terms of 4-brackets that is dual conformal invariant in the sense of eq.~\eqref{eq:dual-conf-sym}. Alternatively, one may introduce the infinity twistors $\mathcal{I}_{1,2}$ spanning the invariant subspace of $P$, i.e., $(1-P)\cdot \mathcal{I}_{1,2} = 0$. With their help, one can express 2-brackets more directly using $\langle ij\rangle \propto \textrm{Det}\, [Z_i Z_j \mathcal{I}_1 \mathcal{I}_2]$. The scale ambiguity associated with their normalization vanishes when considering scale-invariant quantities. Throughout the rest of this section, we will be handling our Wilson loop ratios in this manner, using the specific parametrizations of twistors $Z_i, \mathcal{I}_{1,2}$ and shift operator $P$ given in appendix~\ref{appendix:twistors}.

\subsection{Super pentagons and helicity map}\label{sec:map}

With the half-BPS operators introducing extra units of R-charge to the states that are being absorbed by them, the pentagons that create these states have to be charged accordingly. This procedure can be done with the help of the super-OPE formalism developed in refs.~\cite{Basso:2014hfa,Basso:2015rta}. In a nutshell, one has to equip each pentagon with a Grassmann parameter $\chi^{A}$, in the fundamental representation of $SU(4)$, and replace regular pentagon transitions with their supersymmetric counterparts,
\begin{equation}
\mathbb{P} = P+\chi^A\,P_A+\frac{\chi^A\chi^B}{2}\,P_{AB}+\frac{\chi^A\chi^B\chi^C}{3!}\,P_{ABC}+\chi^1\chi^2\chi^3\chi^4\,P_{1234}\ .
\end{equation}
Grassmann variables $\chi^A$ are responsible for charging the specific pentagon transition, with $P$ denoting the bosonic pentagon, $P_{A}$ its fermionic partner with one unit of charge, etc. The series terminates at $P_{1234}$, which stands for the $\overline{\textrm{MHV}}$ pentagon. Note that to describe MHV form factors, only the chiral half spanned by $\chi^{a}$ with $a=1,2$ is needed; the series terminates then at the $SU(2)$ singlet pentagon $P_{12}$, which produces the scalar excitation of interest.\par
\begin{figure}[t]
\centering
\includegraphics[width=.50\textwidth]{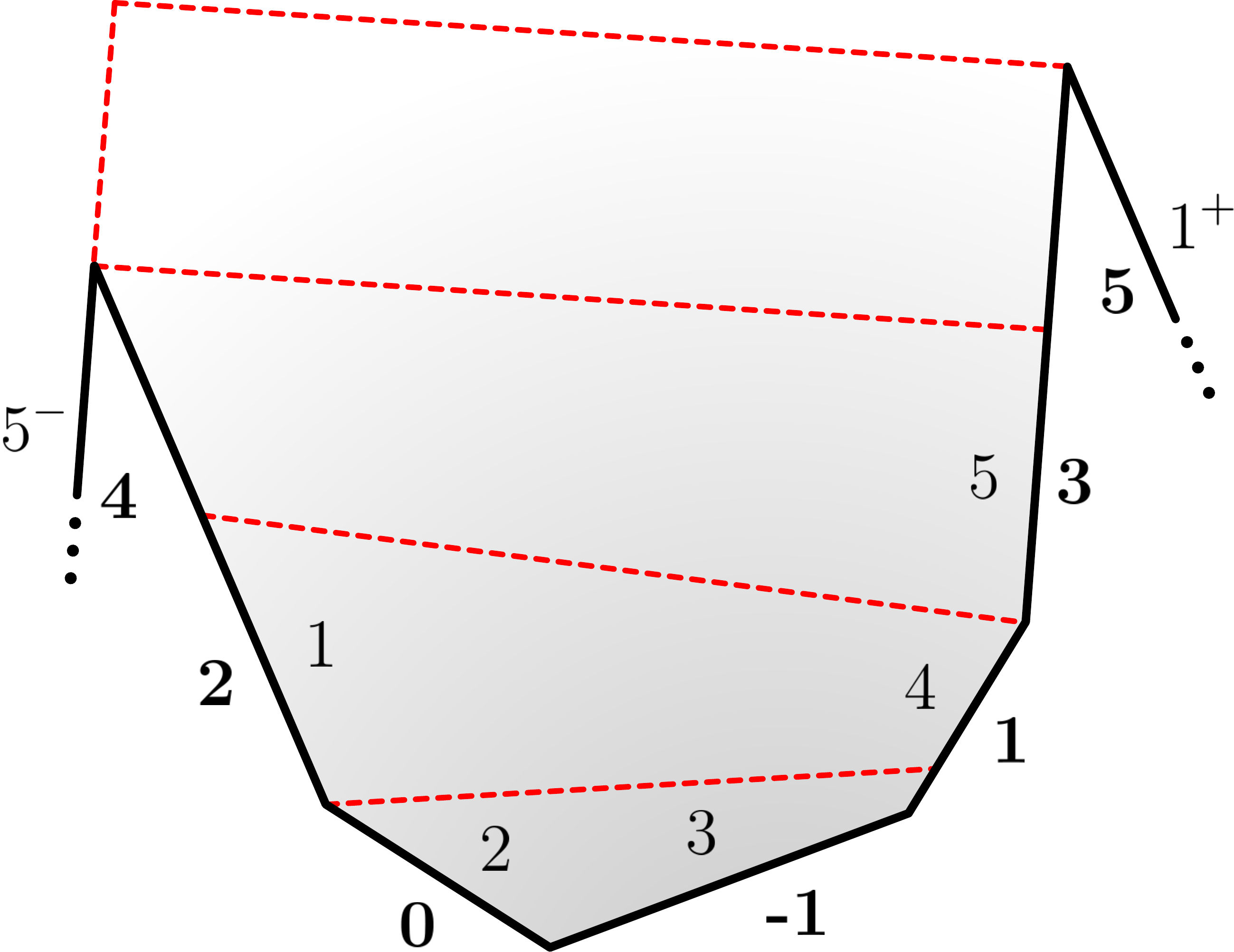}
\caption{OPE-friendly labelling of edges for a five-point form factor. The cyclic edge labelling used throughout the paper is displayed above the Wilson line. The special OPE convention used in equation~(\ref{chi_def}) appears below it in bold. The general map between indexes for an $n$-point form factor is given by $j = \frac{1}{4}+\frac{1}{2}\,n+(-1)^{n+\textbf{j}}\left(\frac{3}{4}+\frac{1}{2}\,\textbf{j}\right)$.}
\label{fig:OPE_labelling}
\end{figure}
In addition, the super-OPE formalism provides us with a map between the variables $\chi_{j}^{A}$ of the super pentagons and the helicity variables $\eta_i^A$ of the super Wilson loop. This map is established by identifying $\partial/\partial \chi_{j}^{A}$ with the action of the supersymmetry generators on all edges $i$ located below the $j$-th pentagon in the OPE sequence. To make the correspondence more precise, it is convenient to use an alternative labelling of the edges, in which they are numbered starting from the bottom of the OPE sequence, with the very bottom edge being labelled as $-\textbf{1}$, and the upper ones as $\textbf{j} = \textbf{0}, \textbf{1}, \ldots\, $, as explained in figure~\ref{fig:OPE_labelling}. In this OPE-friendly labelling, putting a unit of charge onto the $j$-th pentagon results in the following distribution of edge charges,
\begin{align}\label{chi_def}
\frac{\partial}{\partial\chi_j^A} = \frac{1}{\left(\textbf{j}-\textbf{1}\right)_j\left(\textbf{j}\right)_j\left(\textbf{j}+\textbf{1}\right)_j}\sum\limits_{\textbf{i}\, =-1}^{j-2} \langle \textbf{j}-\textbf{1}\, \textbf{j\, }\textbf{j}+\textbf{1}\, \textbf{i}\rangle\,\frac{\partial}{\partial\eta_\textbf{i}^A}\ ,
\end{align}
where the sum is running over the edges located underneath the pentagon. The prefactor is designed to remove the helicity weights of the twistor variables. The operation is performed locally with respect to the twistors of the $j$-th pentagon: Denoting them as $Z_a,Z_b,Z_c,Z_d,Z_e$, one extracts the helicity weight of $Z_a$ using 
\begin{align}\label{pentagon_weights}
\left(\textbf{a}\right)_j = \sqrt[4]{\frac{\langle abcd\rangle\,\langle cdea\rangle\,\langle deab\rangle\,\langle eabc\rangle}{\langle bcde\rangle^3}}\ .
\end{align}
This map allows us to express any product of charged pentagon transitions as a linear combination of helicity components of the super Wilson loop. For example, to describe the $3$-point form factor of $\mathcal{T}_{3}$ in the OPE formalism, one needs a single pentagon carrying two units charge. More precisely,
\begin{equation}
  \langle \mathcal{F}_{3} | P_{12} \rangle  = \frac{\partial}{\partial \chi^{2}_{1}}  \frac{\partial}{\partial \chi^{1}_{1}}\,\langle \mathcal{F}_{3} | \mathbb{P} \rangle = \frac{\langle -\textbf{1} \textbf{0}\textbf{1}\textbf{2} \rangle^2}{(\textbf{0})_{1}^2\,(\textbf{1})_{1}^2\,(\textbf{2})_{1}^2} \frac{\partial}{\partial \eta^{2}_{-\textbf{1}}} \frac{\partial}{\partial \eta^{1}_{-\textbf{1}}}\,\mathcal{W}_{3,3}(\eta)\ ,
\end{equation}
with the map to the OPE-friendly labelling given by $\{3^{-}, 1, 2, 3\}\leftrightarrow \{\textbf{2}, \textbf{0}, -\textbf{1, \textbf{1}}\}$. We will illustrate this procedure further in section~\ref{sec:4pt-T3}, when considering the four-point form factor of $\mathcal{T}_3$, which has more independent helicity components.\par
It is worth pointing out that this procedure gives us access to only a subset of all the helicity components. More precisely, it allows us to reconstruct the components associated with $\eta_{i}$ for $i = 2, \ldots , n-1$ (in the usual cyclic labelling). The Grassmann variables on the uppermost edges, $\eta_{1}$ and $\eta_{n}$, do not enter in the definition of the $\chi$ derivatives. However, since the latter variables can always be eliminated using supertranslation symmetry, there is no loss of information when switching to the $\chi$ description. Alternatively, one may say that the OPE construction is naturally tailored for reconstructing components in the frame where $\eta_{1} = \eta_{n} = 0$.\par
Finally, let us mention that there is a simpler way to proceed for the minimal form factors. In this case, each pentagon is charged with exactly two units of charge, resulting in a single Grassmann component with the maximum number $2(n-2)$ of chiral $\chi$ variables. Due to the triangular nature of the map~\eqref{chi_def}, charging all $(n-2)$ pentagons in the sequence is equivalent to charging all edges from $i = 2, \ldots, n-2$, up to a normalization factor. Namely,
\begin{equation}\label{eq:min-product}
    \prod_{j=1}^{n-2} \frac{\partial}{\partial \chi_{j}^{A}} = \prod_{i=2}^{n-1} \alpha_{i} \frac{\partial}{\partial \eta_{i}^{A}}\ ,
\end{equation}
for any $A$, with $\alpha_{i}$ a coefficient absorbing the helicity weight of the $i$-th edge. The simplification comes from the fact that the same product enters the description of a ``simpler" form factor, namely, the maximally non-MHV component ($\overline{\textrm{MHV}}$) of the $n$-point form factor of $\mathcal{T}_{2}$. This form factor encodes the emission of $n$ negative-helicity gluons by the Lagrangian. It was calculated at tree level in refs.~\cite{Dixon:2004za,Brandhuber:2011tv}. In the OPE formalism, it maps to a sequence of $n-2$ pentagons $P_{1234}$, which is equivalent to a sequence of $n-2$ \textit{uncharged} pentagons $P$ due to parity symmetry, as discussed in ref.~\cite{Basso:2014hfa}. In other words,
\begin{equation}
    1 = \prod_{j=1}^{n-2} \left(\frac{\partial}{\partial \chi_{j}}\right)^4 W^{\overline{\textrm{MHV}}}_{2, n} = \prod_{i=2}^{n-1} \left(\alpha_{i}\frac{\partial}{\partial \eta_{i}}\right)^4 W^{\overline{\textrm{MHV}}}_{2, n}\ ,
\end{equation}
where $W^{\overline{\textrm{MHV}}}_{2, n}$ is the Parke-Taylor-Nair stripped $\overline{\textrm{MHV}}$ form factor at tree level. This condition allows us to fix the normalization factor in eq.~\eqref{eq:min-product} using the formulae given in refs.~\cite{Dixon:2004za,Brandhuber:2011tv}. It yields
\begin{align}
\alpha_2^4\ldots\alpha_{n-1}^4 = \frac{\langle 12\rangle^2\,\ldots\,\langle n1\rangle^2}{\langle n1\rangle^4}\,\frac{s_{12}\,\ldots\, s_{n1}}{q^4}\ ,
\end{align}
where $s_{ii+1} = (p_{i}+p_{i+1})^2$. Equipped with this relation, we may now easily perform the OPE conversion of the universal R-invariant, $W_{n, n}^{\textrm{tree}} = (-1)^{n} \prod_{i=2}^{n-1}(1ii+1)$, multiplying the $n$-point minimal MHV form factor. We get
\begin{equation}
\begin{aligned}\label{eq:omega-n}
\Omega_n &= \prod\limits_{j=1}^{n-2}\left(\frac{\partial}{\partial\chi_j}\right)^2|q|^{2-n}\,W_{n,n}^{\text{tree}} = \prod\limits_{i=2}^{n-1}\left(\alpha_i\,\frac{\partial}{\partial\eta_i}\right)^2|q|^{2-n}\,W_{n,n}^{\text{tree}} \\
&= \sqrt{\prod\limits_{i=1}^{n}\frac{s_{ii+1}}{q^2}} \ ,
\end{aligned}
\end{equation} 
including a factor of $|q|^{2-n}$ to absorb the overall dimension. In summary, instead of acting tediously with the $\chi$ derivatives on the OPE-framed Wilson loop (or its ratio function), in the case of minimal form factors, we may instead strip off the overall R-invariant and replace it by $\Omega_{n}$, which is a simple bosonic function of dimensionless ratios of Mandelstam invariants. This is what we will do in the following, reserving the use of the detailed map~\eqref{chi_def} to non-minimal form factors.

\subsection{Minimal three-point form factor}\label{section:W33}

The simplest test of our conjectures is given by the three-point form factor of $\mathcal{T}_{3}$. This is a minimal form factor with a single $\eta$ structure. Stripping off the R-invariant using eq.~\eqref{eq:omega-n}, one finds
\begin{equation}\label{eq:W33-def}
    \mathcal{W}_{3,3}(u_{1}, u_{2}, u_{3}) = \Omega_{3} \times \exp{\left[\tfrac{1}{4}\,\Gamma_{\textrm{cusp}}\,r_{1} + R_{3,3}\right]}\ ,
\end{equation}
where $u_{1,2,3}$ are the cyclic kinematic invariants of the $n=3$ periodic Wilson loop~\cite{Brandhuber:2014ica,Dixon:2020bbt,Dixon:2021tdw,Dixon:2022rse}
\begin{equation}
    u_{1} = \frac{s_{23}}{q^2}\ , \qquad u_{2} = \frac{s_{13}}{q^2}\ , \qquad u_{3} = \frac{s_{12}}{q^2}\ ,
\end{equation}
subject to the constraint $u_{1}+u_{2}+u_{3} = 1$. In terms of the flux-tube coordinates, $\sigma$ and $\tau$, one has~\cite{Sever:2020jjx}
\begin{equation}
    u_{1} = \frac{1}{1+e^{2\sigma}+e^{-2\tau}}\ , \qquad u_{2} = \frac{e^{2\sigma}}{\left(1+e^{-2\tau}\right)\left(1+e^{2\sigma}+e^{-2\tau}\right)}\ , \qquad u_{3} = \frac{1}{1+e^{2\tau}}\ ,
\end{equation}
and the prefactor $\Omega_{3}$ takes the form
\begin{equation}\label{eq:Omega3-data}
\Omega_{3} = \sqrt{u_{1}u_{2}u_{3}} = \frac{1}{\left(e^{\tau}+e^{-\tau}\right)\left(e^{\sigma}+e^{-\sigma}+e^{-2\tau-\sigma}\right)}\ .
\end{equation}
The function $r_{1}$ multipliying the cusp anomalous dimension, $\Gamma_{\textrm{cusp}} = 4g^2 + \mathcal{O}(g^{4})$, is a finite coupling-independent function of the $u_i$'s, determined from the one-loop form factor and the universal framing factor. It is known explicitly and reads
\begin{equation}\label{eq:r-one}
    r_{1} = 4\left(\tau^2 + \sigma^2\right) +\zeta_{2} - \frac{1}{2} \sum_{i=1}^{3} \log^{2}{(u_{i}/u_{i+1})}\ ,
\end{equation}
where $\zeta_{z} = \sum_{k=1}^{\infty}k^{-z}$ is the Riemann zeta function. Lastly, we have the remainder function $R_{3,3} = R_{3,3}(u_{1}, u_{2}, u_{3})$, introduced in ref.~\cite{Brandhuber:2014ica} after subtracting a suitable BDS-like ansatz. By definition, it is a finite, cyclic-symmetric function of the $u_i$'s,
\begin{equation}
    R_{3,3}(u_{1}, u_{2}, u_{3}) = R_{3,3}(u_{2}, u_{3}, u_{1})\ .
\end{equation}
It receives contributions at every loop order, starting from two loops, $R_{3,3} = \mathcal{O}(g^4)$. Its two-loop expression was worked out in ref.~\cite{Brandhuber:2014ica} and written in terms of weight-4 polylogarithmic functions of the kinematic variables.\par
The collinear limit corresponds to sending $\tau\rightarrow \infty$, that is, $u_{3}\rightarrow 0$ with $u_{1},u_{2}$ fixed. In this limit, the Wilson loop $\mathcal{W}_{3,3} \rightarrow 0$ and its leading behavior is controlled to all loops by the exchange of a scalar flux-tube excitation $\phi$. Our conjecture for it reads
\begin{equation}\label{eq:W33-data}
    \mathcal{W}_{3,3} = \frac{1}{g^{2}}\int \frac{du}{2\pi} \sqrt{\mu_{\phi}(u)\,\nu_{\phi}(u)}\,e^{-E_{\phi}(u)\tau + ip_{\phi}(u) \sigma} +\ldots\ ,
\end{equation}
with the measures $\mu_{\phi}$ and $\nu_{\phi}$ defined in section~\ref{sec:tilted-Pmu}. Here, $E_{\phi}(u) = 1 + 2g^2 (\psi(\tfrac{1}{2}+iu)+\psi(\tfrac{1}{2}-iu)-2\psi(1)) + \mathcal{O}(g^4)$ and $p_{\phi}(u) = 2u - 2\pi g^2\,\textrm{tanh}{(\pi u)} + \mathcal{O}(g^4)$ are the energy and the momentum of a scalar with rapidity $u$, with $\psi$ being the digamma function. The subleading contributions come from heavier states, with three or more flux-tube excitations.\par
The evaluation of this integral at weak coupling is straightforward. In particular, at tree level, the two measures in eq.~\eqref{eq:W33-data} are identical, $\nu^{\textrm{tree}}_{\phi}(u) = \mu^{\textrm{tree}}_{\phi}(u) = g^2 \pi\, \textrm{sech}{(\pi u)}$, and the integral reduces to the scalar component of the NMHV hexagon studied in refs.~\cite{Basso:2013aha,Papathanasiou:2013uoa}. It yields
\begin{equation}
    \mathcal{W}_{3,3} = \frac{e^{-\tau}}{e^{\sigma}+e^{-\sigma}} + \mathcal{O}(g^2)\ ,
\end{equation}
which is in perfect agreement with the large $\tau$ behavior of $\Omega_{3}$, see eq.~\eqref{eq:Omega3-data}.\par
One may proceed similarly at higher loops by expanding the various ingredients in~\eqref{eq:W33-data} in powers of $g^2$. The integrals generated this way fall in the same class as the ones describing the NMHV hexagon. At any loop order, the integrand is a sum of products of derivatives of $\psi$ functions, with arguments $1/2\pm iu$, up to the overall tree-level measure.\par
It is worth noting that the contribution of the measure $\nu_{\phi}(u)$ is particularly simple. This function turns out to be $i$-antiperiodic in $u$,
\begin{equation}
\nu_{\phi}(u+i) = -\,\nu_{\phi}(u)\ ,
\end{equation}
in close analogy with the form-factor transitions in ref.~\cite{Sever:2021xga}.
In particular, at $L$ loops, the ratio $\nu_{\phi}/\nu^{\textrm{tree}}_{\phi}$ may be given as a polynomial of degree $L$ in $c^2 = \pi^2\,\textrm{sech}^2{(\pi u)}$. For illustration, through two loops,
\begin{equation}
  \nu_{\phi}(u)/\nu_{\phi}^{\textrm{tree}}(u) = 1 + (8\zeta_{2} - 2c^2) g^2 + (16\zeta_{4} - 32\zeta_{2}c^2+6 c^4) g^4 + \mathcal{O}(g^6)\ .
\end{equation}
The more detailed expressions for the flux-tube measure, energy and momentum may be found in ref.~\cite{Basso:2013aha} or generated using the Mathematica notebook attached to the latter reference.\par
The resulting integrals may be taken by residues, corresponding to an expansion in the soft limit $u_{1}$ or $u_{2} \rightarrow 0$, or, equivalently, large $|\sigma|$. Alternatively, they may be calculated at finite $\sigma$ and expressed directly in terms of Harmonic Polylogarithms, using the algorithm introduced in ref.~\cite{Papathanasiou:2013uoa}.\par
Going along these lines, we checked the agreement between the integral~\eqref{eq:W33-data} and the form-factor data through two loops. The one-loop result arises entirely from the function $r_{1}$ in eq.~\eqref{eq:r-one}. For the two-loop comparison, we used the remainder function of ref.~\cite{Brandhuber:2014ica}. To be precise, the OPE prediction holds for a remainder function defined in the same convention as the remainder function of the three-point stress-tensor form factor in ref.~\cite{Brandhuber:2012vm}. This choice differs from the one in ref.~\cite{Brandhuber:2014ica} by a constant,
\begin{equation}
    R_{3, 3}\big|_{\textrm{OPE}} = R_{3,3}\big|_{\textrm{\cite{Brandhuber:2014ica}}} - 4\zeta_{4}\ ,
\end{equation}
which has to be included to match precisely with the OPE integral.\par
Higher-loop integrals may be studied similarly, providing powerful boundary data for the form-factor bootstrap program~\cite{to-appear}. 

\subsection{Four-point form factor of $\mathcal{T}_{3}$}\label{sec:4pt-T3}

The structure gets richer at $n=4$. In this case, there are three linearly independent components for the super form factor of $\mathcal{T}_{3}$. They may be chosen arbitrarily from the four R-invariants $(123), (234), (341), (412)$, which, as we recall, are subject to the four-term identity~\eqref{eq:4-term-id}. Alternatively, in analogy with the representation used for the 6-point NMHV amplitudes~\cite{Drummond:2008vq,Dixon:2011pw,Dixon:2014iba}, we may work with an overcomplete cyclic basis and write the super Wilson loop as
\beq\label{eq:W34-V}
W_{3,4} (\eta) = -\left(123\right) V_{4} - \left(234\right) V_{1} - \left(341\right) V_{2} - \left(412\right) V_{3}\ ,
\eeq
where the coefficients $V_{1,2,3,4}$ are independent of the Grassmann variables $\eta_{1,2,3,4}^a$ and are simply functions of the coupling constant $g^2$ and kinematic invariants. As no other $\mathcal{Q}$-invariants can be formed from the Grassmann variables, this decomposition should be valid to all loops.\par
The perturbative results in section~\ref{sec:superFF} may be easily cast into this form. One finds, through one loop,
\begin{equation}\label{eq:Vi-weak}
    V_{i} = \frac{1}{2} + g^2 V^{\textrm{1-loop}}_{i} + \mathcal{O}(g^4)\ ,
\end{equation}
where
\begin{equation}\label{eq:Vi-loop}
V_{i}^{\textrm{1-loop}} = \frac{1}{2}\,\textrm{Div} + \textrm{Li}_{2} \left[1-\frac{v_{i-2}}{u_{i-2}}\right] + \textrm{Li}_{2} \left[1-\frac{v_{i-2}}{u_{i-1}}\right] + \frac{1}{2}\log^{2}{\left[\frac{u_{i-2}}{u_{i-1}}\right]} + \zeta_{2}\ ,
\end{equation}
with $ \textrm{Div} = -\sum_{i=1}^{4} (-s_{ii+1})^{\epsilon}/\epsilon^2$ being the divergent part of the Wilson loop. The kinematic variables $u_i, v_i$ are dimensionless ratios of Mandelstam invariants, defined by
\beq\label{eq:u-and-v}
u_{i} = \frac{(p_{i+1}+p_{i+2})^2}{q^2}\ , \qquad v_{i} = \frac{(p_{i+1}+p_{i+2}+p_{i+3})^2}{q^2}\ ,
\eeq
with $i=1,2,3,4$ and $p_{i} = p_{i+4}$. Note that only 5 of these invariants are linearly independent~\cite{Dixon:2022xqh}, as recalled in appendix~\ref{app:4p-twistors}. Notice also that the functions $V_{1,2,3,4}$ in eq.~\eqref{eq:W34-V} are only defined up to a shift,
\begin{equation}
    V_{1, 3} \rightarrow  V_{1,3} + f\ , \qquad V_{2,4} \rightarrow  V_{2,4} - f\ ,
\end{equation}
with $f$ being an arbitrary function, due to the 4-term identity. When writing eq.~\eqref{eq:Vi-weak}, we picked a particular ``gauge'', which makes cyclic symmetry manifest, $V_{i} \rightarrow V_{i+1}$ upon $u_{i}, v_{i} \rightarrow u_{i+1}, v_{i+1}$.\par
\begin{figure}[t]
  \centering
  \begin{minipage}[b]{0.32\textwidth}
    \includegraphics[width=\textwidth]{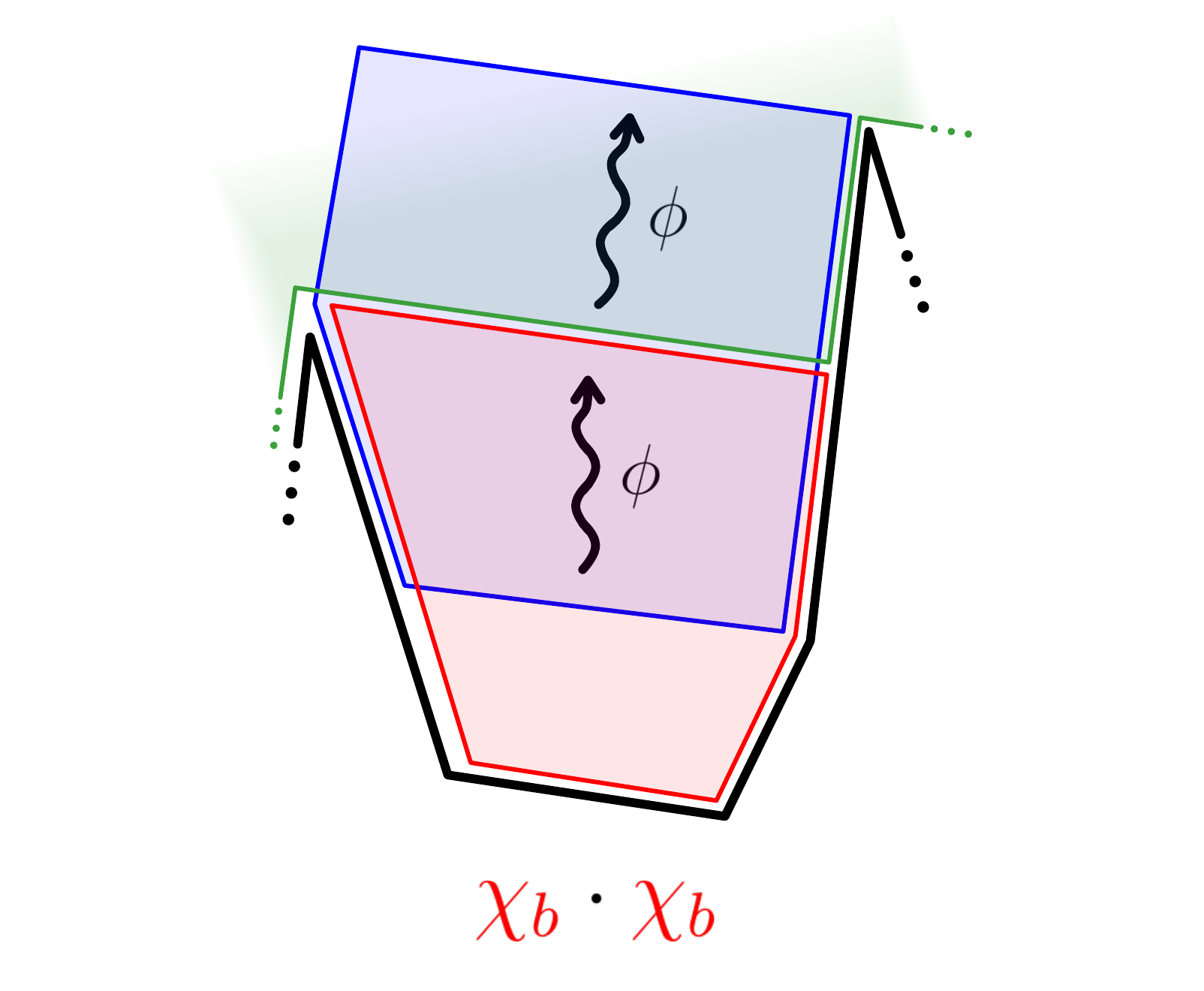}
  \end{minipage}
  \hfill
  \begin{minipage}[b]{0.32\textwidth}
    \includegraphics[width=\textwidth]{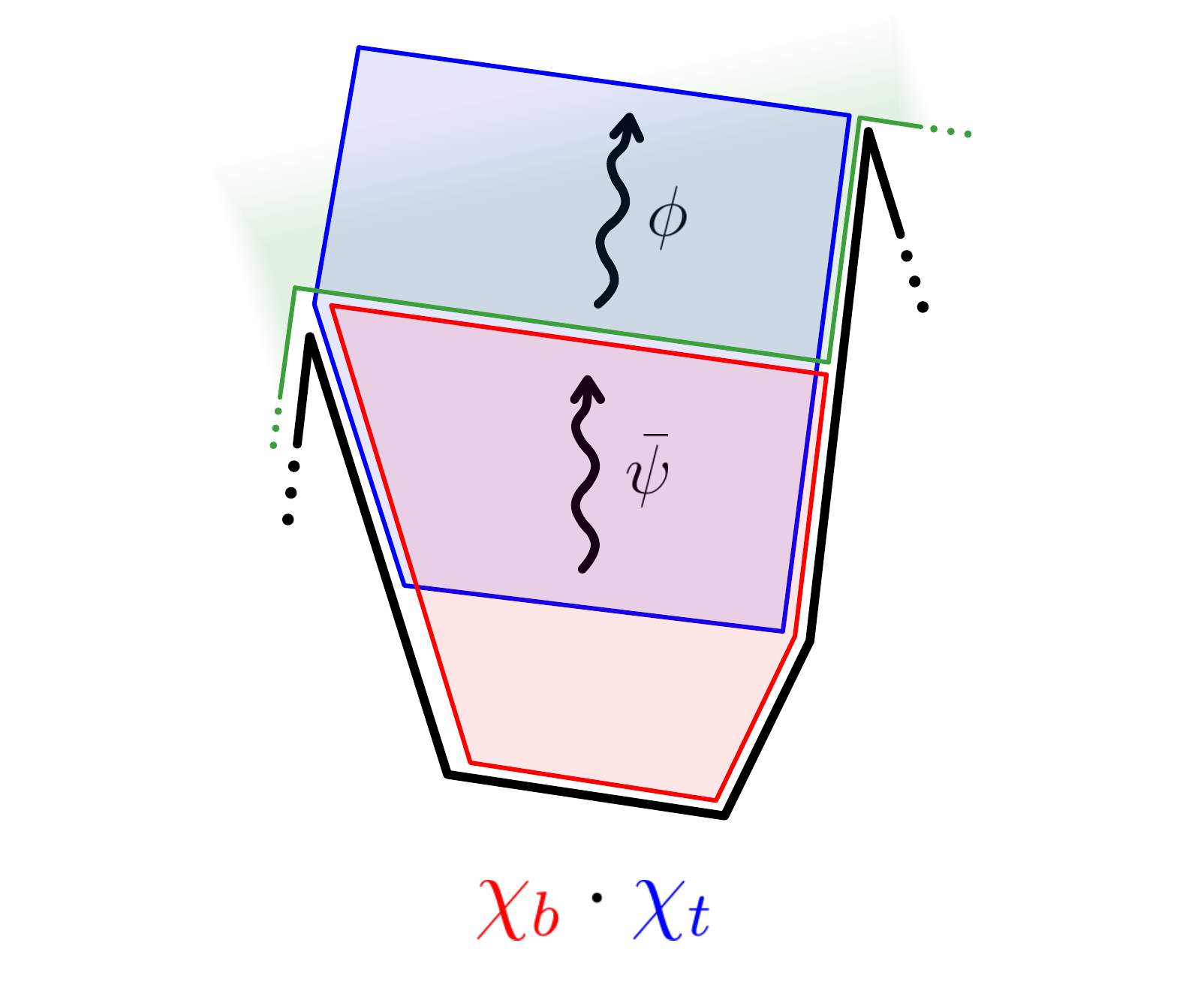}
  \end{minipage}
  \hfill
  \begin{minipage}[b]{0.32\textwidth}
    \includegraphics[width=\textwidth]{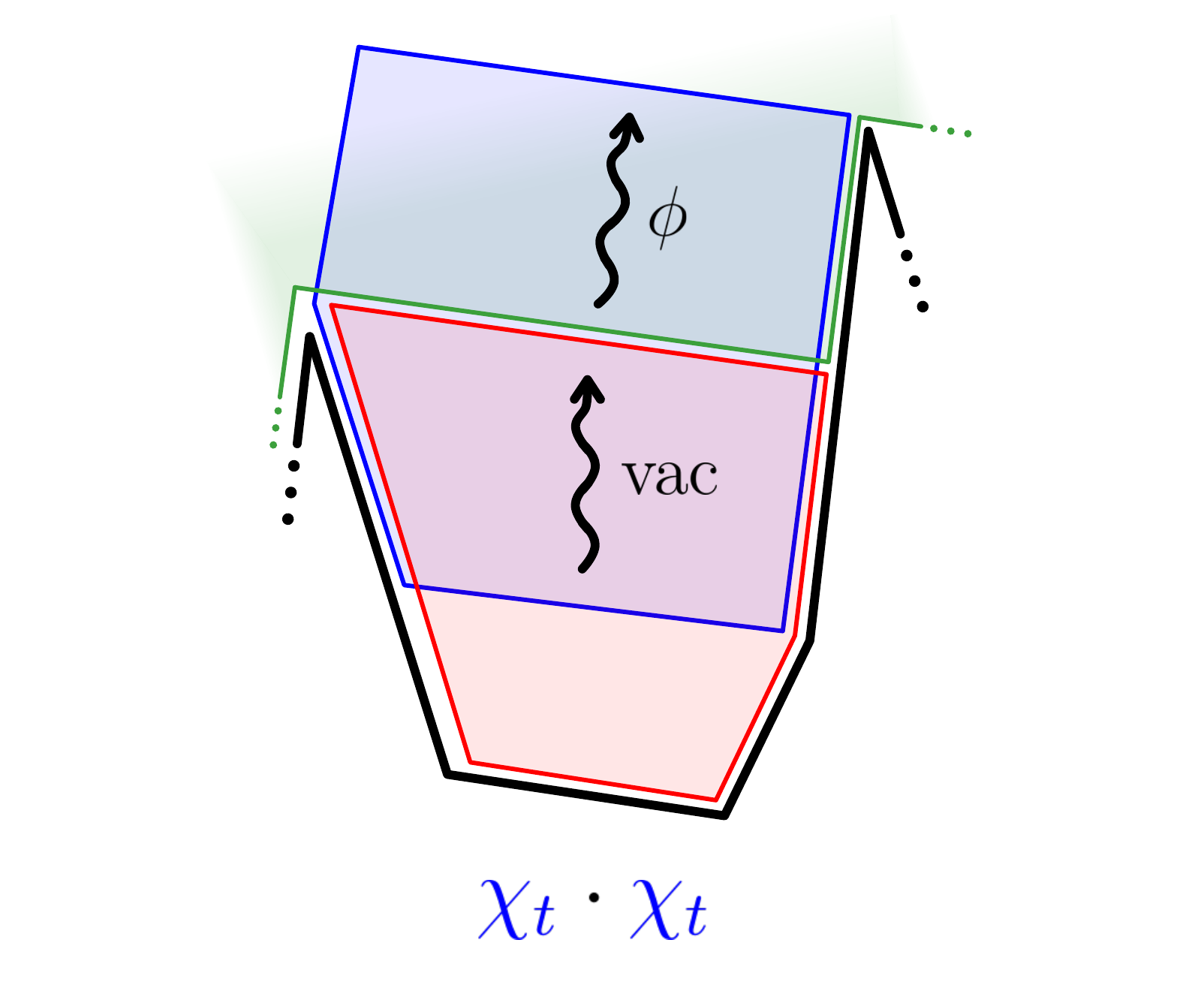}
  \end{minipage}
  \caption{Flux-tube representation of the 3 OPE components of the $n=4$ super form factor of $\mathcal{T}_{3}$ at leading twist. Left panel: a scalar $\phi$ is produced by the bottom pentagon; it propagates freely through the next pentagon before being absorbed at the top. Central panel: a fermion $\bar{\psi}$ is produced at the bottom, converted into a scalar $\phi$ in the second pentagon and absorbed at the top. The right panel corresponds to the strict collinear limit to the $n=3$ form factor, with the scalar $\phi$ produced in the second pentagon.}
  \label{fig:phi3_4pt}
\end{figure}

We are interested in comparing these expressions with the OPE representation in the limit where edges $1,2,3$ become collinear. In the OPE parametrization of appendix~\ref{appendix:twistors}, this double collinear limit corresponds to
\begin{equation}
    \tau, \tau_{b} \rightarrow \infty\ ,
\end{equation}
where $\tau$ and $\tau_{b}$ are, respectively, the OPE time variables in the top and bottom channels, shown in fig.~\ref{fig:phi3_4pt}. As alluded to before, in this limit, the OPE-framed Wilson loop $\mathcal{W}_{3,4}$ is identical to the ratio function $\mathcal{R}_{3,4} = \mathcal{W}_{3,4}/\mathcal{W}_{2,4}$, to the leading order in the flux-tube expansion. As the latter is slightly easier to construct, we will focus on it here.\par
At one loop, the ratio function is obtained by subtracting one-half of the expression for the $n=4$ form factor of the stress-tensor from each $V$ component. In other words, $\mathcal{R}_{3,4}$ admits the same expansion as the bare Wilson loop~\eqref{eq:W34-V} with the one-loop coefficients
\begin{equation}\label{eq:coeff-Ri}
    R_i = \frac{1}{|q|}\left[\frac{1}{2} + g^2\left(V_i^{\text{1-loop}} - \tfrac{1}{2}V_0^{\text{1-loop}}\right)+\mathcal{O}(g^4)\right],
\end{equation}
where
\begin{equation}\label{eq:V0}
V_0^{\text{1-loop}} = \textrm{Div} -2 \sum_{i=1}^{4} \textrm{Li}_{2}\left[1-v_i\right] + \sum_{i=1}^{4} \textrm{Li}_{2}\left[1-\frac{v_i v_{i-1}}{u_i}\right] + \frac{1}{2}\sum_{i=1}^{4} \log{u_i} \log{\left[\frac{u_i}{u_{i+1}u_{i-1}}\right]} +4\zeta_2\ .
\end{equation}
One verifies, for each coefficient $R_{i}$, that the divergent parts cancel out between $V_{i}$ and $V_{0}$, leaving a finite function of the kinematic invariants. Following our previous discussion, we also introduced a suitable power of $q^2$ to cancel the scaling dimension of the R-invariants and make the ratio function dimensionless. In what follows, we will drop the $|q|$ dependence and work with the normalization $|q|=1$.\par
The second important ingredient for the comparison is the mapping between the $\eta$ variables on the edges of the super Wilson loop and the $\chi$ variables carried by the superpentagons,
\begin{equation}
\begin{aligned}\label{eq:PPs}
\mathcal{W}_{3,4}(\eta) &= \langle \mathcal{F}_{3} | \mathbb{P}_{t} \circ  \mathbb{P}_{b}\rangle \\
&= \langle \mathcal{F}_{3} | P \circ  P_{12}\rangle \, \chi_{b}\cdot \chi_{b} + 2\langle \mathcal{F}_{3} | P_{1} \circ  P_{2}\rangle \, \chi_{t}\cdot \chi_{b} + \langle \mathcal{F}_{3} | P_{12} \circ  P\rangle \, \chi_{t}\cdot \chi_{t}\ ,
\end{aligned}
\end{equation}
with $t, b$ referring to the top and bottom pentagon in the sequence and with $\chi\cdot \chi' = \frac{1}{2}\epsilon_{ab}\chi^{a}\chi'^{b}$. As there are more edges than pentagons, the mapping between $\eta$'s and $\chi$'s is not one-to-one. It only becomes one-to-one after setting the $\eta$'s to zero on two edges, using the supertranslation symmetry of the super loop. In the OPE frame where $\eta^{a}_{1} = \eta^{a}_{4} = 0$, with $a=1,2$, the $3$ components in the right-hand side of eq.~\eqref{eq:PPs} can be put in correspondence with the $3$ invariants $\eta_{i}\cdot \eta_{j}$, with $1< i\leqslant j< 4$, standing in the left-hand side.\par
The precise dictionary reads, see section~\ref{sec:map},
\beq
\begin{aligned}
&\frac{\partial}{\partial \chi_{b}} = \alpha_{2}\,\frac{\partial}{\partial \eta_{2}} \ , \\
&\frac{\partial}{\partial \chi_{t}} = \alpha_{3}\left[\frac{\partial}{\partial \eta_{3}} + \frac{\langle 4^- 124\rangle}{\langle 4^- 134\rangle}\frac{\partial}{\partial \eta_{2}}\right] \ , \\
\end{aligned}
\eeq
where $Z_i$ is the momentum twistor of edge $i$ and $Z_{i^{-}}$ the twistor of the same edge in the preceding period. Going to the OPE edge labelling, $\{4^{-}, 1, 2, 3, 4\} \rightarrow \{\textbf{3}, \textbf{1}, -\textbf{1}, \textbf{0}, \textbf{2}\}$, and using the OPE parametrization of the momentum twistors in appendix~\ref{app:4p-twistors}, we find
\begin{equation}
\begin{aligned}
    \alpha_{2}^4 = \frac{\langle-\textbf{1} \textbf{0}\textbf{1}\textbf{2} \rangle^4}{(\textbf{0})_{b}^4\,(\textbf{1})_{b}^4\,(\textbf{2})_{b}^4} = -1\ , \qquad 
    \alpha_{3}^4 = \frac{\langle \textbf{0123} \rangle^4}{(\textbf{1})_{t}^4\,(\textbf{2})_{t}^4\,(\textbf{3})_{t}^4} = -\,e^{4\tau_{b} + 2i\phi_{b}}\ ,
\end{aligned}
\end{equation}
with $\phi_{b}$ being the OPE angle variable in the bottom channel. Additionally, in this parametrization, $\langle 4^- 124\rangle = \langle 4^- 134\rangle$ and $q^2=-1$.

Applying these derivatives to the left-hand side of eq.~\eqref{eq:PPs} and taking the collinear limit $\tau, \tau_b\rightarrow \infty$, we arrive at
\begin{equation}
\begin{aligned}\label{eq:W-to-data}
\frac{\partial}{\partial \chi_{b}^{2}} \frac{\partial}{\partial \chi_{b}^{1}}\,\mathcal{W}_{3,4} &= \frac{e^{-\tau-\tau_b}}{e^{\sigma-\sigma_b}+e^{\sigma_b-\sigma}+e^{\sigma+\sigma_b}}\left(R_{1}+R_{4}\right) + \ldots\ , \\
\frac{\partial}{\partial \chi_{b}^{2}} \frac{\partial}{\partial \chi_{t}^{1}}\,\mathcal{W}_{3,4} &= \frac{e^{-\tau-\tau_b-\frac{i\phi_b}{2}}}{e^{\sigma-\sigma_b}+e^{\sigma_b-\sigma}+e^{\sigma+\sigma_b}}\,\left[\frac{e^{\sigma-\sigma_b}}{e^{\sigma}+e^{-\sigma}}\left(R_{3}+R_{4}\right) + e^{\sigma_b}\left(R_{3}-R_{1}\right)\right] + \ldots\ ,
\end{aligned}
\end{equation}
with $R_i$ defined earlier, see eq.~\eqref{eq:coeff-Ri}, and with dots standing for subleading terms. Lastly, we have
\beq
\frac{\partial}{\partial \chi_{t}^{2}} \frac{\partial}{\partial \chi_{t}^{1}}\,\mathcal{W}_{3,4} = \frac{1}{\left(e^{\tau}+e^{-\tau}\right)\left(e^{\sigma}+e^{-\sigma}+e^{-2\tau-\sigma}\right)}\left(R_{2}+R_{3}\right) + \mathcal{O}(e^{-\tau_b})\ ,
\eeq
which holds for any $\tau$ at large $\tau_b$. This contribution corresponds to the strict collinear limit $3||2$, onto the 3pt form factor of $\mathcal{T}_{3}$. In this limit, $(123), (234) \rightarrow 0$ and $(341) = (412)$, meaning that the bottom pentagon is empty, as depicted in fig.~\ref{fig:phi3_4pt}. This collinear limit is, of course, built into the perturbative data, so the comparison with the flux-tube formula for this component reduces to the analysis carried out previously.\par
The other two components are more interesting. They correspond, respectively, to the OPE processes
\begin{equation}\label{eq:processes}
\textrm{vacuum} \rightarrow \phi \rightarrow \phi \rightarrow \mathcal{F}_{3}\ , \qquad \textrm{vacuum} \rightarrow \bar{\psi} \rightarrow \phi \rightarrow \mathcal{F}_{3}\ .
\end{equation}
In both cases, the sequences end with a single scalar $\phi$ at the top, absorbed by the half-BPS operator $\mathcal{T}_{3}$. However, whereas in the first case the scalar is produced from the onset, using the scalar component $P_{12}$ of the superpentagon, in the second case it is created through a fermion $\bar{\psi}$ propagating in the bottom channel. See figure~\ref{fig:phi3_4pt}.\par
The flux-tube representations of these two processes in~\eqref{eq:processes} are expressed in terms of the (charged) pentagon transitions $P_{\phi\phi}$ and $P_{\bar{\psi}\phi}$ worked out in refs.~\cite{Basso:2013aha,Basso:2014koa,Belitsky:2014lta,Belitsky:2014sla,Basso:2015rta}. In the normalization used in ref.~\cite{Basso:2015rta}, they read
\beq
\begin{aligned}\label{eq:int-4pt}
&\frac{\partial}{\partial \chi_{b}^{2}}\frac{\partial}{\partial \chi_{b}^{1}}\,\mathcal{W}_{3, 4} = \langle \mathcal{F}_{3}| P \circ P_{12} \rangle = g^{-1} \int \frac{du dv}{(2\pi)^2} \, \hat{\mu}_{\phi}(u)\,P_{\phi \phi} (u|v)\,\tilde{\mu}_{\phi}(v) + \ldots \ , \\
&\frac{\partial}{\partial \chi_{b}^{2}}\frac{\partial}{\partial \chi_{t}^{1}}\,\mathcal{W}_{3, 4} = \langle \mathcal{F}_{3}| P_{2} \circ P_{1} \rangle = g^{-\frac{5}{4}}e^{-i\phi_{b}/2}\int \frac{du dv}{(2\pi)^2} \,  \hat{\mu}_{\bar{\psi}}(u)\,P_{\bar{\psi}\phi} (u|v)\,\tilde{\mu}_{\phi}(v) + \ldots \ ,
\end{aligned}
\eeq
with the overall coupling factors determined by the requirement that the processes start at order $\mathcal{O}(g^0)$ at weak coupling. Here, in order to avoid cluttering the formulae, we introduced notations for the integration measures,
\beq\label{eq:effective-mu}
\hat{\mu}_{X}(u) = \mu_{X}(u)\,e^{-E_{X}(u)\tau_b-ip_{X}(u) \sigma_b}\ , \qquad \tilde{\mu}_{\phi} (v) = \sqrt{\mu_{\phi}(v)\,\nu_{\phi}(v)}\,e^{-E_{\phi}(v) \tau + ip_{\phi}(v)\sigma}\ ,
\eeq
with $X = \phi, \bar{\psi}$ and $\tau, \sigma, \tau_{b}, \sigma_{b}$ denoting the OPE spacetime variables in the top and bottom channels.\footnote{Note that in order to match the conventions in ref.~\cite{Basso:2015rta} we must rescale the scalar measure and transition, using $\mu^{[\alpha]}_{\phi} \rightarrow \mu^{[\alpha]}_{\phi}/g$ and $P^{[\alpha]}_{\phi\phi}\rightarrow gP^{[\alpha]}_{\phi\phi}$.} The rapidity $u$ is integrated along the real axis with a small \textit{negative} imaginary part whereas the rapidity $v$ is given a small \textit{positive} imaginary part. This prescription is needed to avoid the pole at $u=v$ in the first integral, coming from the Gamma function in eq.~\eqref{tiltedP}, and to deal with the pole at $v=0$ in the second integral, coming from the fermion measure, see refs.~\cite{Basso:2014koa,Basso:2015rta}.\par
To check our conjecture for $\nu_{\phi}$, we only need to compare the outcome of integrations in eqs.~\eqref{eq:int-4pt} with the corresponding coefficients $R_i$.
The analysis presents no major difficulty. We illustrate it here for the first integral.\par
As in the three-point case, one may simplify the analysis by exploiting a connection to NMHV amplitudes. The main observation is that the first integral in eq.~\eqref{eq:int-4pt} is almost identical to the one introduced in ref.~\cite{Basso:2013aha} to study the double collinear limit of the scalar component of the NMHV heptagon Wilson loop ``$\mathcal{W}_{\textrm{hept}}^{(7145)}$". The only difference is that for the heptagon one should replace $\nu_{\phi}$ with $\mu_{\phi}$ in the first integral of eq.~\eqref{eq:int-4pt}. To be precise, one has
\begin{equation}\label{eq:WW-Delta}
   \frac{\partial}{\partial \chi_{b}^{2}}\frac{\partial}{\partial \chi_{b}^{1}}\,\mathcal{W}_{3,4} = \mathcal{W}_{\textrm{hept}}^{(7145)} + \int \frac{du dv}{(2\pi)^2}\,\hat{\mu}_{\phi}(u)\,P_{\phi\phi}(u|v)\,\Delta\tilde{\mu}_{\phi}(v) + \mathcal{O}(e^{-3\tau})\ ,
\end{equation}
with
\begin{equation}
    \Delta \tilde{\mu}_{\phi}(v) = \left[\sqrt{\mu_{\phi}(v)\,\nu_{\phi}(v)}-\mu_{\phi}(v)\right] e^{-E_{\phi}(v)\tau + ip_{\phi}(v)\sigma}\ .
\end{equation}
As stressed earlier, at weak coupling, $\nu_{\phi}/\mu_{\phi} = 1 + \mathcal{O}(g^2)$. Hence, the $\Delta$ term vanishes at tree level, giving~\cite{Basso:2013aha}
\begin{equation}
\begin{aligned}
    \frac{\partial}{\partial \chi_{b}^{2}}\frac{\partial}{\partial \chi_{b}^{1}}\,\mathcal{W}_{3,4} & = e^{-\tau-\tau_{b}} \int \frac{dudv}{(2\pi)^2}\,\Gamma(\tfrac{1}{2}-iu)\,\Gamma(iu-iv)\,\Gamma(\tfrac{1}{2}+iv)\,e^{-2iu\sigma_{b} + 2iv\sigma} + \mathcal{O}(g^2) \\
    &= \frac{e^{-\tau-\tau_{b}}}{e^{\sigma-\sigma_{b}} + e^{\sigma_b-\sigma} + e^{\sigma+\sigma_b}} + \mathcal{O}(g^2)\ .
\end{aligned}
\end{equation}
It agrees precisely with the first line of eq.~\eqref{eq:W-to-data}, using $R_{i} = 1/2+\mathcal{O}(g^2)$. At one loop, one may refer to formulae in ref.~\cite{Basso:2013aha} to evaluate the heptagon integral. The extra bit is easily computed,
\begin{equation}
\begin{aligned}
   &\int \frac{du dv}{(2\pi)^2}\,\hat{\mu}_{\phi}(u)\,P_{\phi\phi}(u|v)\,\Delta \tilde{\mu}_{\phi}(v) = \frac{g^2\,e^{-\tau-\tau_{b}}}{e^{\sigma-\sigma_{b}} + e^{\sigma_b-\sigma} + e^{\sigma+\sigma_b}}\\
   &\qquad \times \left[\zeta_{2} + \log{\left[\frac{e^{\sigma_{b}-\sigma}}{e^{\sigma-\sigma_{b}} + e^{\sigma_b-\sigma} + e^{\sigma+\sigma_b}}\right]}\log{\left[\frac{e^{\sigma-\sigma_{b}}+e^{\sigma+\sigma_{b}}}{e^{\sigma-\sigma_{b}} + e^{\sigma_b-\sigma} + e^{\sigma+\sigma_b}}\right]} \right]  + \mathcal{O}(g^4)\ .
\end{aligned}
\end{equation}
Combining all the pieces together, one easily verifies agreement with the perturbative result in eqs.~\eqref{eq:Vi-loop} and~\eqref{eq:V0}. A similar analysis applies to the second integral in eq.~\eqref{eq:int-4pt}, with a different component of the NMHV heptagon Wilson loop.\par
One could proceed similarly at higher loops using the all-order OPE conjectures. The connection to scattering amplitudes indicates that the contributions to the form factor should live in the same function space as the NMHV heptagon in the double collinear limit, facilitating their exploration at higher loops. It would be interesting to perform this analysis explicitly at two loops, using known result for the heptagon OPE integral~\cite{Basso:2013aha}. It would also be fascinating to see if one could combine insights from the leading OPE integrals and the structure of the NMHV heptagon amplitude~\cite{Caron-Huot:2011dec,Drummond:2018caf,Dixon:2016nkn,Dixon:2021nzr} to bootstrap the $n=4$ super form factor of $\mathcal{T}_{3}$ in general kinematics at higher loops, as done recently through two loops for $\mathcal{T}_{2}$ in refs.~\cite{Dixon:2022xqh,Guo:2022qgv}.

\subsection{Minimal four-point form factor}

\begin{figure}[t]
  \centering
  \begin{minipage}[b]{0.45\textwidth}
    \includegraphics[width=\textwidth]{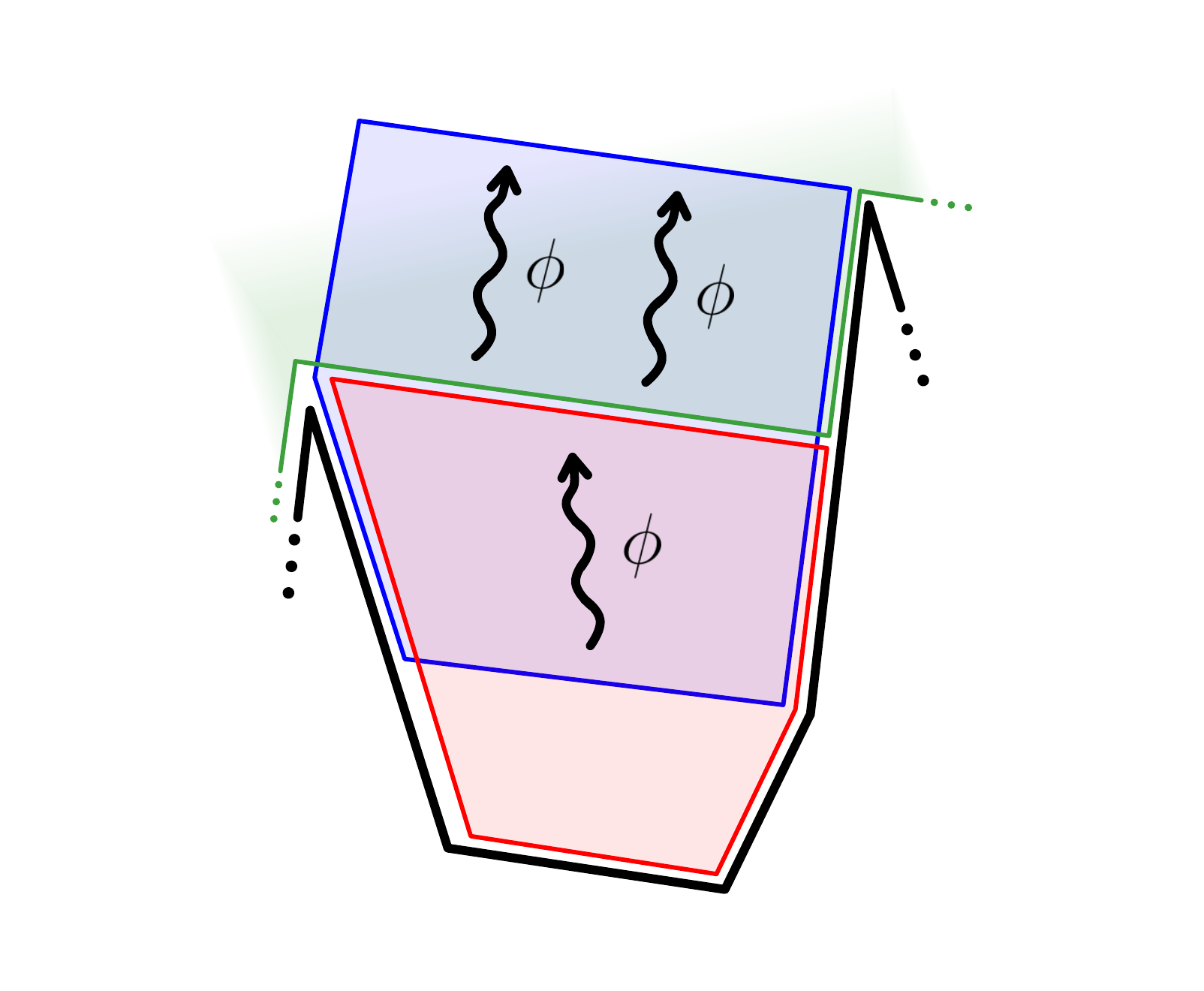}
  \end{minipage}
  \caption{Leading flux-tube contribution to the form factor of $\mathcal{T}_{4}$. A scalar is produced by each pentagon. The two scalars are then absorbed by the operator. Only one scalar can be created by a single pentagon, due to the fact that the pentagons can only be charged under a chiral half of the R-symmetry group.}
  \label{fig:phi4_leading}
\end{figure}

We conclude our perturbative checks with a brief discussion of the four-point form factor of $\mathcal{T}_{4}$. This is a minimal form factor with a single $\eta$ stucture, namely, $(123)\,(341)$. Stripping off the overall factor $\chi_{b}^{1}\chi_{b}^{2}\chi_{t}^{1}\chi_{t}^{2}$, one gets for the ratio function,
\begin{equation}\label{eq:R441loop}
 \mathcal{R}_{4,4} = \Omega_4\times \left[1 - g^2\,V'_0 + \mathcal{O}(g^4)\right]\ ,
\end{equation}
where $V'_0$ denotes the finite part of $V_0$ in eq.~\eqref{eq:V0} and with
\begin{equation}
\Omega_{4} =  \sqrt{u_{1}u_{2}u_{3}u_{4}}\ .
\end{equation}
In the collinear limit, $\tau, \tau_{b}\rightarrow \infty$, one finds
\begin{equation}
\Omega_4 = \frac{e^{-2\tau-\tau_b}}{\left(e^{\sigma}+e^{-\sigma}\right)\left(e^{\sigma-\sigma_b}+e^{\sigma_b-\sigma}+e^{\sigma+\sigma_b}\right)} + \mathcal{O}(e^{-2\tau-2\tau_b},e^{-4\tau-\tau_b})\ ,
\end{equation}
using the OPE parametrization given in appendix~\ref{app:4p-twistors}.\par
The leading OPE process on the flux-tube side is depicted in figure~\ref{fig:phi4_leading}. It is described by a sequence of two $P_{12}$ pentagon transitions, producing a two-scalar state in the top OPE channel. The corresponding integral formula reads
\begin{equation}
\mathcal{R}_{4,4} = g^{-6}\int \frac{du\,dv_1 dv_{2}}{2\,(2\pi)^3}\,\hat{\mu}_{\phi}(u)\,P_{\phi|\phi\phi}(u|v_{1},v_{2})\,F_{\phi\phi}(v_{1},v_{2})\,\tilde{\mu}_{\phi}(v_{1})\,\tilde{\mu}_{\phi}(v_{2}) + \ldots \ ,
\end{equation}
with the overall power of $g$ adjusted such that $\mathcal{R}_{4,4} = \mathcal{O}(g^0)$ at weak coupling. The measures are defined in eq.~\eqref{eq:effective-mu} and the $u$ integral is taken along $\mathbb{R}-i0$. The multi-particle pentagon transition can be written concisely in terms of fundamental pentagon transitions~\cite{Basso:2013aha,Belitsky:2014rba}
\begin{equation}
P_{\phi|\phi\phi}(u|v_{1},v_{2}) = \frac{P_{\phi\phi}(u|v_{1})\,P_{\phi\phi}(u|v_{2})}{P_{\phi\phi}(v_{1}|v_{2})}\ ,
\end{equation}
and our conjecture~\eqref{FFtransgeneral} for the 2-scalar form factor reads
\begin{equation}
    F_{\phi\phi}(v_{1},v_{2}) = \frac{1}{Q_{\phi\phi}(v_{2}|v_{1})}\ ,
\end{equation}
with $Q_{\phi\phi}$ the tilted transition at $\alpha=0$, see eqs.~\eqref{tiltedP} and~\eqref{QandNu}. It yields
\begin{equation}\label{eq:final-R44}
  \mathcal{R}_{4,4} =  g^{-6}\int\frac{du\,dv_1 dv_2}{2(2\pi)^3}\,\hat{\mu}_{\phi}(u)\,\frac{P_{\phi\phi}(u|v_1)\,P_{\phi\phi}(u|v_2)}{P_{\phi\phi}(v_1|v_2)\,Q_{\phi\phi}(v_2|v_1)}\,\tilde{\mu}_\phi(v_1)\,\tilde{\mu}_\phi(v_2) + \ldots \ ,
\end{equation}
with all ingredients determined to all loops using formulae in section~\ref{sec:tilted-Pmu}.\par
This integral is significantly harder to evaluate than the previous ones. In order to simplify the analysis, we will consider it in the regime $\sigma_{b}, \sigma \gg 1$, corresponding to the double soft limit $p_{2}, p_{3}\rightarrow 0$.%
\footnote{In terms of the kinematic invariants~\eqref{eq:u-and-v}, $u_{4}, u_{1}, u_{2}, v_{4}, v_{1} \rightarrow 0$ and $u_{3}, v_{2}, v_{3} \rightarrow 1$.
}
In this regime, the measures in eq.~\eqref{eq:final-R44} oscillate wildly and the integral is dominated by the singularities in the complex planes of $u$ and $v_{1,2}$. At weak coupling, through any loop order $L$, these singularities are poles located at $u = -i/2-in$ and $v_{1,2} = i/2+im_{1,2}$, with $n, m_{1,2} = 0, 1, 2, \ldots\,$. They generate contributions that can be easily compared with the perturbative data~\eqref{eq:R441loop}; they are of the type
\begin{equation}
    \sum_{m_{1}+m_{2} = m}\textrm{res}\,_{n, m_{1,2}} = e^{-2\tau-\tau_b-2(m+1)\sigma-(2n+1)\sigma_{b}} \times P_{m, n}^{(2L)}(\sigma, \sigma_{b}, \tau, \tau_{b})\ ,
\end{equation}
where $P_{n, m}^{(2L)}$ is a polynomial of degree $2L$ in the OPE spacetime variables. For illustration, after plugging inside eq.~\eqref{eq:final-R44} the one-loop expressions for $P, Q, \mu, \nu$, together with the energy and momentum of a scalar, one finds
\begin{equation}
    \mathcal{R}_{4,4} = e^{-2\tau-\tau_{b}-2\sigma-\sigma_{b}} \bigg[1 + 2g^2 (\zeta_{2}+2\sigma \sigma_{b}-2\sigma \tau-2\sigma_{b}\tau_{b}) + \mathcal{O}(g^4)\bigg] + \ldots\ ,
\end{equation}
from the leading singularity, $n = m_{1,2} = 0$. It agrees perfectly with the Feynman diagrammatic formula~\eqref{eq:R441loop} in the limit of interest. We proceeded similarly for a few higher residues, finding agreement in every cases. Finally, we performed checks at two loops using the symbol in refs.~\cite{Brandhuber:2014ica,Loebbert:2015ova}.\footnote{Note that the matching requires flipping the overall sign of the two-loop remainder function in eq.~(5.9) of ref.~\cite{Brandhuber:2014ica}, as was done in ref.~\cite{Loebbert:2015ova}.}    
\section{Conclusion}\label{section:conclusion}

In this paper, we proposed a dual description for MHV form factors of half-BPS operators in terms of matrix elements of infinite periodic super Wilson loops with states composed of zero-momentum scalars. Based on this description, we generalized the OPE program, recently developed for the form factors of the stress-tensor multiplet, to the super form factors of half-BPS operators. The form factor transitions, which describe the absorption of the flux-tube state by the operator, were found to be controlled by the same building blocks that appeared previously in the case of the stress-tensor multiplet. We performed a series of checks of our construction at weak coupling, finding agreement for the three- and four-point form factors of $\mathcal{T}_3$ and the four-point form factor of $\mathcal{T}_4$. Higher-loop data may be easily produced, allowing one to set up a function bootstrap for the three-point form factor of $\mathcal{T}_{3}$~\cite{to-appear}, along the lines of what has been done for $\mathcal{T}_{2}$~\cite{Dixon:2020bbt,Dixon:2021tdw,Dixon:2022rse}. It would also be interesting to use the same methods for studying the four-point form factors of $\mathcal{T}_3$ and $\mathcal{T}_4$ at higher loops. This analysis could shed light on the origins of the recently discovered antipodal (self) duality \cite{Dixon:2021tdw,Dixon:2022xqh}, currently only observed for the form factors of $\mathcal{T}_2$.\par
While we have only considered the MHV case, it would be interesting to generalize our results to the N$^{k'}$MHV components. There is a natural proposal for how this generalization could work. The MHV super Wilson loop only depends on the $\eta^{-}$ components of the on-shell supermultiplets. At higher values of $k'$, $\eta^{+}$ components will enter in the construction. This implies that the super Wilson loop edge and vertex insertions, eqs.~(\ref{eq:calE-i}) and~(\ref{scalar_vertex}), must be promoted to their unrestricted values defined in refs.~\cite{Caron-Huot:2010ryg,Groeger:2012xz,Groeger:2012hqk}. 
Recall, however, that the $\eta^{+}$ components are not periodic, but instead get shifted by $\gamma^+$, the supersymmetric analog of $q$: $\eta_{i+n}^{+} = \eta_{i}^{+} + \lambda_i\gamma^+$. While this feature may be swiftly accommodated in the periodic super Wilson loop set-up, see ref.~\cite{Bork:2014eqa}, it is not immediately clear how to incorporate it into the OPE framework. At tree level, we expect the $m=2$ amplituhedron picture to be broken by the N$^{k'}$MHV corrections, since these ones depend on all components of the supertwistors, including the non-periodic ones. It would be interesting to study the tree-level form factors for all values of $k$ and $k'$ using techniques like the BCFW recursion relations \cite{Britto:2004ap,Britto:2005fq} or the MHV diagrams in momentum twistor space \cite{Bullimore:2010pj,Adamo:2011pv}, and compare the results with other twistorial approaches~\cite{Koster:2016fna,Koster:2016loo,Koster:2017fvf}.\par
Lastly, it would be interesting to extend the Wilson loop and OPE descriptions to form factors of unprotected operators, studied, for instance, in refs.~\cite{Loebbert:2015ova,Brandhuber:2016fni,Loebbert:2016xkw,Brandhuber:2018xzk}. Although the problem looks daunting at the level of the Wilson loop, one may hope that progress can be made on the OPE side, by combining the integrable description of the local operators with the Form Factor OPE. To do so, one could draw inspiration from the integral representations obtained for the correlation functions of local operators using the method of the Separation of Variables (SoV), see e.g.~refs.~\cite{Derkachov:2002tf,Jiang:2016ulr,Bercini:2022jxo}, which share many similarities with the OPE formulae. They suggest that a general solution to the form factor axioms~\cite{Sever:2020jjx,Sever:2021xga} might be found in terms of the Baxter Q-functions describing the SoV wave functions of the operators. It would be fascinating to see if this analogy can be made precise and turned into an OPE-SoV framework for calculating form factors of generic operators. Reciprocally, it would be interesting to see if the tilted transitions, and notably the $\alpha = 0$ one, play a role in the definition of the SoV measures at higher loops~\cite{Bercini:2022jxo}. It would also be natural to seek connections with the correlation functions of large-spin operators, studied recently in refs.~\cite{Bercini:2021jti,Bercini:2020msp}.

\subsection*{Acknowledgments}

It is a pleasure to thank Lance Dixon, Lionel Mason, Amit Sever and Matthias Wilhelm for inspiring discussions and comments on the manuscript. We are also grateful to Amit Sever for several clarifications on periodic Wilson loops and to Lance Dixon for collaboration on a related project. AT was supported by the Institut Philippe Meyer at the Ecole Normale Sup\'erieure in Paris.

\appendix

\section{IR divergences}\label{app:IR-div}
\begin{figure}[t]
  \centering
  \begin{minipage}[b]{0.30\textwidth}
    \includegraphics[width=\textwidth]{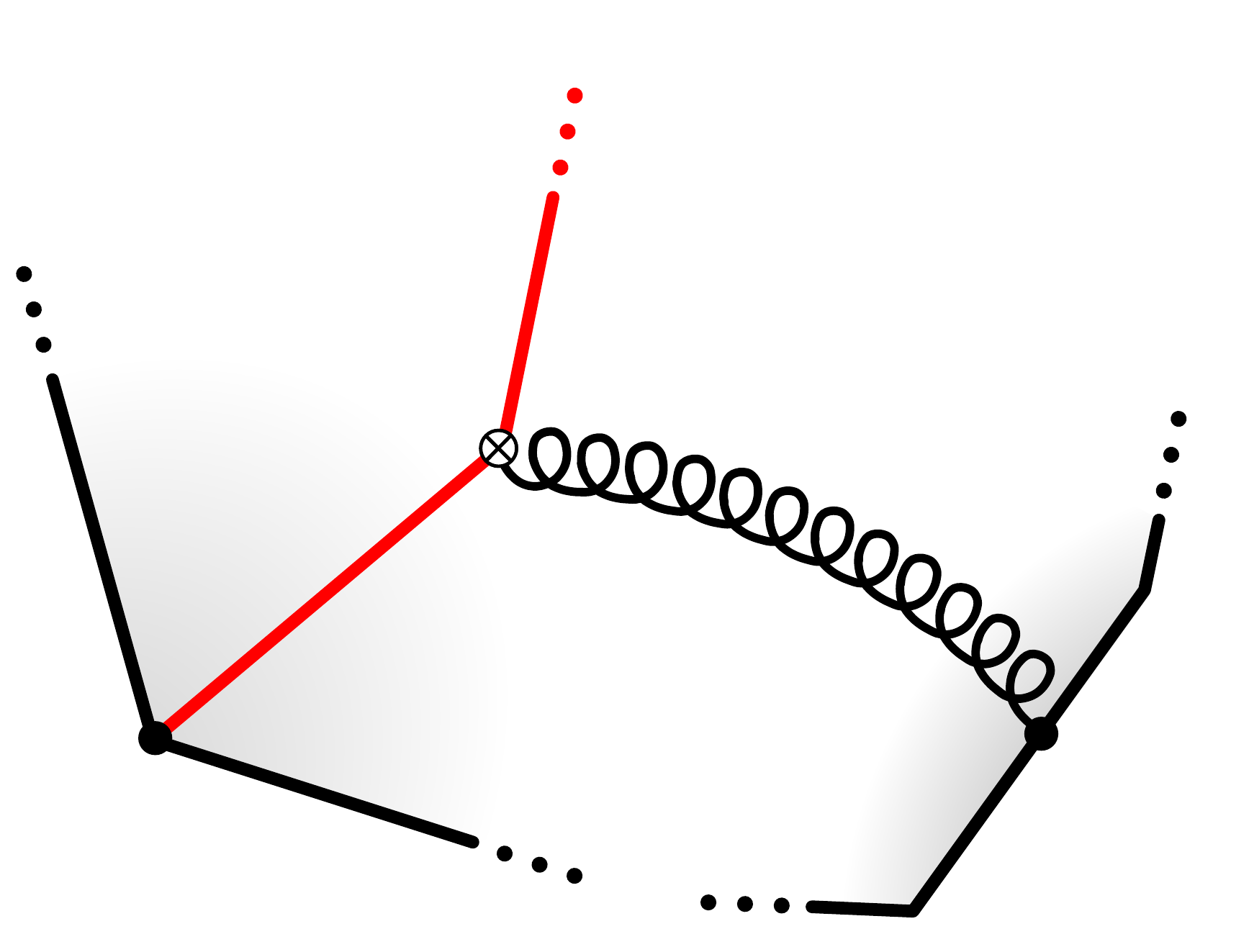}
  \end{minipage}
  \hfill
  \begin{minipage}[b]{0.30\textwidth}
    \includegraphics[width=\textwidth]{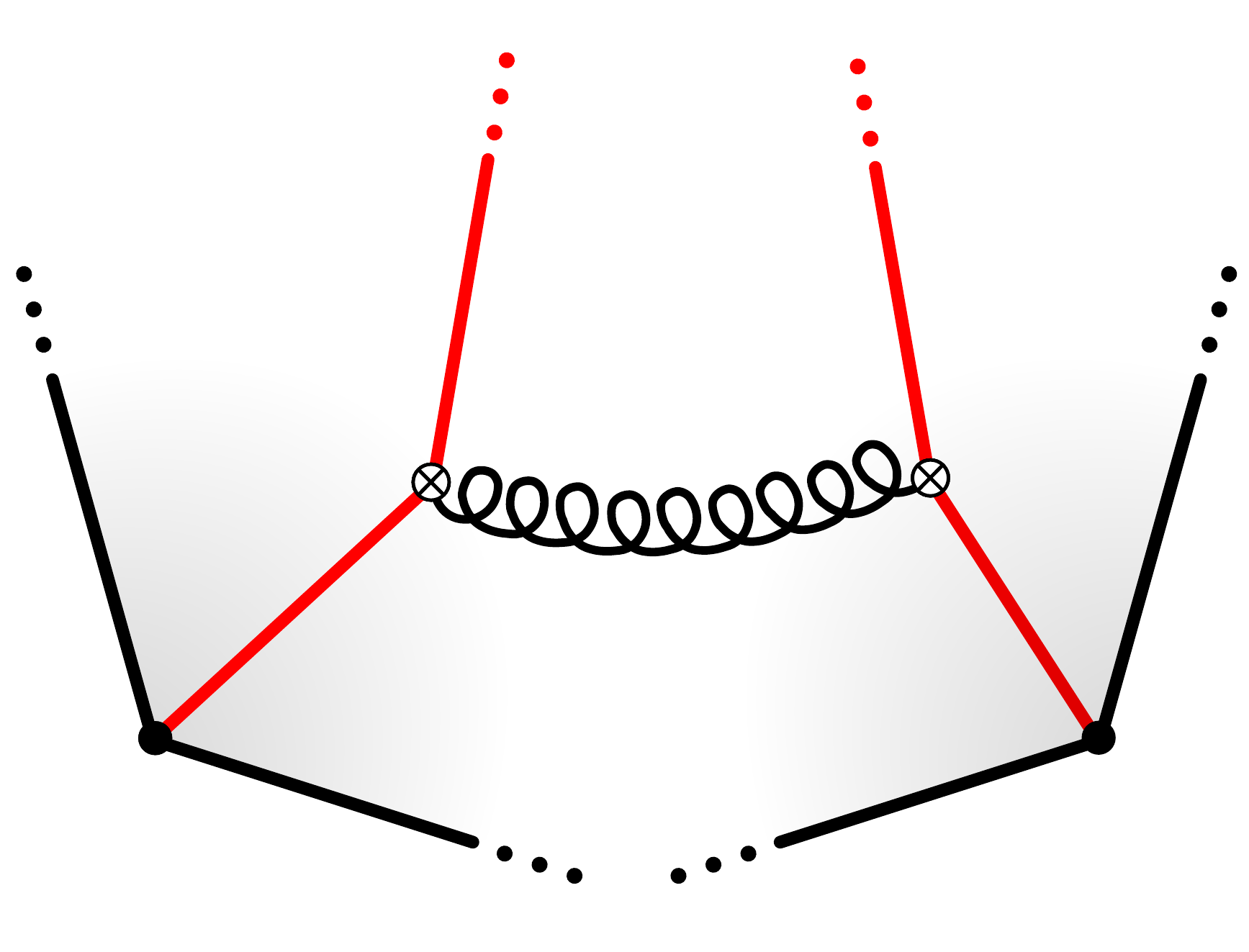}
  \end{minipage}
  \hfill
  \begin{minipage}[b]{0.30\textwidth}
    \includegraphics[width=\textwidth]{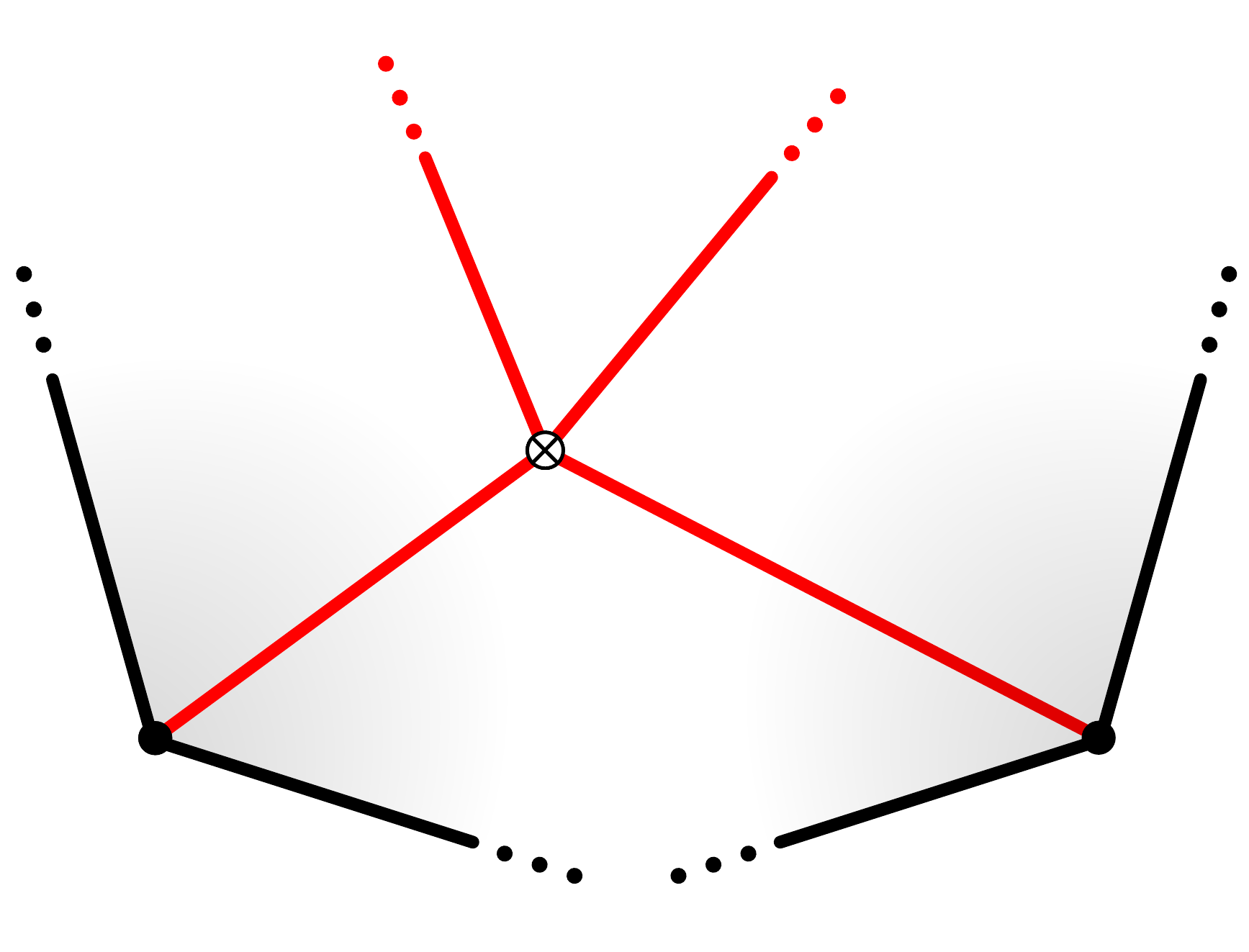}
  \end{minipage}
  \caption{Diagrams renormalizing the outgoing scalars at one loop.}
  \label{fig:one_loop_extra}
\end{figure}

To complete the equivalence between form factors and Wilson loops at one loop, we should address the few extra diagrams describing the interactions between the outgoing scalars. Examples of such diagrams are shown in figure~\ref{fig:one_loop_extra}. Their specificity is that they contain IR divergences, which are dual to the UV divergences of the local operators. Since we are considering states dual to half-BPS operators, these extra diagrams should cancel out. We explain below how this is happening.

To handle the IR divergences, we will use dimensional regularization, $D = 4-2\epsilon$, and give small non-zero momenta, $P_1, \ldots , P_{k-2}$, to all the outgoing scalars. These momenta should not be confused with the form-factor momenta, $p_1, \ldots , p_n$. The latter are encoded in the contour of the Wilson loop, whereas the former are just regulators for the sate. A proper treatment of the self-energy diagrams also requires working with off-shell particles. Hence, we will take $P_i^2\neq 0$ and consider the limit
\begin{equation}
    |P_i^2| \ll |P_i \cdot P_j| \rightarrow 0\ .
\end{equation}
We also assume $Pq =0$, where $P=\sum_{i=1}^{k-2} P_i$, for consistency with the periodicity condition.\par
Essentially, all the diagrams in figure~\ref{fig:one_loop_extra} can be accounted for by considering a single simple building block, which consists of two scalars inserted at points $X_1$ and $X_2$, with the segment of the Wilson loop between them. This object can be calculated as a Sudakov form factor of a bilocal operator,
\begin{equation}\label{eq:FFxP}
\langle \phi(P_1)\phi(P_2) | \bar{\phi}(X_1)\left[X_1,X_2\right]\bar{\phi}(X_2) |0\rangle = e^{i\sum_{j=1,2}P_{j}X_j} \times F \ ,
\end{equation}
with $\phi = \phi^{12}$, $\bar{\phi} = \bar{\phi}_{12}$, and with $[X_1,X_2]$ denoting the Wilson line between $\phi(X_1)$ and $\phi(X_2)$. Our normalization is such that $F = 1$ at tree level. The scalar fields can be inserted anywhere along the loop, at the cusps or on the edges, but they are assumed to be next to each other as far as the color ordering is concerned.%
\footnote{When $\phi(X_i)$ is on an edge, one should take a derivative $\partial_{X_i}$ of the form factor. However, this distinction proves immaterial, since, as we will see, the one-loop correction to form factor~\eqref{eq:FFxP} vanishes for any $X_1, X_2$ when $P_{1,2}\rightarrow 0$.}\par
\begin{figure}[t]
  \centering
  \begin{minipage}[b]{0.30\textwidth}
    \includegraphics[width=\textwidth]{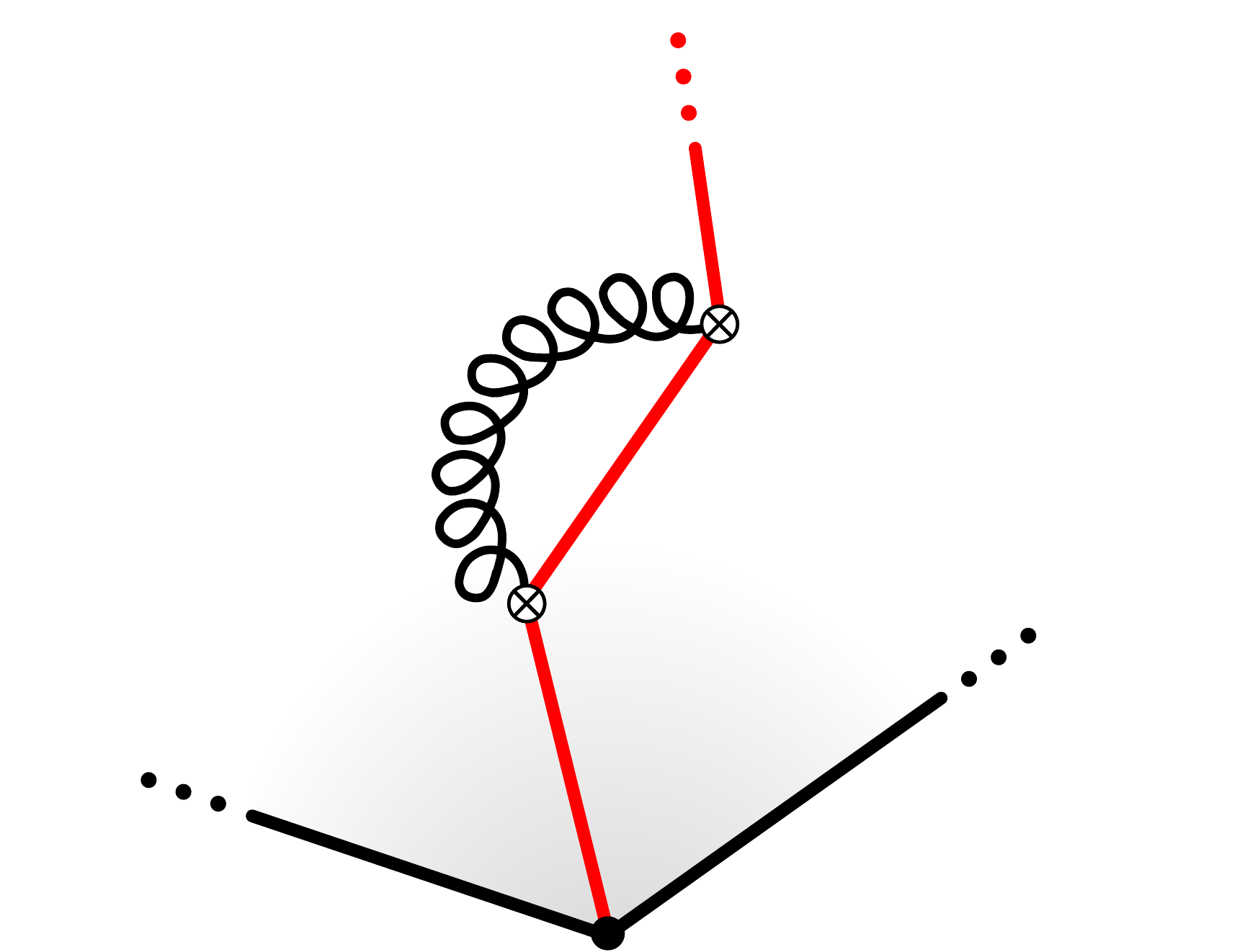}
  \end{minipage}
  \hfill
  \begin{minipage}[b]{0.30\textwidth}
    \includegraphics[width=\textwidth]{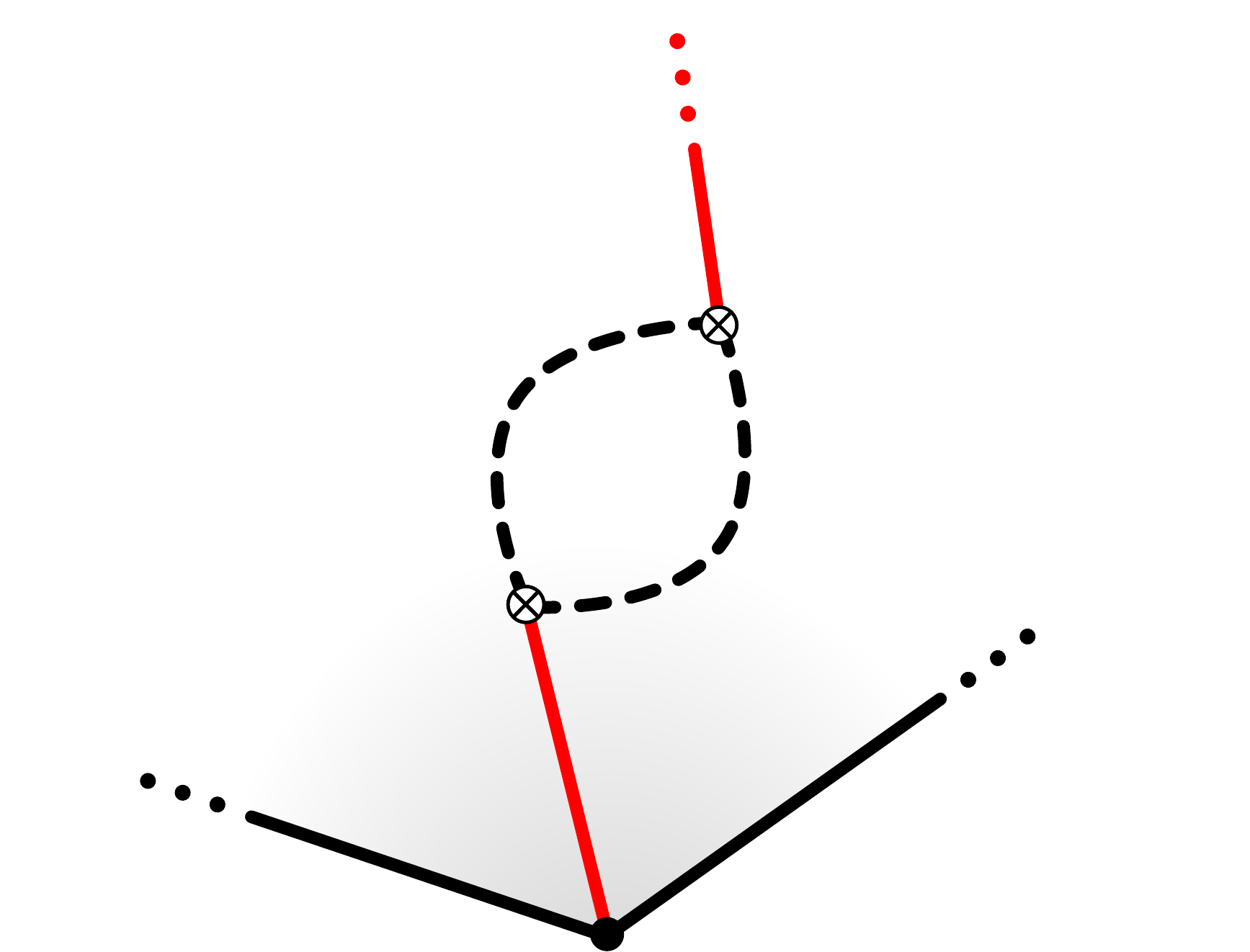}
  \end{minipage}
  \hfill
  \begin{minipage}[b]{0.30\textwidth}
    \includegraphics[width=\textwidth]{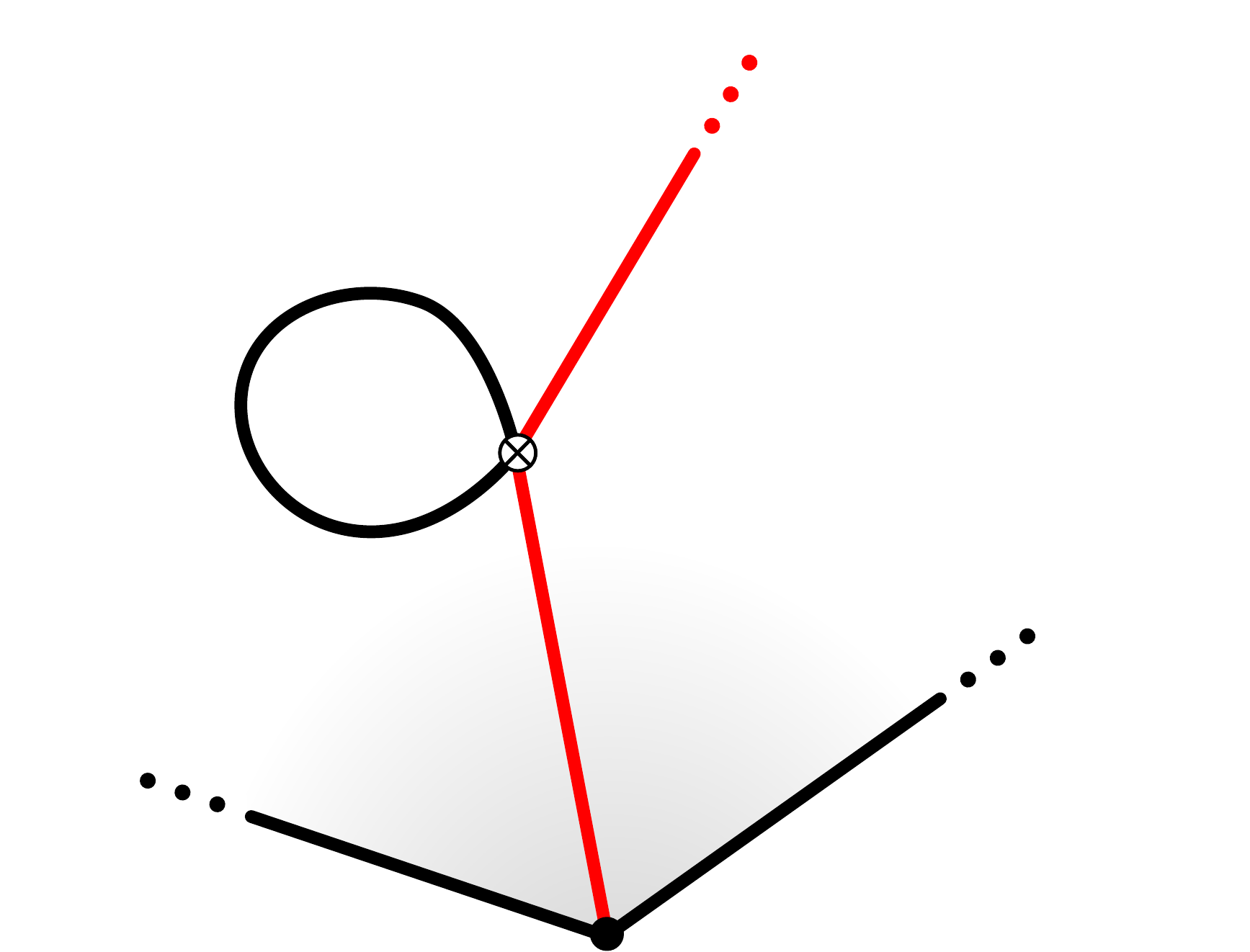}
  \end{minipage}
  \\\vspace{15pt}
  \begin{minipage}[b]{0.30\textwidth}
    \includegraphics[width=\textwidth]{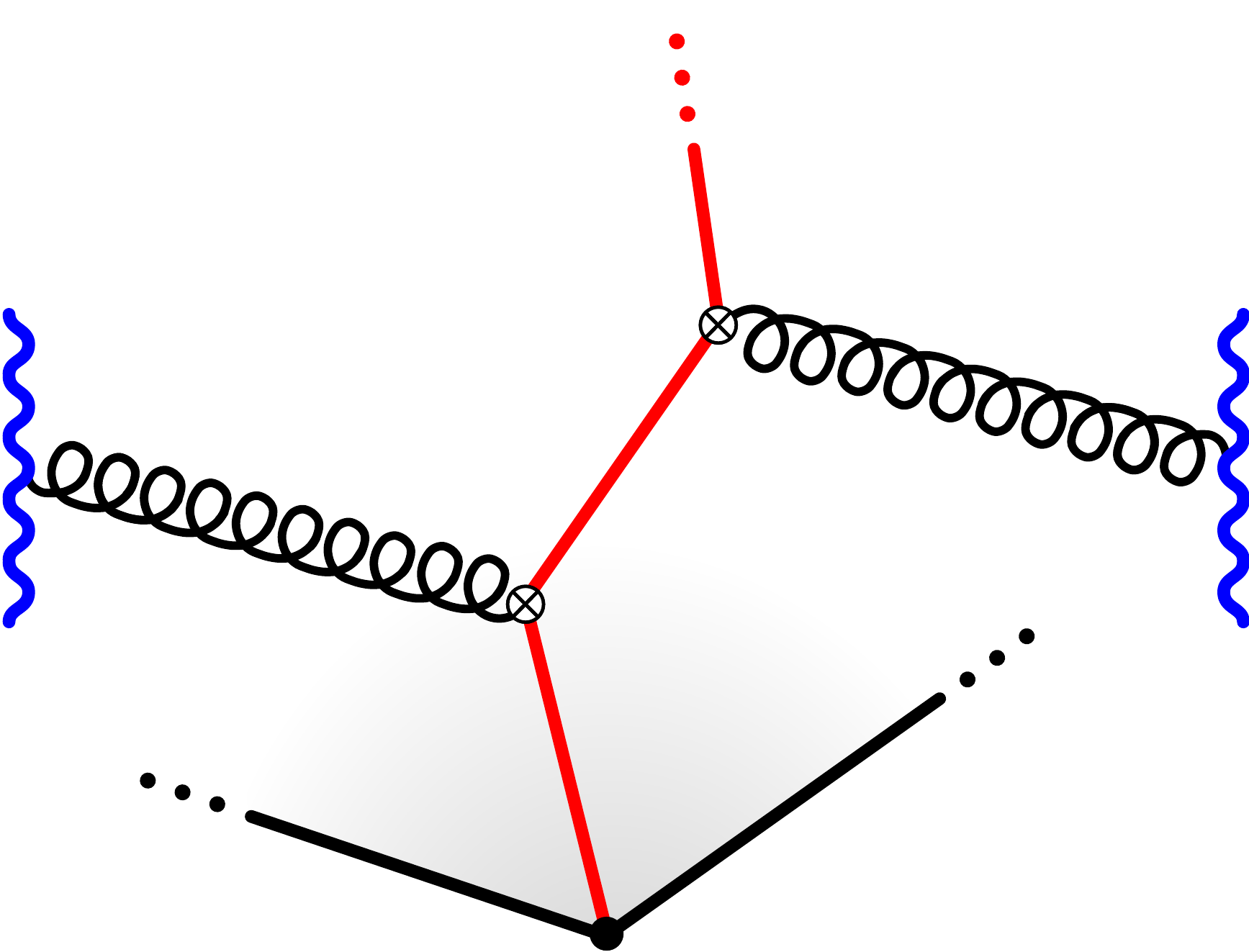}
  \end{minipage}
  \hfill
  \begin{minipage}[b]{0.30\textwidth}
    \includegraphics[width=\textwidth]{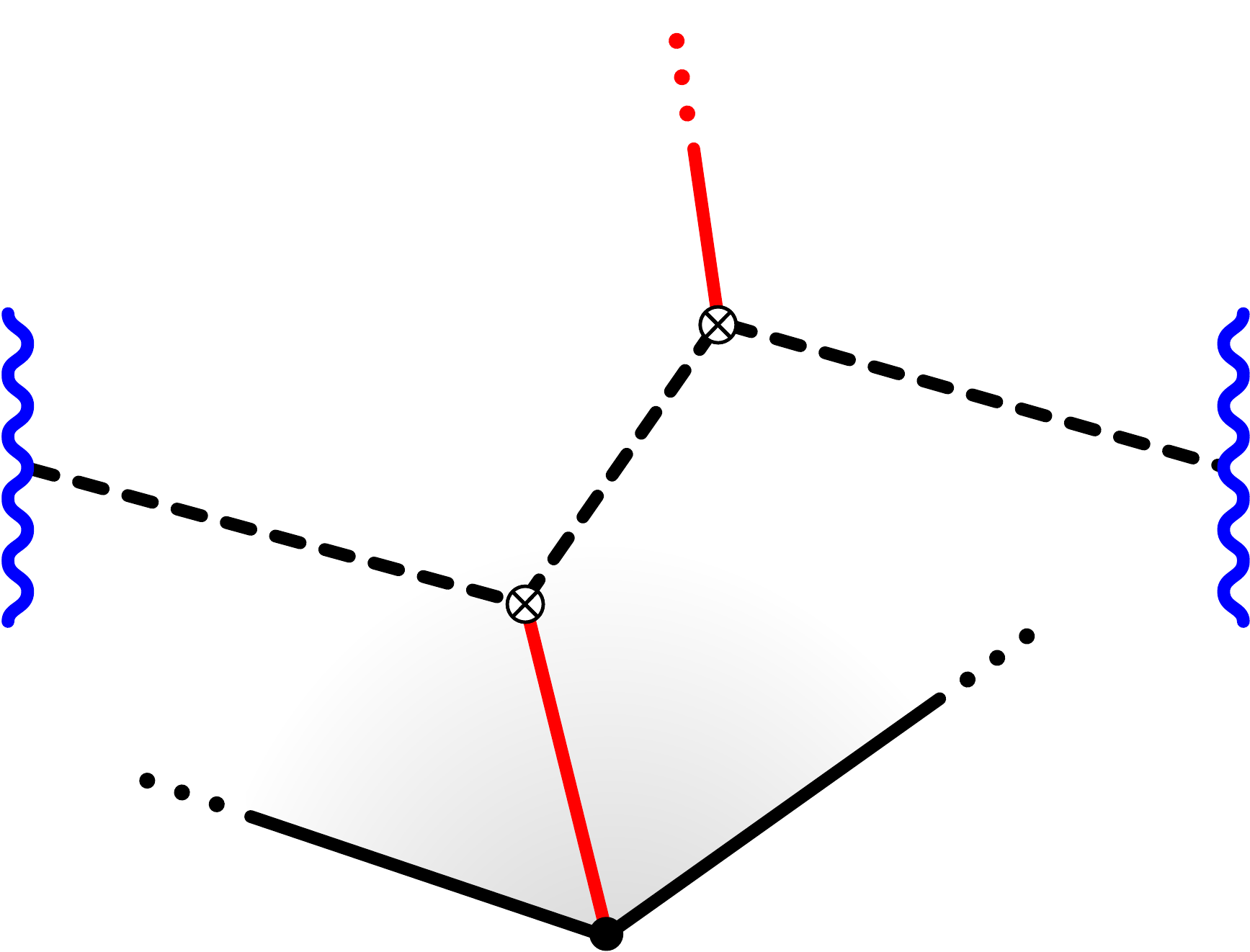}
  \end{minipage}
  \hfill
  \begin{minipage}[b]{0.30\textwidth}
    \includegraphics[width=\textwidth]{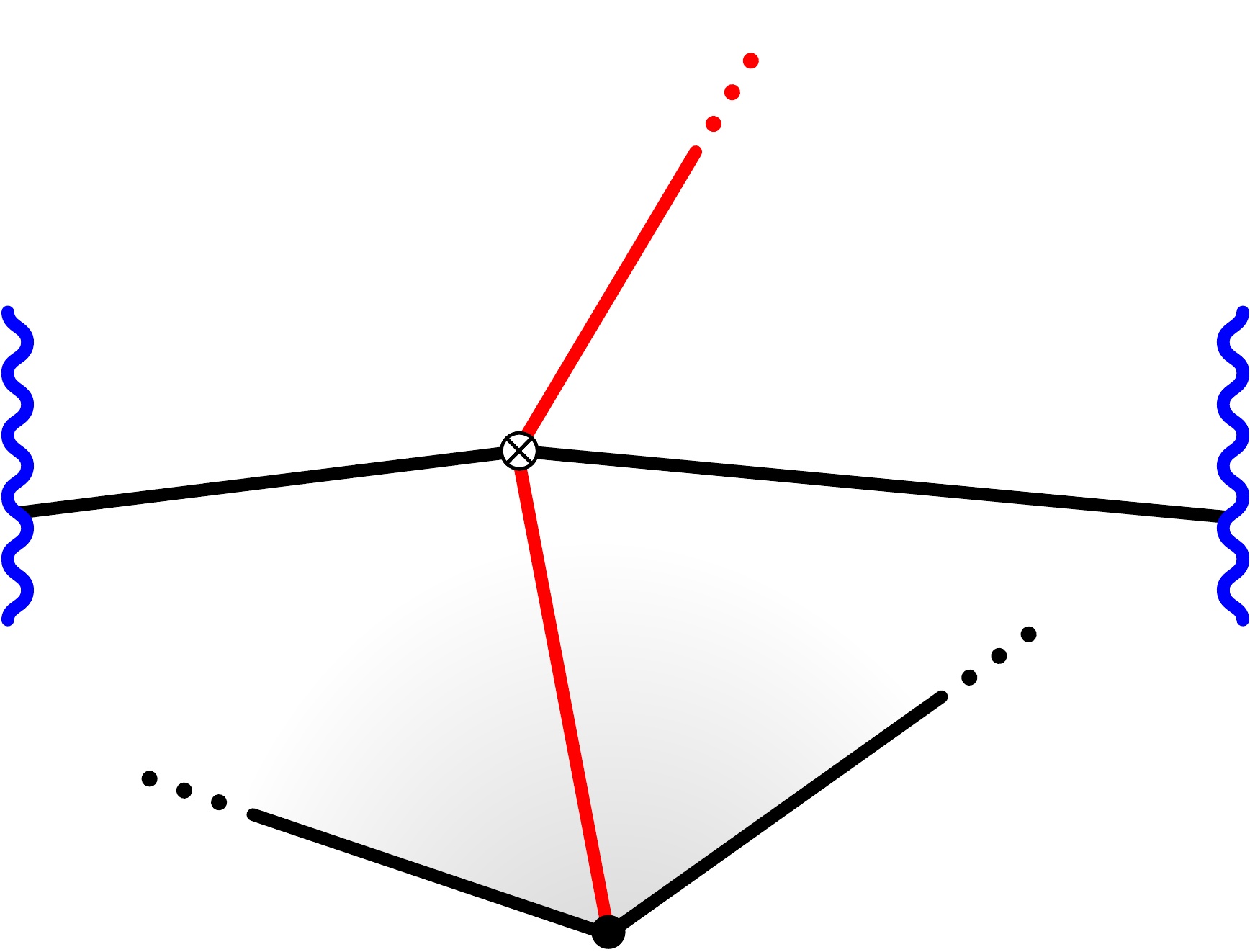}
  \end{minipage}
  \caption{Gluon, fermion and scalar contributions to the self energy of a scalar. (The gluon contribution also contains a quartic diagram not represented here.) Because the diagrams in the bottom row wrap around the cylinder, they are exclusive to the $k=3$ case.}
  \label{fig:self-energy}
\end{figure}

To avoid double counting, we mod out by the diagrams connecting the Wilson line to itself. Hence, the only diagrams in which the Wilson line plays a role are the ones in which a gluon connects it to an outgoing particle, as shown in the left panel of figure~\ref{fig:one_loop_extra}. We get two such diagrams, for each particle in~\eqref{eq:FFxP}. Importantly, in the limit, $P_{1,2} \rightarrow 0$, these diagrams do not depend on the contour. They only depend on the distance $\Delta X = X_1-X_2$ and read
\begin{equation}\label{eq:I-W-line}
    F_{\textrm{W-line}} = I(\Delta X)-I(0) +\mathcal{O}(P_{i})\ ,
\end{equation}
where $I$ are the so-called bubble integrals, with two propagators,
\begin{equation}\label{eq:bubble-I}
    I(\Delta X) = \frac{i\lambda}{2} \int\frac{d^D\ell}{(2\pi)^D} \frac{e^{i\ell \Delta X}}{(P_1+\ell)^2\,\ell^2} + \frac{i\lambda}{2} \int\frac{d^D\ell}{(2\pi)^D} \frac{e^{i\ell \Delta X}}{\ell^2\,(P_2-\ell)^2}\ .
\end{equation}
The two other diagrams in figure~\ref{fig:one_loop_extra} stand for the scalar quartic interactions,
\begin{equation}
        F_{\textrm{quartic}} = -\frac{i\lambda}{2}\int \frac{d^{D}\ell}{(2\pi)^{D}} \frac{e^{i\ell\Delta X}}{(P_{1}+\ell)^2\,(P_{2}-\ell)^2}\ ,
\end{equation}
and the gluon exchange,
\begin{equation}\label{eq:FF-gluon}
        F_{\textrm{gluon}} = \frac{i\lambda}{2} \int \frac{d^{D}\ell}{(2\pi)^{D}} \frac{(2P_1+\ell)\,(2P_{2}-\ell)}{(P_{1}+\ell)^2\,\ell^2\,(P_{2}-\ell)^2}\,e^{i\ell\Delta X} \ .
\end{equation}
Lastly, we have the self-energy diagrams shown in the top row of figure~\ref{fig:self-energy}. We should include half the self energy for each leg, giving%
\footnote{In fact, only a quarter of the self energy is being used here, since the self-energy diagrams are shared between two planar form factors.}
\begin{equation}
    F_{\textrm{self}} = I(0)\ .
\end{equation}
A great simplification occurs when we put all these integrals together. Namely, the bubble integrals cancel out. For instance, the integral $I(0)$ in eq.~\eqref{eq:I-W-line} is cancelled by the self energy. To eliminate the rest, one should massage the numerator in eq.~\eqref{eq:FF-gluon}, using
\begin{equation}
(2P_{1}+\ell)\,(2P_{2}-\ell) = 4P_{1} P_{2} +P_{1}^2+P_{2}^2 -(P_1+\ell)^2 -(P_2-\ell)^2 +\ell^2\ ,
\end{equation}
and observe that the last three terms in this equation remove $I(\Delta X)$ and $F_{\textrm{quartic}}$.
As a consequence, the form factor is proportional to a triangle integral, with three propagators,
\begin{equation}\label{eq:one-loop-F}
    F^{\textrm{1-loop}} = \frac{i\lambda}{2} \int \frac{d^D \ell}{(4\pi)^D} \frac{4P_1 P_2 + P_1^2 +P_2^2}{(P_1+\ell)^2\,\ell^2\,(P_2-\ell)^2}\,e^{i\ell \Delta X}\ .
\end{equation}
It is then straightforward to take the zero-momentum limit. Continuing the integral to $D = 4-2\epsilon$, we bring the limit $P_{i} \rightarrow 0$ below the integral sign and conclude that
\begin{equation}\label{eq:limit-P}
    \lim_{P_{1,2}\rightarrow 0} F^{\textrm{1-loop}} = 0\ ,
\end{equation}
no matter the distance $\Delta X$.\par
Let us stress that this cancellation heavily depends on the vanishing of the bubble integrals. Unlike the integral~\eqref{eq:one-loop-F}, these integrals are logarithmically divergent and stay non-trivial when $P_{i}\rightarrow 0$  and $(\Delta X)^2 \neq 0$. E.g.,
\begin{equation}
   \lim_{P_{i}\rightarrow 0} I(\Delta X) = i\lambda \int \frac{d^{D}\ell}{(2\pi)^{D}} \frac{e^{i\ell \Delta X}}{(\ell^2)^2} =  g^2\,\frac{[-(\Delta X)^{2}]^\epsilon}{\epsilon} \,\pi^{\epsilon}\,\Gamma(1-\epsilon)\ ,
\end{equation}
with $g^2 = \lambda/(4\pi)^2$, and similarly for $F_{\textrm{scalar}}$. These integrals drop out in our case because we are considering symmetric scalar states, dual to half-BPS local operators. For scalar states in other representations, extra bubbles will contribute to $F_{\textrm{quartic}}$ and the cancellation will be incomplete. This pattern is reminiscent of the one found for the UV divergences of local scalar operators~\cite{Minahan:2002ve} and it would be very interesting to perform a careful comparison of the two analyses for generic scalar states/operators. It may give important clues for extending the state-operator dictionary to non-BPS operators.\par
Our analysis above may appear a bit fast given that the integral in eq.~\eqref{eq:one-loop-F} also contains infrared divergences in $D = 4$ when the momenta are small. In fact, when $P^2_{i}=0$ and $\Delta X = 0$, our integral coincides with the traditional Sudakov form factor, with well-known double pole,%
\footnote{Up to an irrelevant factor $\tilde{c} = (4\pi)^\epsilon \Gamma(1+\epsilon)\Gamma(1-\epsilon)^2/\Gamma(1-2\epsilon) = 1 + \mathcal{O}(\epsilon)$.}
\begin{equation}
    F^{\textrm{1-loop}} = -\,g^2\,\frac{(-2P_{1} P_{2})^{-\epsilon}}{\epsilon^2}\ .
\end{equation}
Consistency with~\eqref{eq:limit-P} requires that the limit $P_{i}\rightarrow 0$ is taken for $\epsilon < 0$, that is, $D>4$. With this choice, the interactions die off at large distances and with them all the dangerous terms at small $P$.

There may still be a pitfall in analyzing divergences this way at higher loops. The problem is that the condition $\epsilon<0$ is \textit{opposite} to the one used to regularize the cusp divergences of the Wilson loop. The latter give rise to contributions $\sim 0^{\epsilon}$ which are hard, if not impossible, to interpret if $\epsilon <0$. This dual use of the dimensional regulator may be an issue for diagrams containing both UV and IR divergences, for which it may not be obvious which condition to use and how. In such cases, one may need to separate the problematic regions more carefully. However, at one loop, we see nothing wrong in using different $\epsilon$'s for different diagrams, depending on the nature of their divergences.\par
The above analysis explains the vanishing of the extra diagrams for generic $k$. However, it does not extend to $k=3$, where there is only a single scalar in the state. In this case, the relevant form factor is
\begin{equation}
\langle \phi(P) | \textrm{Tr}\, \bar{\phi}(X) [X, X+q] |0\rangle = e^{iP X} \, F(P, q)\ ,
\end{equation}
with $\bar{\phi}(X)$ connected to its image at $X+q$ through a Wilson line `going around the world'. As before, this line generates the interaction
\begin{equation}\label{eq:int-Iq}
   F_{\textrm{W-line}}(P, q) =  I(q)-I(0)\ ,
\end{equation}
at low momentum, with $I$ given by eq.~\eqref{eq:bubble-I} with $P_1 = P_2 = P$. The self energy still takes care of the $I(0)$ part but, in the lack of additional diagrams, it may not be clear how to cancel $I(q)$. The resolution is that for a single particle we must also include \textit{wrapping diagrams} with virtual particles traveling around the period $q$. These diagrams produce double-trace corrections for $k>3$, which are subleading at large $N$. However, the very same diagrams become planar for a single outgoing scalar. Their structure mirrors the one for the self energy, and for each contribution to the usual self energy we should include a diagram in which the virtual particle is wrapping around the period, as shown in figure~\ref{fig:self-energy}. There is also a minus sign coming from the different color ordering. Adding up all diagrams, one finds
\begin{equation}
F_{\textrm{self}} \rightarrow I(0) - I(q)\ ,
\end{equation}
which perfectly cancels the Wilson loop contribution~\eqref{eq:int-Iq}.

\section{Twistor parametrization}\label{appendix:twistors}

\begin{figure}[t]
  \centering
  \begin{minipage}[b]{0.3\textwidth}
    \includegraphics[width=\textwidth]{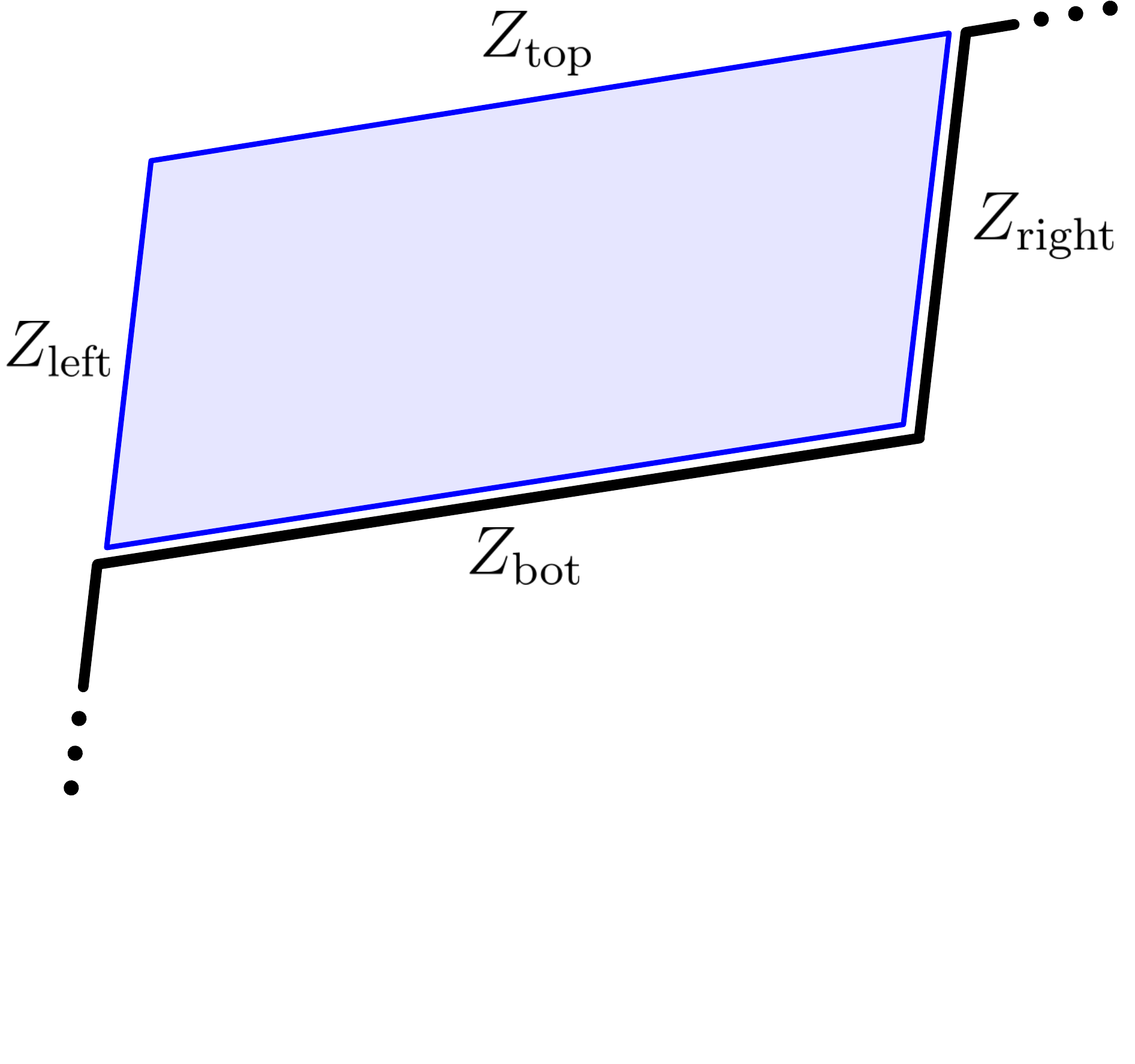}
  \end{minipage}
  \hfill
  \begin{minipage}[b]{0.3\textwidth}
    \includegraphics[width=\textwidth]{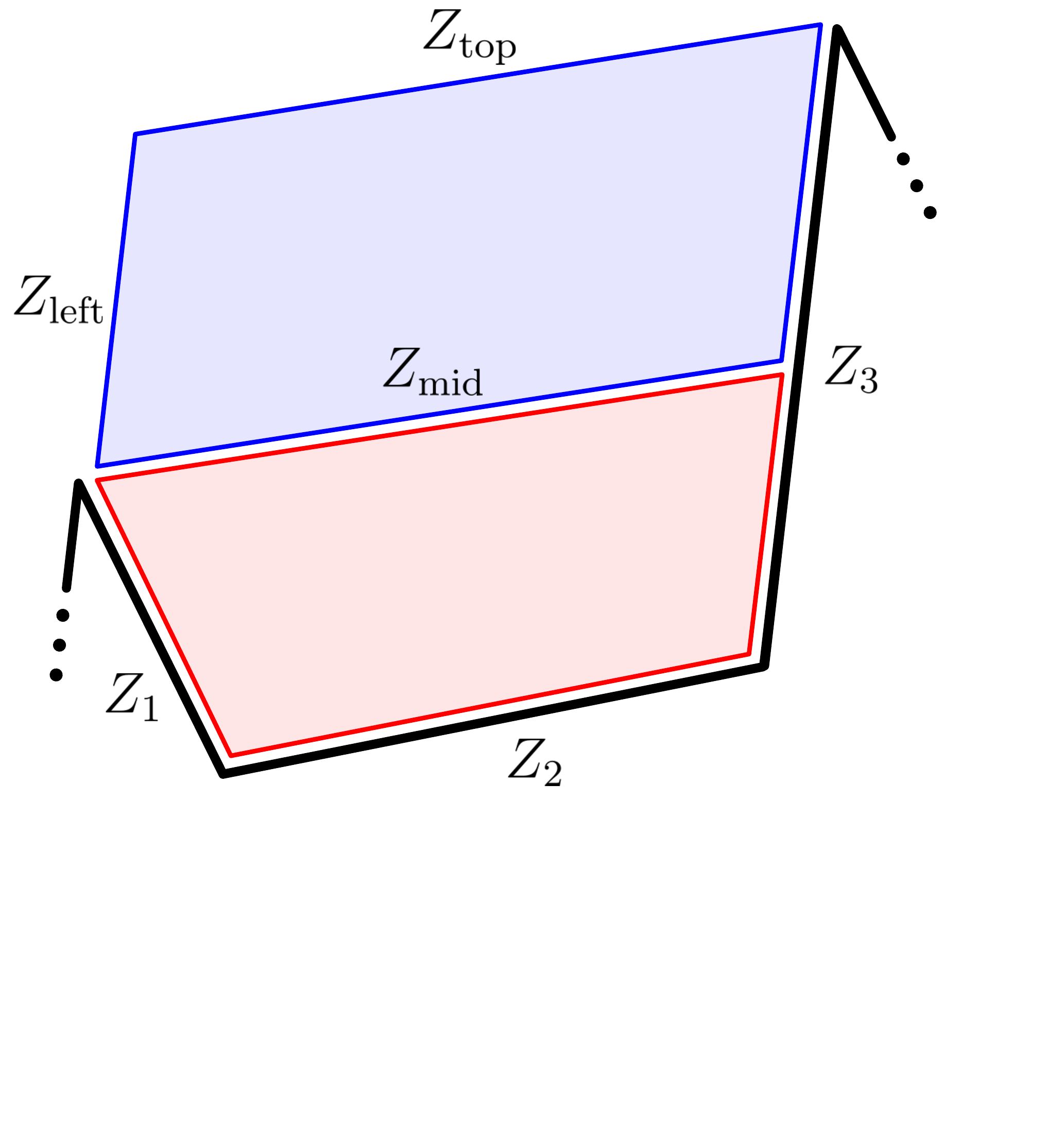}
  \end{minipage}
  \hfill
  \begin{minipage}[b]{0.3\textwidth}
    \includegraphics[width=\textwidth]{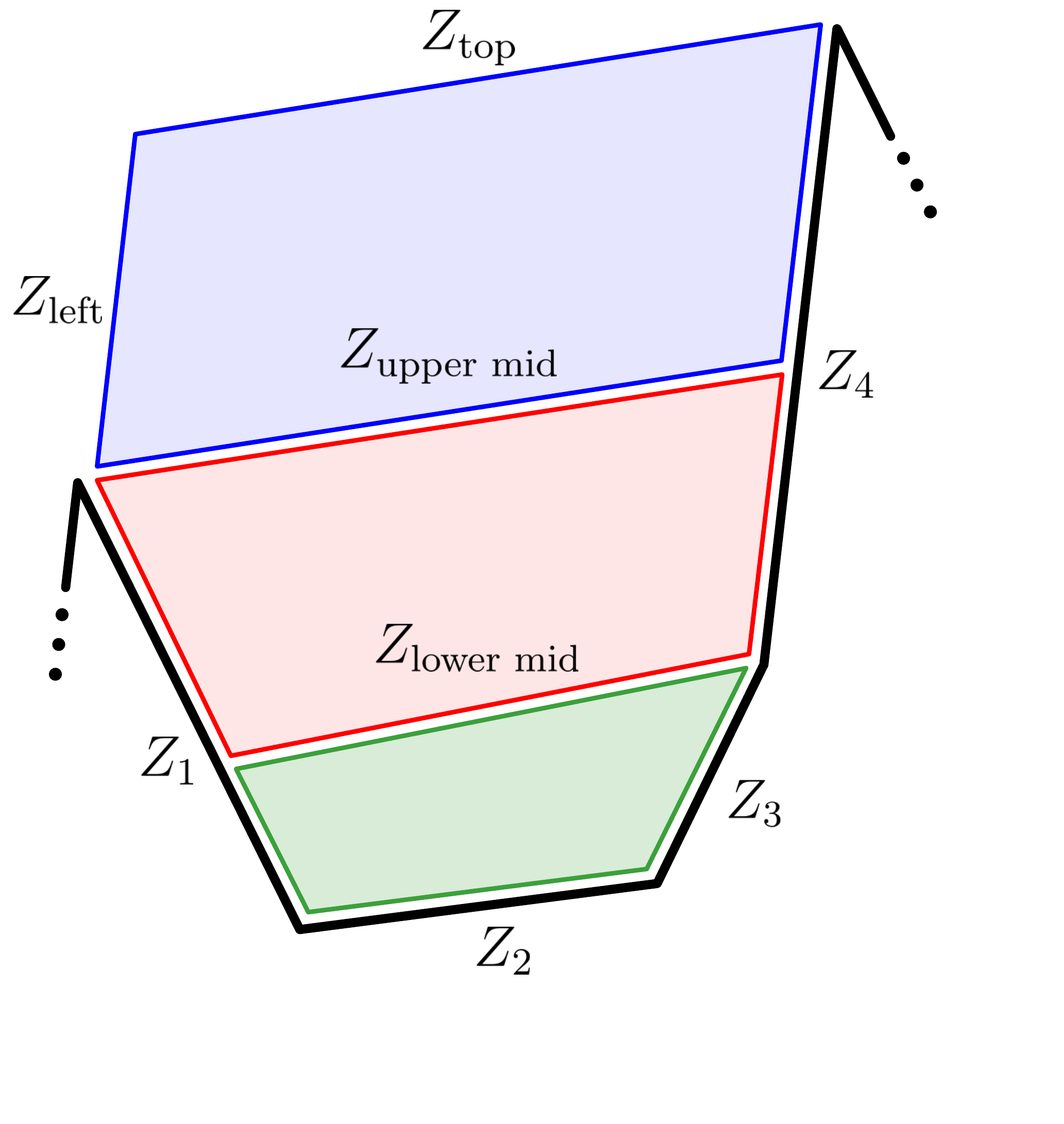}
  \end{minipage}
  \caption{OPE parametrizations of two-, three-, and four-point form factors. The blue square is the reference square. The red square is deformed with an arbitrary group element that leaves the blue square intact, while the green square is deformed by both the symmetries of the blue and the red squares.}\label{fig:squares}
  \label{twistparam}
\end{figure}

In this appendix we give explicit parametrizations of the momentum twistors in terms of the OPE variables in two-, three- and four-point cases. The geometry of the corresponding Wilson loop configurations is presented in figure \ref{twistparam}.
\subsection{Two-point form factor}
The centerpiece of the construction is the reference square, formed by four twistors that can be arbitrarily chosen to have the following form,
\begin{equation}\label{refsq}
\begin{aligned}
&Z_\text{left} =\left(0,0,0,1\right), && Z_\text{right} =\left(0,0,1,0\right),\\
&Z_\text{top} =\left(1,0,0,0\right), && Z_\text{bot} =\left(0,1,0,0\right).
\end{aligned}
\end{equation}
One segment of the Wilson loop is formed by the $Z_\text{bot}$ and $Z_\text{right}$ twistors, while twistors in the subsequent periods can be obtained by acting with the periodic shift operator $P$: $Z^{[n]}_i = P^n\cdot Z_i$. This, in particular, implies that
\begin{align}
Z_\text{right} = P\cdot Z_\text{left}\ ,\qquad Z_\text{top} = P\cdot Z_\text{bot}\ .
\end{align}
One particular choice of $P$ that is consistent with this constraint is
\begin{align}
P=\left(\begin{array}{cccc}
2 & 1 & 0 & 0 \\
-1 & 0 & 0 & 0 \\
0 & 0 & 2 & 1 \\
0 & 0 & -1 & 0
\end{array}\right)\ .
\end{align}
The two-dimensional space of twistors invariant under the action of $P$ represents the point at infinity and, therefore, enters the definition of the two-bracket,
\begin{align}\label{Pmatrix}
P\cdot\mathcal{I}_i = \mathcal{I}_i\ , \qquad \langle ij\rangle \equiv \langle \mathcal{I}_1\mathcal{I}_2ij\rangle\ .
\end{align}
Explicitly,
\begin{align}\label{Itwistor}
\mathcal{I}_1 = \left(1,-1,0,0\right),\qquad \mathcal{I}_2 = \left(0,0,1,-1\right).
\end{align}
Under this definition, the two-bracket is manifestly shift invariant: $\langle i^{[n]}j^{[m]}\rangle = \langle ij\rangle$.
\subsection{Three-point form factor}
The three conformal symmetries that leave the reference square (\ref{refsq}) invariant can be encoded into the following matrix
\begin{equation}
M = \left(\begin{array}{cccc}
   e^{-\tau+\frac{i\phi}{2}} & 0 & 0 & 0 \\
   0 & e^{\tau+\frac{i\phi}{2}} & 0 & 0 \\
   0 & 0 & e^{\sigma-\frac{i\phi}{2}} & 0 \\
   0 & 0 & 0 & e^{-\sigma-\frac{i\phi}{2}}
\end{array}\right).
\end{equation}
In order to obtain the most general three-point configuration, one has to deform the twistors of the red square in figure \ref{twistparam} with this matrix. We arrive at the following expressions for the twistors that form the Wilson loop,
\begin{equation}
\begin{aligned}
Z_1 = M\cdot\left(0,1,0,1\right),\quad Z_2 = M\cdot\left(1,1,-1,0\right),\quad Z_3 = \left(0,0,1,0\right),
\end{aligned}
\end{equation}
while the internal twistor $Z_\text{mid}$ is simply equal to $Z_\text{bot}$.\par
The three OPE variables that parametrize $M$ can be recast in the form that is common in the amplitudes literature,
\begin{equation}\label{3D_u}
\begin{aligned}
u_1 &= \frac{\langle1231^{+}\rangle\,\langle31\rangle}{\langle3^{-}131^{+}\rangle\,\langle12\rangle} = \frac{1}{1+e^{2\sigma}+e^{-2\tau}}\ ,\\
u_2 &= \frac{\langle231^{+}2^{+}\rangle\,\langle31\rangle^2}{\langle3^{-}131^{+}\rangle\,\langle12\rangle\,\langle23\rangle} = \frac{e^{2\sigma}}{\left(1+e^{-2\tau}\right)\left(1+e^{2\sigma}+e^{-2\tau}\right)}\ ,\\
u_3 &= \frac{\langle31^+2^+3^+\rangle\,\langle31\rangle}{\langle3^{-}131^{+}\rangle\,\langle23\rangle} = \frac{1}{1+e^{2\tau}}\ .
\end{aligned}
\end{equation}
As expected, the dependence on the angular variable $\phi$ does not enter any of the cross-ratios.
\subsection{Four-point form factor}\label{app:4p-twistors}
In the four-point case, we require an additional matrix that encodes the symmetries of the red square,
\begin{equation}
M_b = \left(\begin{array}{cccc}
   e^{\tau_b+\frac{i\phi_b}{2}} & 0 & 0 & 0 \\
   e^{\tau_b+\frac{i\phi_b}{2}} - e^{-\tau_b+\frac{i\phi_b}{2}} & e^{-\tau_b+\frac{i\phi_b}{2}} & 0 & e^{\sigma_b-\frac{i\phi_b}{2}} - e^{-\tau_b+\frac{i\phi_b}{2}} \\
   e^{-\sigma_b-\frac{i\phi_b}{2}} - e^{\tau_b+\frac{i\phi_b}{2}} & 0 & e^{-\sigma_b-\frac{i\phi_b}{2}} & 0 \\
   0 & 0 & 0 & e^{\sigma_b-\frac{i\phi_b}{2}}
\end{array}\right).
\end{equation}
The Wilson line is formed by the following four twistors and their periodic images,
\begin{equation}
\begin{aligned}
&Z_1 = M\cdot\left(0,1,0,1\right),&&Z_2 = M\cdot M_b\cdot\left(1,3,-1,1\right),\\
&Z_3 = M\cdot M_b\cdot\left(1,1,-2,0\right), &&Z_4 = \left(0,0,1,0\right),
\end{aligned}
\end{equation}
while the twistors in the middle, which separate the squares, are given by
\begin{equation}
\begin{aligned}
Z_\text{upper mid} = \left(0,1,0,0\right),\quad Z_\text{lower mid} = M\cdot\left(1,1,-1,0\right).
\end{aligned}
\end{equation}
Given this parametization we can introduce a set of kinematical variables $u_i$, $v_i$,
\begin{equation}\label{4D_u_and_v}
\begin{aligned}
u_1 &= \frac{\langle1234\rangle\,\langle41\rangle^2}{\langle4^{-}141^{+}\rangle\,\langle12\rangle\,\langle34\rangle} = \frac{e^{-2\tau_b}}{\left(1+e^{2\tau}\right)\left(1+e^{2\sigma}+e^{-2\tau}+e^{-2\tau_b}\right)}\ ,\\
u_2 &= \frac{\langle2341^{+}\rangle\,\langle41\rangle}{\langle4^{-}141^{+}\rangle\,\langle23\rangle}\\
&= \left[1+e^{-2\tau}+e^{2(\sigma-\sigma_b)}\left(1+e^{2\sigma_b}+e^{-2\tau}+e^{-2\tau_b}+2\,e^{\sigma_b-\tau_b}\cos{\phi_b}\right)\right]^{-1}\ ,\\
u_3 &= \frac{\langle341^{+}2^{+}\rangle\,\langle41\rangle^2}{\langle4^{-}141^{+}\rangle\,\langle12\rangle\,\langle34\rangle} = e^{2(\sigma+\tau+\tau_b)}\,u_1\ ,\\
u_4 &= \frac{\langle41^{+}2^{+}3^{+}\rangle\,\langle41\rangle}{\langle4^{-}141^{+}\rangle\,\langle23\rangle
} = e^{2(\sigma-\sigma_b-\tau)}\,u_2\ ,
\end{aligned}
\end{equation}
as well as
\begin{equation}
\begin{aligned}
&v_1 = \frac{\langle1241^{+}\rangle\,\langle41\rangle}{\langle4^{-}141^{+}\rangle\,\langle12\rangle} = 1+u_1-u_3-v_4\ ,\\
&v_2 = \frac{\langle231^{+}2^{+}\rangle\,\langle41\rangle}{\langle4^{-}141^{+}\rangle\,\langle23\rangle} = 1+u_2-u_4-v_1\ ,\\
&v_3 = \frac{\langle342^{+}3^{+}\rangle\,\langle41\rangle^2}{\langle4^{-}141^{+}\rangle\,\langle23\rangle\,\langle34\rangle} = 1+u_3-u_1-v_2\ ,\\
&v_4 = \frac{\langle41^{+}3^{+}4^{+}\rangle\,\langle41\rangle}{\langle4^{-}141^{+}\rangle\,\langle34\rangle} = \frac{1}{1+e^{2\tau}}\ .
\end{aligned}
\end{equation}

\section{Mirror and crossing}\label{appendix: crossing}

In this appendix, we summarize the properties of the tilted functions $f_{a,s}^{[\alpha]}$, defined in (\ref{tiltedf}), under the mirror transformations and illustrate how they translate into the crossing properties of the tilted transitions. Here, we closely follow similar analyses performed in refs.~\cite{Basso:2013pxa,Sever:2021xga}. The mirror transformation shifts the rapidity $u$ of a scalar excitation to $u^{\pm\gamma} = u\pm i$. Performing this deformation involves crossing one of the cuts that stretch between $-2g\pm\frac{i}{2}$ and $2g\pm\frac{i}{2}$, as a result of which we find ourselves on a different sheet of the Riemann surface after the transformation.\par
In order to perform this transformation, it is convenient to introduce an integral representation for $f^{[\alpha]}_{a,s}$ in terms of auxiliary functions $\gamma^v_{\alpha}$ and $\tilde{\gamma}^v_{\alpha}$,
\begin{align}\label{eq:gamma_def}
\gamma_\alpha^v = \frac{1}{1+\mathbb{K}(\alpha)}\,\kappa^{v}_\alpha\ , \qquad \tilde{\gamma}_\alpha^v = \frac{1}{1+\mathbb{K}(\alpha)}\,\tilde{\kappa}^{v}_\alpha\ .
\end{align}
These equations naturally define these objects as infinite-dimensional vectors in Bessel function space, but we can conveniently repackage them into functions,
\begin{align}
\gamma_\alpha^v(2gt) = \sum\limits_{n=1}^\infty 2n\,\gamma_{\alpha,n}^v\,J_n(2gt)\ ,\qquad \tilde{\gamma}_\alpha^v(2gt) = \sum\limits_{n=1}^\infty 2n\,\tilde{\gamma}_{\alpha,n}^v\,J_n(2gt)\ .
\end{align}
Due to the kernel acting differently on even and odd components, we are splitting these functions into the corresponding parts,
\begin{align}
\gamma_\alpha^v = \gamma_{\alpha,\bullet}^v + \gamma_{\alpha,\circ}^v\ ,\quad \tilde{\gamma}_\alpha^v = \tilde{\gamma}_{\alpha,\bullet}^v + \tilde{\gamma}_{\alpha,\circ}^v\ .
\end{align}
Equations (\ref{eq:gamma_def}) can be rewritten as the following set of integral equations,
\begin{equation}
\begin{aligned}
\gamma^v_{\alpha,2n+1} &+ \int\frac{dt}{t}\,J_{2n+1}(2gt)\,\frac{2\sin{\alpha}\cos{\alpha}\,\gamma_{\alpha,\bullet}^v(2gt) + 2\cos^2{\alpha}\,\gamma_{\alpha,\circ}^v(2gt)}{e^t-1} = \kappa_{\alpha, 2n+1}^v\ ,\\
\gamma^v_{\alpha,2n} &+ \int\frac{dt}{t}\,J_{2n}(2gt)\,\frac{2\cos^2{\alpha}\,\gamma_{\alpha,\bullet}^v(2gt) - 2\sin{\alpha}\cos{\alpha}\,\gamma_{\alpha,\circ}^v(2gt)}{e^t-1} = \kappa_{\alpha, 2n}^v\ ,
\end{aligned}
\end{equation}
and an identical one for $\tilde{\gamma}^v_\alpha$. These equations can be repackaged into the following set of cut-crossing identities,
\begin{equation}\label{cutcrossing}
\begin{aligned}
&\int\frac{dt}{t}\,\sin{(ut)}\left[\gamma^v_{\alpha,\circ}(2gt)\,\frac{e^t+\cos{2\alpha}}{e^t-1} + \gamma^v_{\alpha,\bullet}(2gt)\,\frac{\sin{2\alpha}}{e^t-1} + \sin{2\alpha}\,\frac{\cos{(vt)}\,e^{t/2}-J_0(2gt)}{e^t-1}\right]=0\ ,\\
&\int\frac{dt}{t}\cos_J{(ut)}\left[\gamma^v_{\alpha,\bullet}(2gt)\,\frac{e^t+\cos{2\alpha}}{e^t-1} - \gamma^v_{\alpha,\circ}(2gt)\,\frac{\sin{2\alpha}}{e^t-1} + 2\cos^2{\alpha}\,\frac{\cos{(vt)}\,e^{t/2}-J_0(2gt)}{e^t-1}\right]=0\ ,\\
&\int\frac{dt}{t}\,\sin{(ut)}\left[\tilde{\gamma}^v_{\alpha,\circ}(2gt)\,\frac{e^t+\cos{2\alpha}}{e^t-1} + \tilde{\gamma}^v_{\alpha,\bullet}(2gt)\,\frac{\sin{2\alpha}}{e^t-1} - 2\cos^2{\alpha}\,\frac{\sin{(vt)}\,e^{t/2}}{e^t-1}\right]=0\ ,\\
&\int\frac{dt}{t}\cos_J{(ut)}\left[\tilde{\gamma}^v_{\alpha,\bullet}(2gt)\,\frac{e^t+\cos{2\alpha}}{e^t-1} - \tilde{\gamma}^v_{\alpha,\circ}(2gt)\,\frac{\sin{2\alpha}}{e^t-1} + \sin{2\alpha}\,\frac{\sin{(vt)}\,e^{t/2}}{e^t-1}\right]=0\ ,
\end{aligned}
\end{equation}
which are valid inside the interval $u^2\leq4g^2$. Here, $\cos_J{(ut)}$ is a shorthand notation for $\cos{(ut)}-J_0(2gt)$.\par
Instead of using the symmetric and antisymmetric $f$-functions in eq.~(\ref{tiltedf}), for the purposes of the finite coupling analysis of this section we will instead use,
\begin{equation}\label{tiltedf1234}
\begin{aligned}
&f_1^{[\alpha]}(u,v)=\frac{1}{\cos^2{\alpha}}\,\tilde{\kappa}^u_\alpha\,\mathbb{Q}\,\frac{1}{1+\mathbb{K}(\alpha)}\,\kappa^v_\alpha\ ,\,\,\,
f_2^{[\alpha]}(u,v)=\frac{1}{\cos^2{\alpha}}\,\kappa^u_\alpha\,\mathbb{Q}\,\frac{1}{1+\mathbb{K}(\alpha)}\,\tilde{\kappa}^v_\alpha\ ,\\
&f_3^{[\alpha]}(u,v)=\frac{1}{\cos^2{\alpha}}\,\tilde{\kappa}^u_\alpha\,\mathbb{Q}\,\frac{1}{1+\mathbb{K}(\alpha)}\,\tilde{\kappa}^v_\alpha\ ,\,\,
f_4^{[\alpha]}(u,v)=\frac{1}{\cos^2{\alpha}}\,\kappa^u_\alpha\,\mathbb{Q}\,\frac{1}{1+\mathbb{K}(\alpha)}\,\kappa^v_\alpha\ ,
\end{aligned}
\end{equation}
with
\begin{align}
f_a^{[\alpha]}(u,v) = f_2^{[\alpha]}(u,v) - f_1^{[\alpha]}(u,v)\ , \qquad f_s^{[\alpha]}(u,v) = f_4^{[\alpha]}(u,v) - f_3^{[\alpha]}(u,v)\ .
\end{align}
Under the flip of the arguments, $f_{1,2,3,4}(v,u) = f_{2,1,3,4}(u,v)$. These functions admit the following representations in terms of $\gamma^v_\alpha$ and $\tilde{\gamma}^v_\alpha$,
\begin{equation}
\begin{aligned}
f_1^{[\alpha]}(u,v) &= \frac{1}{\cos{\alpha}}\int\frac{dt}{t}\,\frac{\sin{(ut)}\,e^{t/2}}{e^t-1}\left(\cos{\alpha}\,\gamma^v_{\alpha,\circ}(2gt) + \sin{\alpha}\,\gamma^v_{\alpha,\bullet}(2gt)\right),\\
f_2^{[\alpha]}(u,v) &= \frac{1}{\cos{\alpha}}\int\frac{dt}{t}\,\frac{\cos{(ut)}\,e^{t/2}-J_0(2gt)}{e^t-1}\left(\cos{\alpha}\,\tilde{\gamma}^v_{\alpha,\bullet}(2gt) - \sin{\alpha}\,\tilde{\gamma}^v_{\alpha,\circ}(2gt)\right),\\
f_3^{[\alpha]}(u,v) &= \frac{1}{\cos{\alpha}}\int\frac{dt}{t}\,\frac{\sin{(ut)}\,e^{t/2}}{e^t-1}\left(\cos{\alpha}\,\tilde{\gamma}^v_{\alpha,\circ}(2gt) + \sin{\alpha}\,\tilde{\gamma}^v_{\alpha,\bullet}(2gt)\right),\\
f_4^{[\alpha]}(u,v) &= \frac{1}{\cos{\alpha}}\int\frac{dt}{t}\,\frac{\cos{(ut)}\,e^{t/2}-J_0(2gt)}{e^t-1}\left(\cos{\alpha}\,\gamma^v_{\alpha,\bullet}(2gt) - \sin{\alpha}\,\gamma^v_{\alpha,\circ}(2gt)\right).
\end{aligned}
\end{equation}
Now, to perform the mirror transformation, we follow the standard procedure of shifting the rapidity $u$ all the way to the cut, crossing the cut using the identities (\ref{cutcrossing}), and finally shifting $u$ the rest of the way to $u^{\pm\gamma}$. As a result of this procedure, we find the following transformation properties for the functions $f_{1,2,3,4}^{[\alpha]}$,
\begin{align}
f_1^{[\alpha]}(u^{\pm\gamma}&,v) = - \cos{2\alpha}\,f_1^{[\alpha]}(u,v) \mp i \sin{2\alpha}\,f_4^{[\alpha]}(u,v)\nonumber\\ 
&\mp i \sin{2\alpha} \int\frac{dt}{t}\,\frac{\cos{(vt)}\,e^{t/2}-J_0(2gt)}{e^t-1}\left(e^{\mp iu^\pm t}-J_0(2gt)\right),\nonumber\\
f_2^{[\alpha]}(u^{\pm\gamma}&,v) = - \cos{2\alpha}\,f_2^{[\alpha]}(u,v) \mp i \sin{2\alpha}\,f_3^{[\alpha]}(u,v)\nonumber\\
&- \sin{2\alpha} \int\frac{dt}{t}\,\frac{\sin{(vt)}\,e^{t/2}}{e^t-1}\left(e^{\mp iu^\pm t}-J_0(2gt)\right),\\
f_3^{[\alpha]}(u^{\pm\gamma}&,v) = - \cos{2\alpha}\,f_3^{[\alpha]}(u,v) \mp i \sin{2\alpha}\,f_2^{[\alpha]}(u,v)\nonumber\\
&\mp 2i\int\frac{dt}{t}\,\frac{\sin{(vt)}\,e^{t/2}}{e^t-1}\left[\left(\cos{(u^\pm t)} - J_0(2gt)\right)\sin^2{\alpha} \pm i\sin{(u^\pm t)}\cos^2{\alpha}\right],\nonumber\\
f_4^{[\alpha]}(u^{\pm\gamma}&,v) = - \cos{2\alpha}\,f_4^{[\alpha]}(u,v) \mp i \sin{2\alpha}\,f_1^{[\alpha]}(u,v)\nonumber\\
&- 2\int\frac{dt}{t}\,\frac{\cos{(vt)}\,e^{t/2}-J_0(2gt)}{e^t-1}\left[\left(\cos{(u^\pm t)} - J_0(2gt)\right)\cos^2{\alpha} \pm i\sin{(u^\pm t)}\sin^2{\alpha}\right],\nonumber
\end{align}
where in the right-hand side $u^{\pm} = u\pm i/2$. In terms of the antisymmetric and symmetric scattering phases,
\begin{align}
f_a^{[\alpha]}(u^{\pm\gamma}&,v) = - \cos{2\alpha}\,f_a^{[\alpha]}(u,v) \pm i \sin{2\alpha}\,f_s^{[\alpha]}(u,v)\pm i \sin{2\alpha}\,\mathcal{J}(\pm u,\pm v)\ ,\\
f_s^{[\alpha]}(u^{\pm\gamma}&,v) = - \cos{2\alpha}\,f_s^{[\alpha]}(u,v) \pm i \sin{2\alpha}\,f_a^{[\alpha]}(u,v) - \mathcal{J}(\mp u^{[\pm 2]},\mp v) -\cos{2\alpha}\,\mathcal{J}(\pm u,\pm v)\ .\nonumber
\end{align}
with $u^{[\pm 2]} = u\pm i$ and
\begin{equation}
\begin{aligned}
\mathcal{J}(u,v) &= \int\frac{dt}{t}\,\frac{\left(e^{t/2- iut}-J_0(2gt)\right)\left(e^{t/2+ ivt}-J_0(2gt)\right)}{e^t-1}\\
&=\log{\left[\frac{\Gamma(iu-iv)}{\Gamma(\frac{1}{2}+iu)\Gamma(\frac{1}{2}-iv)}\right]} + J_{\phi}(u) + J_{\phi}(-v)\ ,
\end{aligned} 
\end{equation}
which satisfies $\mathcal{J}(-u,-v)=\mathcal{J}(v,u)$. It also happens to coincide with the logarithm of the simple part of the transition (\ref{tiltedP}), not captured by $f_{a}$ and $f_s$. One can see that up to inhomogeneous terms $\propto \mathcal{J}$, the mirror transformation is equivalent to the combination of a rotation by $2\alpha$ and a parity transformation in the plane $(f_s,if_a)$,
\begin{equation}
\left(\begin{array}{c}f_s\\if_a\end{array}\right)\overset{\pm \gamma}{\longrightarrow}-\left(\begin{array}{cc}
    \cos{2\alpha} & \mp\sin{2\alpha} \\
    \pm\sin{2\alpha} & \cos{2\alpha}
\end{array}\right)\left(\begin{array}{c}f_s\\if_a\end{array}\right) +\ldots\ .
\end{equation}
When performing mirror transformations on the tilted transition (\ref{tiltedP}), the simple part of it, given by the exponent of $\mathcal{J}(u,v)$, also needs to be transformed. The simplest move to consider is the inverse mirror rotation $u \rightarrow u^{-\gamma}$. As one can see from its integral representation, $\mathcal{J}(u, v)$ is an analytic function of $u$ for $\textrm{Im}\,  (u) < 1/2$. Hence, there is no cut to cross and we simply get%
\footnote{Moving the rapidity in the opposite direction is more complicated, due to the presence of a cut at $\textrm{Im}\, u = 1/2$. It results in the more involved relation
\begin{equation}
    \mathcal{J}(u^{\gamma}, v) = \mathcal{J}(u, v) - \mathcal{J}(v, u) + \mathcal{J}(v,u^{[+2]}) + \log{\left[\frac{g^2}{(u-v)(u-v+i)}\right]}\ .\nonumber
\end{equation}
}
\begin{equation}
    \mathcal{J}(u^{-\gamma}, v) = \mathcal{J}(u^{[-2]}, v)\ .
\end{equation}
This term neatly cancels when combined with $f_s^{[\alpha]}(u^{-\gamma}, v)$. Putting all pieces together, we arrive at the following compact representation for the mirror transformation of the tilted transition~\eqref{tiltedP}
\begin{equation}\label{mirror_Palpha}
\begin{aligned}
    &\log\left[g^2 P^{[\alpha]}(u^{-\gamma}|v)\right] = if_{a}^{[\pi/4]}(u^{-\gamma}, v) + f_{s}^{[\alpha]}(u^{-\gamma}, v) + \mathcal{J}(u^{-\gamma}, v) \\
    & \,\, = f_{s}^{[\pi/4]}(u, v)-\cos{2\alpha}\,f_{s}^{[\alpha]}(u, v) -i\sin{2\alpha}\,f_{a}^{[\alpha]}(u, v) + 2\sin^2\alpha\,\mathcal{J}(v, u)\ ,
\end{aligned}
\end{equation}
using $\mathcal{J}(-u, -v)= \mathcal{J}(v, u)$. We may also consider the transformation under $v \rightarrow v^{\gamma}$. This transformation follows from the above equations after recalling the permutation properties of $f_{a}$ and $f_{s}$,
\begin{equation}
    f_{s}^{[\alpha]} (u, v^{\gamma}) = f_{s}^{[\alpha]} (v^{\gamma}, u)\ , \qquad f_{a}^{[\alpha]} (u, v^{\gamma}) = -f_{a}^{[\alpha]} (v^{\gamma}, u)\ .
\end{equation}
Using these, one can verify that the $v^{+\gamma}$ move is equivalent to the $u^{-\gamma}$ one,
\begin{equation}
    P^{[\alpha]}_{\phi\phi}(u^{-\gamma}|v) = P^{[\alpha]}_{\phi\phi}(u|v^{\gamma})\ .
\end{equation}
This equation holds true for any rapidities $u, v$ and may be equivalently written as
\begin{equation}\label{mirror_genral_alpha}
    P^{[\alpha]}_{\phi\phi}(u|v) = P^{[\alpha]}_{\phi\phi}(u^{\gamma}|v^{\gamma})\ ,
\end{equation}
after replacing $u \rightarrow u^{\gamma}$ on both sides. This is the mirror symmetry axiom. Notice that it holds for any $\alpha$.\par
In the particular case of $\alpha = \pi/4$, mirror transformation (\ref{mirror_Palpha}) takes the transition back to its original form, up to a permutation of the rapidities. That is, we verify that
\begin{equation}
    P_{\phi\phi}(u^{-\gamma}|v) = P_{\phi\phi}(v|u)\ ,
\end{equation}
for $P_{\phi\phi} = P^{[\pi/4]}_{\phi\phi}$, as it should be for a pentagon. The other case of interest corresponds to $\alpha = 0$, for which the transition does not map back into itself. Yet, its mirror image takes a very simple form
\begin{equation}
    Q_{\phi\phi}(u^{-\gamma}|v) = g^{-2} \exp{\left[f_{s}^{[\pi/4]}(u, v) -f_{s}^{[0]}(u, v)\right]}\ . 
\end{equation}
Importantly, this function is symmetric under permutation of $u$ and $v$,
\begin{equation}\label{app:mirror-Q}
    Q_{\phi\phi}(u^{-\gamma}|v) = Q_{\phi\phi}(v^{-\gamma}|u) \ .
\end{equation}
Combined with eq.~\eqref{mirror_genral_alpha} it leads to the crossing relation
\begin{equation}\label{app:crossing-Q}
    Q_{\phi\phi}(u^{-2\gamma}|v) = Q_{\phi\phi}(v^{-\gamma}|u^{-\gamma}) = Q_{\phi\phi}(v|u)\ .
\end{equation}
A slightly more complicated relation may be written for the crossing transformation in the opposite direction, $Q_{\phi\phi}(u^{2\gamma}|v)$. To derive it, one may first apply Watson equation,
\begin{equation}
    Q_{\phi\phi}(u^{2\gamma}|v) = S_{\phi\phi}(u^{2\gamma}, v)\,Q_{\phi\phi}(v|u^{2\gamma})\ .
\end{equation}
Taking then into account mirror symmetry, $Q_{\phi\phi}(v|u^{2\gamma}) = Q_{\phi\phi}(v^{-2\gamma}|u)$, eq.~\eqref{app:crossing-Q}, and the crossing-unitary relations of the flux-tube S-matrix,
\begin{equation}
    S_{\phi\phi}(u, v)\,S_{\phi\phi}(v, u) = 1\ , \qquad S_{\phi\phi}(u^{2\gamma}, v)\,S(u, v) = \frac{u-v}{u-v+2i}\ ,
\end{equation}
one finds
\begin{equation}
\begin{aligned}
    Q_{\phi\phi}(u^{2\gamma}|v) = \frac{u-v}{u-v+2i}\,Q_{\phi\phi}(v|u)\ .
\end{aligned}
\end{equation}
This identity is the key formula derived in ref.~\cite{Sever:2021xga} for proving the crossing symmetry of the singlet form factor transition in eq.~\eqref{eq:singlet-F-to-Q}.

\bibliographystyle{utphys2}
\bibliography{biblio}

\end{document}